\PassOptionsToPackage{table}{xcolor}
\documentclass[acmsmall]{acmart}

\AtBeginDocument{%
  \providecommand\BibTeX{{%
    \normalfont B\kern-0.5em{\scshape i\kern-0.25em b}\kern-0.8em\TeX}}}

\setcopyright{acmcopyright}
\copyrightyear{2023}
\acmYear{2023}
\acmDOI{XXXXXXX.XXXXXXX}

\acmJournal{CSUR}
\acmVolume{1}
\acmNumber{1}
\acmArticle{1}
\acmMonth{6}

\usepackage{amsmath}
\usepackage[]{graphicx}
\DeclareGraphicsExtensions{.pdf,.jpeg,.png}
\usepackage{subfig}
\usepackage{xspace}
\usepackage{paralist}
\usepackage{multirow}
\usepackage{enumitem}
\usepackage{forest}
\usepackage{adjustbox}
\usepackage{array}
\usepackage{makecell}
\usepackage{xcolor}

\newcommand{\fakeparagraph}[1]{\vspace{1mm}\noindent\textit{#1.}}

\newcommand{\fig}[1]{Fig.~\ref{fig:#1}}
\newcommand{\tab}[1]{Table~\ref{tab:#1}}
\newcommand{\s}[1]{Sec.~\ref{sec:#1}}
\newcommand{\secref}[1]{\S\ref{sec:#1}}

\newcolumntype{L}{>{\raggedright\arraybackslash}m{1.7cm}}
\newcolumntype{C}{>{\centering\arraybackslash}m{2cm}}
\newcolumntype{M}{>{\raggedright\arraybackslash}m{2.7cm}}
\newcolumntype{D}{>{\centering\arraybackslash}m{3cm}}

\newcommand{\sysdep}{sys dep}
\newcommand{\classdep}{class dep}
\newcommand{\exc}{$\text{*}$}
\newcommand{\stproc}{$^{\text{SP}}$}
\newcommand{\batch}{$^{\text{B}}$}
\newcommand{\stream}{$^{\text{S}}$}
\newcommand{\detjob}{DC\xspace}
\newcommand{\seq}{SEQ\xspace}
\newcommand{\onewrite}{1W\xspace}
\newcommand{\oneportion}{1P\xspace}
\newcommand{\durstore}{STOR\xspace}
\newcommand{\durrepl}{REPL\xspace}
\newcommand{\replaysource}{REPLAY\xspace}

\newcommand{\extendedonly}[1]{#1}

\begin{document}

\title{A Model and Survey of Distributed Data-Intensive Systems}

\author{Alessandro Margara}
\email{alessandro.margara@polimi.com}
\affiliation{%
  \institution{Politecnico di Milano}
  \country{Italy}
}
\author{Gianpaolo Cugola}
\email{gianpaolo.cugola@polimi.com}
\affiliation{%
  \institution{Politecnico di Milano}
  \country{Italy}
}
\author{Nicol\`{o} Felicioni}
\email{nicolo.felicioni@polimi.com}
\affiliation{%
  \institution{Politecnico di Milano}
  \country{Italy}
}
\author{Stefano Cilloni}
\email{stefano.cilloni@mail.polimi.com}
\affiliation{%
  \institution{Politecnico di Milano}
  \country{Italy}
}

\renewcommand{\shortauthors}{Margara et al.}

\begin{abstract}
  Data is a precious resource in today's society, and is generated at an
  unprecedented and constantly growing pace.  The need to store, analyze, and
  make data promptly available to a multitude of users introduces formidable
  challenges in modern software platforms.
  These challenges radically impacted the research fields that gravitate
  around data management and processing, with the introduction of
  \emph{distributed data-intensive systems} that offer innovative programming
  models and implementation strategies to handle data characteristics such as
  its volume, the rate at which it is produced, its heterogeneity, and its
  distribution.
  Each data-intensive system brings its specific choices in terms of data
  model, usage assumptions, synchronization, processing strategy, deployment,
  guarantees in terms of consistency, fault tolerance, ordering.
  Yet, the problems data-intensive systems face and the solutions they propose
  are frequently overlapping.
  This paper proposes a unifying model that dissects the core functionalities
  of data-intensive systems, and discusses alternative design and
  implementation strategies, pointing out their assumptions and implications.
  The model offers a common ground to understand and compare highly
  heterogeneous solutions, with the potential of fostering cross-fertilization
  across research communities.
  We apply our model by classifying tens of systems: an exercise that brings to
  interesting observations on the current trends in the domain of
  data-intensive systems and suggests open research directions.
\end{abstract}



\received{19 March 2022}
\received[revised]{25 November 2022}
\received[accepted]{31 May 2023}

\maketitle

\section{Introduction}
\label{sec:intro}

As data guides the decision-making process of increasingly many human
activities, software applications become
\emph{data-intensive}~\cite{kleppmann:2016:data-intensive}.
They handle large amounts of data produced by disparate sources.  They perform
complex data analysis to extract valuable knowledge from the application
environment.  They take automated decisions in near real-time.  They serve
content to a multitude of users, spread over wide geographical areas.
The challenges for these applications come from data characteristics such as
its volume, the rate at which it is generated, its heterogeneity. This demands
for \emph{distributed software systems} that could exploit the resources of
interconnected computers to efficiently store, query, analyze, and serve data
to customers at scale.
Distribution brings along issues related to communication, concurrency and
synchronization, deployment of data and computational tasks on physical nodes,
replication and consistency, handling of partial
failures~\cite{kleppmann:2016:data-intensive}.
To address these issues, the last two decades have seen a flourishing of
systems and execution models that abstract away some of the concerns related
to distributed data management and processing.
We collectively denote them \emph{distributed data-intensive systems}: they
originate from research and development efforts in various communities, in
particular those working on database and distributed systems.

\fakeparagraph{Background: different research lines addressing overlapping
  problems}
In the database community, the mutating requirements brought by data-intensive
applications put the traditional (relational) data model and implementation
strategies under question.
The increasing complexity and volume of data demanded for flexibility and
scalability.  Internet-scale applications demanded for
geographically-replicated stores, supporting a multitude of users concurrently
reading and updating data~\cite{ajoux:HOTOS:2015:Challenges}.  In these
contexts, communication and synchronization costs may become the main
bottleneck~\cite{bailis:2013:VLDB:HAT}.  Workload characteristics changed as
well: analytical tasks emerged and complemented query
tasks~\cite{stonebraker:CACM:2010:SQLvsNoSQL}.
In response to these challenges, researchers first investigated so called
NoSQL solutions~\cite{davoudian:CSur:2018:NoSQL} that trade strong consistency
and transactional guarantees in favor of flexibility and scalability.
More recently, the complexity of writing applications with weak guarantees
inspired a renaissance of transactional
semantics~\cite{stonebraker:CACM:2012:NewSQL}, coupled with programming,
design, and implementation approaches to make transactions management more
scalable~\cite{stonebraker:VLDB:2007:hstore, corbett:TOCS:2013:Spanner,
  stonebraker:IEEEB:2013:voltdb, thomson:SIGMOD:2012:Calvin}.

In parallel, within the distributed systems research community, the increasing
centrality of data fostered the development of new systems that exploit the
compute capabilities of cluster infrastructures to extract valuable
information from this large amount of data.
Pioneered by MapReduce~\cite{dean:CACM:2008:mapreduce}, they organize the
computation into a dataflow graph of operators that apply functional
transformations on their input data.  This dataflow model promotes distributed
computations: operators may run simultaneously on different machines (task
parallelism), and multiple instances of each operator may process independent
portions of the input data in parallel (data parallelism).  Developers only
specify the behavior of operators, while the system automates their
deployment, the exchange of data, and the re-execution of lost computations
due to failures.
Over the years, a multitude of systems adopted and revised this processing
model in terms of programming abstractions (e.g., support for streaming data
and iterative computations), as well as design and implementation choices
(e.g., strategies to associate operators to physical machines and to exchange
data across operators)~\cite{carbone:IEEEB:2015:flink,
  zaharia:CACM:2016:spark}, while some systems brought alternative programming
models suited to specific domains, such as graph
processing~\cite{mcCune:CSur:2015:TLaV}.

In general, the challenging demands of data-intensive applications are
continuously pushing researchers and practitioners to build novel solutions
that go beyond traditional
categories~\cite{stonebraker:ICDE:2005:one_size_fits_all}.
For instance, many dataflow platforms offer abstractions to process data
through complex relational queries, thus crossing the boundaries of database
technologies.
Data stores such as VoltDB~\cite{stonebraker:IEEEB:2013:voltdb} and
S-Store~\cite{cetintemel:VLDB:2014:sstore} aim to support near real-time
processing of incoming data within a relational database core.
Messaging services such as Kafka~\cite{kreps:NetDB:2011:kafka} offer
persistency, querying, and processing
capabilities~\cite{bejeck:2018:KafkaStreams}.

\fakeparagraph{Motivations}
In summary, a multitude of distributed data-intensive systems proliferated
over the years.
The problems they face and the solutions they propose are frequently
overlapping, but the commonalities in design principles and implementation
choices often hide behind concrete realizations and heterogeneous
terminologies adopted by different research communities.
In this scenario, it is difficult for users to grasp the subtle differences
between systems, evaluate their benefits and limitations, capture their
assumptions and guarantees, and select the most suitable ones for a given
application.
Also, it is hard for researchers to get a coherent and comprehensive view of
the area, identify salient design choices, and understand their consequences
in terms of functionalities and performance.

\fakeparagraph{Contributions and methodological approach}
To address these issues, this paper proposes \emph{a model} for data-intensive
systems that integrates key design and implementation choices into a unifying
view.  This model captures our view of a data-intensive system as a collection
of abstract components that together provide the system functionalities.  For
each component, we define:
\begin{inparaenum}[(i)]
\item the assumptions it relies upon;
\item the functionalities it provides;
\item the guarantees it offers;
\item the possible strategies for its design and implementation, highlighting
  their implications.
\end{inparaenum}
From these characteristics, we derive \emph{a list of classification criteria}
that we use to survey tens of state-of-the-art data-intensive systems,
organized into \emph{a taxonomy} that highlights their similarities and
differences.

To develop our model and compile our taxonomy and survey, we started from
works appearing in top tier conferences and journals in the area of database
and distributed systems in the last twenty years.  We then checked the systems
they were citing, including commercial products not presented in scientific
publications.  To remain focused on our goal of surveying existing systems, we
skipped works only presenting individual algorithms or mechanisms, and we
concentrated on systems addressing data management and processing problems,
not compute-intensive problems such as computer simulations.  From this large
background of material and our own past experience in the area, we derived our
model and the relevant dimensions to classify systems.
Without pretending to embrace all possible systems and issues, we are
confident that this survey captures the key design strategies adopted in
currently available distributed data-intensive systems.

Our work contributes to the research areas that gravitate around
data-intensive systems in various ways:
\begin{inparaenum}[(i)]
\item it enables an unbiased comparison of their features;
\item it offers a broad view of the research in the field
\item it promotes cross-fertilization between communities, defining a
  vocabulary and conceptual framework that they can use to exchange ideas and
  solutions to common problems;
\item it highlights consolidated results and open challenges, and helps
  identifying promising research directions.
\end{inparaenum}

\paragraph{Paper outline}
The paper is organized as follows: \s{model} presents our model and list of
classification criteria.  Based on these criteria, \s{systems} proposes a
taxonomy to organize data-intensive systems in a small set of classes.  Next,
we apply the model to describe these classes: pure data management systems in
\s{systems_db}, pure data processing systems in \s{systems_proc}, and other
systems that propose new or hybrid programming and execution models in
\s{systems_other}.  \s{discussion} discusses key aspects that emerge from our
analysis and points out future research directions.  Finally, \s{conclusion}
concludes the paper.  In the appendices, we describe individual systems in
detail, we provide a summary of the terms and a map of the concepts that
appear in the paper, together with their mutual relations, and we report
related surveys and studies.


\section{A Unifying Model for Data-Intensive Systems}
\label{sec:model}

This section presents a model that captures the core functionalities of
data-intensive systems within a collection of abstract components.
From the model, we derive fine-grained classification criteria that we
summarize in Tables~\ref{tab:criteria:functional}--\ref{tab:criteria:reconf}.
We organize data-intensive systems into a taxonomy in \s{systems}, and we use
the classification criteria to describe the systems within this taxonomy in
Sec.~\ref{sec:systems_db}--\ref{sec:systems_other}.
In this section, coherently with the taxonomy in \s{systems}, when providing
concrete examples of systems to explain the concepts in the model, we denote
\emph{data management systems} (DMSs) those that are primarily designed to
store some state and expose functions to query and mutate such state, while we
denote \emph{data processing systems} (DPSs) those that enable expensive
transformation and analysis of static (batch) or dynamic (streaming) data.

\subsection{Functional model}
\label{sec:model:functional}

\begin{figure}[tb]
  \begin{minipage}{0.6\textwidth}
    \centering
    \includegraphics[width=0.94\textwidth]{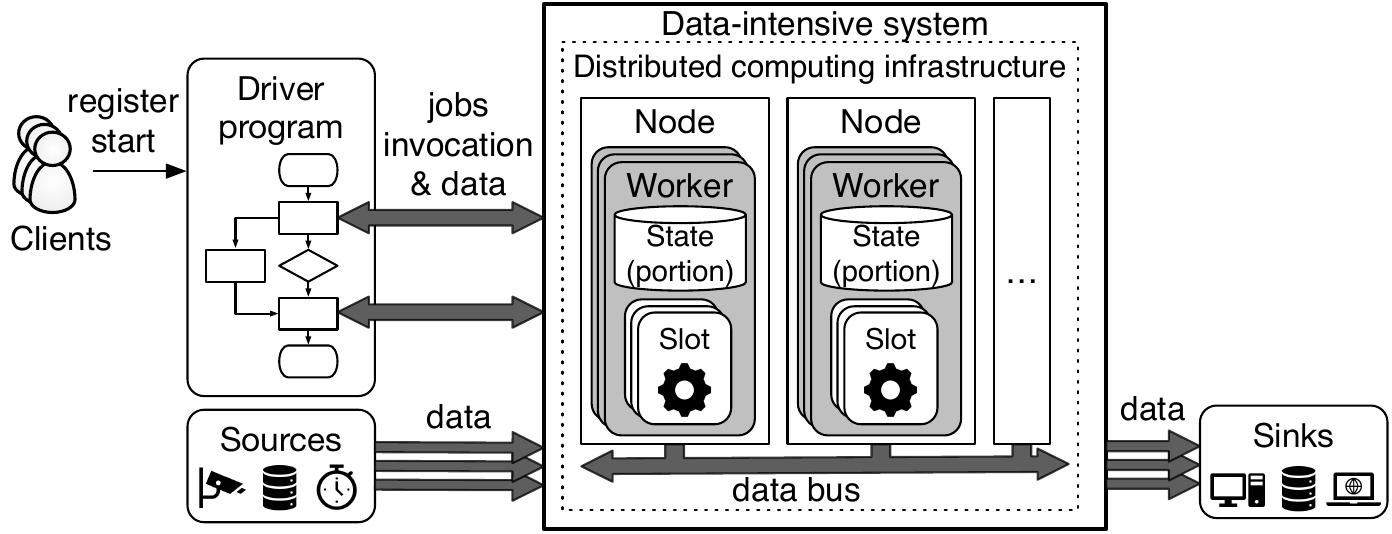}
    \captionof{figure}{Functional model of a data-intensive system.}
    \label{fig:model:functional}
  \end{minipage}
  \hfill
  \begin{minipage}{0.34\textwidth}
    \centering
    \tiny
    \rowcolors{1}{gray!25}{white}
    \renewcommand{\arraystretch}{0.94}
    \begin{tabular}[t]{|l|c|}
      \hline
      Driver exec. & client-side / system-side \\
      Driver exec. time & on registration / on start \\
      Invocation of jobs & synchronous / asynchronous \\
      Sources & no / passive / active / both \\
      Sinks & no / yes \\
      State & no / yes \\
      Deployment & cluster/wide area/hybrid \\
      \hline
    \end{tabular}
    \captionof{table}{Classification criteria: functional model.}
    \label{tab:criteria:functional}
  \end{minipage}
\end{figure}

\fig{model:functional} depicts the functional model of a data-intensive
system.  The resulting classification criteria are in
\tab{criteria:functional}.
A data-intensive system offers data management and processing functionalities
to external \emph{clients}.  Clients can register and start \emph{driver
  programs}, which are the parts of the application logic that interact with
the data-intensive system and exploit its functionalities.  Specifically,
during its execution, a driver program can \emph{invoke} one or more
\emph{jobs}, which are the largest units of execution that can be offloaded
onto the \emph{distributed computing infrastructure} made available by the
data-intensive system.
Depending on the specific system, a driver program may execute
\emph{client-side} or \emph{system-side}. Some systems decouple activation of
driver programs from their registration, in this case we say that driver
execution time is \emph{on start}, otherwise we say that it is \emph{on
  registration}.
To exemplify, in a DMS a driver program may be a stored procedure that
combines code expressed in some general purpose programming language with one
or more queries (the jobs) expressed in the language offered by the engine
(e.g., SQL). Stored procedures typically run system-side every time a client
activates them (on start).
%
%
Similarly, in a DPS, the driver program may be a piece of Java code that
spawns one or more distributed computations (the jobs) written using the API
offered by the data processing engine.  In this context, the driver program
will typically run system-side on registration.

The data-intensive system runs on the distributed computing infrastructure as
a set of \emph{worker} processes, hosted on the same or different \emph{nodes}
(physical or virtual machines).  We model the processing resources offered by
workers as a set of \emph{slots}.
Jobs are compiled into elementary units of execution that we denote as
\emph{tasks} and run sequentially on slots.
Jobs consume input \emph{data} and produce output data.  Some systems also
store some \emph{state} within the distributed computing infrastructure: in
this case, jobs may access (read and modify) the state during their execution.
When present, state can be split (partitioned and replicated) across workers,
such that each of them is responsible for a state \emph{portion}.

In our model, data elements are immutable and are distributed through
communication channels (dark gray arrows in \fig{model:functional} and
\fig{model:jobs_lifecycle}) that we collectively refer to as the \emph{data
  bus}.
Notice that the data bus also distributes \emph{jobs invocations}.  Indeed,
our model emphasizes the dual nature of invocations and data, which can both
carry information and trigger jobs execution: invocations may transport data
in the form of input parameters and return values, while the availability of
new data may trigger the activation of jobs.  Our model exploits this duality
to capture the heterogeneity in activating jobs and exchanging data of the
systems we surveyed.

Jobs invocations may be either \emph{synchronous}, if the driver program waits
for jobs completion before making progress, or \emph{asynchronous}, if the
driver program continues to execute after submitting the invocation.  In both
cases, invocations may return some result to the driver program, as indicated
by the bidirectional arrows in \fig{model:functional}.
In some systems, jobs also consume data from external \emph{sources} and
produce data for external \emph{sinks}.  We distinguish between \emph{passive}
sources, which consist of static datasets that jobs can access during their
execution (for instance, a distributed filesystem), and
\emph{active} sources, which produce new data dynamically and may trigger job
execution (for instance, a messaging system).

To exemplify, stored procedures (the driver programs) in a DMS invoke
(synchronously or asynchronously, depending on the specific system) one or
more queries (the jobs) during their execution.  Invocations carry input data
in the form of actual parameters.  Queries can access (read-only queries) and
modify (read-write queries) the state of the system, and return query results.
In batch DPSs such as MapReduce, jobs read input data from passive sources
(for instance, a distributed filesystem), apply functional transformations
that do not involve any mutable state, and store the resulting data into sinks
(for instance, the same distributed filesystem).
In stream DPSs, jobs run indefinitely and make progress when active sources
provide new input data. We say that input data \emph{activates} a job, and in
this case jobs may preserve some state across activations.

We characterize the distributed computing infrastructure based on its
\emph{deployment}.  In a \emph{cluster} deployment all nodes belong to the
same cluster or data center, which provides high bandwidth and low latency for
communication.  Conversely, in a \emph{wide area} deployment nodes can be
spread in different geographical areas, a choice that increases the latency of
communication and may impact the possibility of synchronizing and coordinating
tasks.  For this reason we also consider \emph{hybrid} deployments, when the
system adopts a hierarchical approach, exploiting multiple fully-functional
cluster deployments that are loosely synchronized with each other.

\subsection{Jobs and their lifecycle: from definition to execution}
\label{sec:model:jobs}

\begin{figure}[hptb]
\begin{minipage}{0.56\textwidth}
  \centering
  \includegraphics[width=0.98\textwidth]{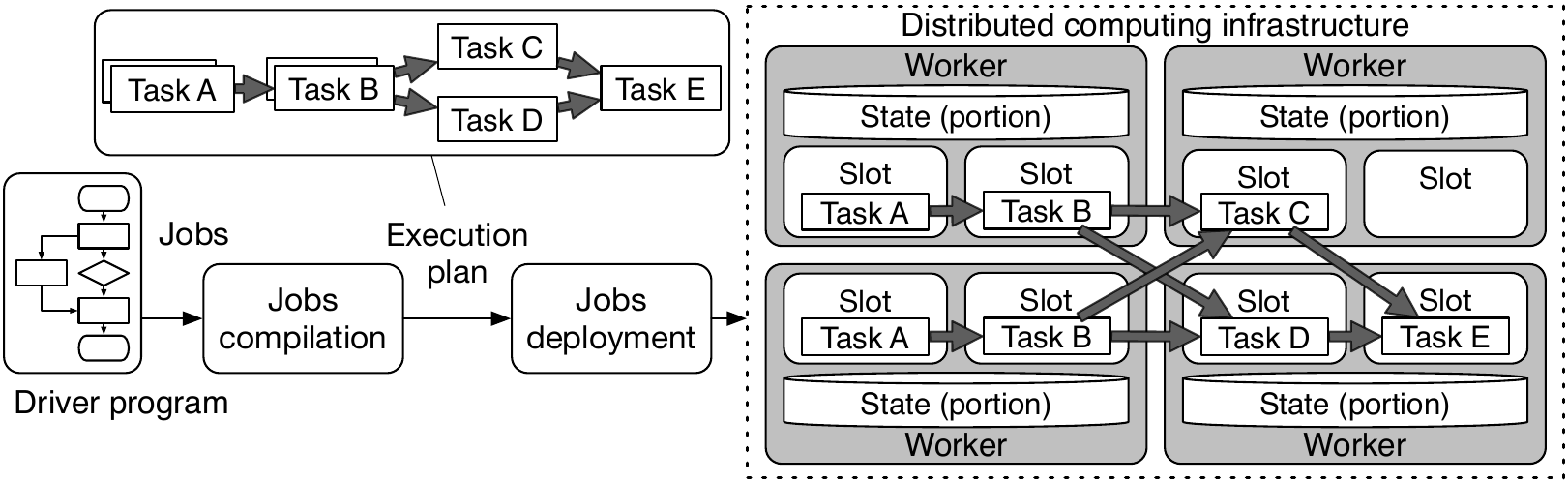}
  \captionof{figure}{Jobs definition, compilation, deployment, and execution.}
  \label{fig:model:jobs_lifecycle}
\end{minipage}
\hfill
\begin{minipage}{0.42\textwidth}
  \centering
  \tiny
  \rowcolors{2}{gray!25}{white}
  \renewcommand{\arraystretch}{0.94}    
  \begin{tabular}[t]{|l|C|}
    \hline
    \rowcolor{gray!50}
    \multicolumn{2}{|c|}{\bf Jobs definition} \\ \hline
    Jobs definition API & library / DSL \\
    Exec. plan definition & explicit / implicit \\
    Task communication & explicit / implicit \\
    Exec. plan structure & dataflow / workflow \\
    Iterations & no / yes \\
    Dynamic creation & no / yes \\
    Nature of jobs & one-shot / continuous \\
    State management & absent / explicit / implicit \\
    Data parallel API & no / yes \\
    Placement-aware API & no / yes \\
    \hline
    \rowcolor{gray!50}
    \multicolumn{2}{|c|}{\bf Jobs compilation} \\ \hline
    Jobs compil. time & on driver registration / on driver execution \\
    Use resources info & no / static / dynamic \\
    \hline
    \rowcolor{gray!50}
    \multicolumn{2}{|c|}{\bf Jobs deployment and execution} \\ \hline
    Granularity of deployment & job-level / task-level \\
    Deployment time & on job compilation / on task activation \\
    Use resources info & static / dynamic \\
    Management of res. & system-only / shared \\
    \hline
  \end{tabular}
  \captionof{table}{Classification criteria: jobs definition, compilation,
    deployment, and execution.}
  \label{tab:criteria:jobs}
\end{minipage}
\end{figure}

This section concentrates on jobs, following their lifecycle from definition
to execution (see \fig{model:jobs_lifecycle}).  Jobs are defined inside a
driver program (\s{model:jobs:definition}), compiled into an execution plan of
elementary tasks (\s{model:jobs:compilation}), which are deployed and executed
on the distributed computing infrastructure (\s{model:jobs:exec}).  The
resulting classification criteria are in \tab{criteria:jobs}.

\subsubsection{Jobs definition}
\label{sec:model:jobs:definition}

Jobs are defined inside driver programs.  Frequently, driver programs include
multiple jobs and embed the logic that coordinates their execution.
For instance, stored procedures (driver programs) in DMSs may embed multiple
queries (jobs) within procedural code.  Similarly, DPSs invoke analytic jobs
from a driver program written in a standard programming language with a
fork-join execution model.  Notably, some systems implement iterative
algorithms by spawning a new job for each iteration and by evaluating
termination criteria within the driver program.

Jobs are expressed using programming primitives (\emph{jobs definition API})
with heterogeneous forms.  For instance, relational DMSs rely on SQL (a domain
specific language - \emph{DSL}), while DPSs usually offer \emph{libraries} for
various programming languages. Some systems support both forms.

Jobs are compiled into an \emph{execution plan}, which defines the computation
as a set of elementary units of deployment and execution called \emph{tasks}.
Tasks:
\begin{inparaenum}[(i)]
\item run on slots;
\item exchange data over the data bus (dark gray arrows in
  \fig{model:jobs_lifecycle}) according to the communication schema defined in
  the execution plan;
\item may access the state portion of the worker they are deployed on.
\end{inparaenum}
We say that the \emph{execution plan definition} is \emph{explicit} if the
programming primitives directly specify the individual tasks and their logical
dependencies.  The definition is instead \emph{implicit} if the logical plan
is compiled from a higher-level, declarative specification of the job.
To exemplify, the dataflow formalism adopted in many DPSs provides an explicit
definition of the logical plan, while SQL, and most query languages, provide
an implicit definition.
%
%
With an explicit definition of the logical plan, the \emph{communication}
between tasks can itself be \emph{explicit} or \emph{implicit}.  In the first
case, the system APIs include primitives to send and receive data across
tasks, while in the latter case the exchange of data is implicit.
%
%
The execution plan \emph{structure} can be a generic \emph{workflow}, where
there are no restrictions to the pattern of communication between tasks, or a
\emph{dataflow}, where tasks need to be organized into an acyclic graph and
data can only move from upstream tasks to downstream tasks.  When present, we
also highlight further structural constraints.  For instance, the execution
plan of the original MapReduce system forces data processing in exactly two
phases: a map and a reduce.

\emph{Iterations} within the execution plan may or may not be allowed.
We say that a system supports \emph{dynamic creation} of the plan if it
enables spawning new tasks during execution.  Dynamic creation gives the
flexibility of defining or activating part of the execution plan at runtime,
which may be used to support control flow constructs.

Jobs can be either \emph{one-shot} or \emph{continuous}.  One-shot jobs are
executed once and then terminate.  We use the term \emph{invoke} here: as
invoking a program twice leads to two distinct processes, invoking a one-shot
job multiple times leads to separate executions of the same code.  For
instance, queries in DMSs are typically one-shot jobs and indeed multiple
invocations of the same query lead to independent executions.
Instead, continuous jobs persist across invocations.  In this case we use the
term \emph{activate} to highlight that the same job is repeatedly activated by
the arrival of new data.  This happens in stream DPSs, where continuous jobs
are activated when new input data comes from active sources.
As detailed in \s{model:state}, the key distinguishing factor of continuous
jobs is their ability to persist some private task state across activations.
By definition, this option is not available for one-shot jobs, since each
invocation is independent from the other.

State management in jobs may be \emph{absent}, \emph{explicit}, or
\emph{implicit}.  For instance, state management is absent in batch DPSs,
which define jobs in terms of functional transformations that solely depend on
the input data.
State management is explicit when the system provides constructs to directly
access state elements to read and write them.  For instance, queries in
relational DMSs provide \texttt{select} clauses to retrieve state elements and
\texttt{insert} and \texttt{update} clauses to store new state elements and
update them.
State management is implicit when state accesses are implicit in job
definition.  For instance, stream DPSs manage state implicitly through ad-hoc
operators such as windows that record previously received data and use it to
compute new output elements.

Another relevant characteristic of the programming model is the support for
\emph{data parallelism}. It allows defining computations for a single element
and automatically executing them on many elements in parallel.
Data parallelism is central in many systems, and in particular in DPSs, as it
simplifies the definition of jobs by letting developers focus on individual
elements.  It promotes parallel execution, as the tasks operating on different
elements are independent.
Systems supporting data parallelism apply partitioning strategies to both data
and state (when available), as we discuss in \s{model:data} and
\s{model:state}.
As inter-task communication and remote data access may easily become
performance bottlenecks, some systems aim to reduce inter-task communication
and to promote local access to data and state by offering \emph{placement
  aware} API that enable developers to suggest suitable placement strategies
based on the expected workload.

\subsubsection{Jobs compilation}
\label{sec:model:jobs:compilation}

The process of compiling jobs into an execution plan may either start \emph{on
  driver registration} or \emph{on driver execution}.  The first case models
situations where the driver program is registered in the system and can be
executed multiple times, as in the case of stored procedures.  The second case
happens in DPSs, which usually offer a single command to submit and execute a
program.
Jobs compilation may \emph{use resources information}, that is, information
about the resources of the distributed computing infrastructure.
The information is \emph{static} if it only considers the available
resources. For instance, data parallel operators are converted into multiple
tasks that run the same logical step in parallel: the concrete number of tasks
is typically selected depending on the available processing resources (overall
number of CPU cores).
The information is \emph{dynamic} if it also considers the actual use of
resources.  For instance, a join operation may be compiled into a distributed
hash join or into a sort-merge join depending on the current cardinality and
distribution of the elements to join.

\subsubsection{Jobs deployment and execution}
\label{sec:model:jobs:exec}

Jobs deployment is the process of allocating the tasks of a job execution plan
onto slots.  For instance, the execution plan in \fig{model:jobs_lifecycle}
consists of seven tasks and each of them is deployed on a different slot.
Tasks tagged A and B exemplify data-parallel operations, each executed by two
tasks in parallel.
Deployment can be performed with \emph{job-level} or with \emph{task-level}
\emph{granularity}.  Job-level granularity is common when the deployment takes
place \emph{on job compilation}, while task-level granularity is used when the
deployment (of individual tasks) takes place \emph{on task activation}.
It is important to note that the above classification is orthogonal to the
nature of jobs (one-shot or continuous) as defined above.
One-shot jobs may be:
\begin{inparaenum}[(i)]
\item entirely deployed on job compilation, or
\item progressively, as their input data is made available by previous tasks
  in the execution plan.
\end{inparaenum}
The first choice is frequent in DMSs, while the latter characterizes several
DPSs.
Similarly, continuous jobs may be:
\begin{inparaenum}
\item fully deployed on compilation, with their composing tasks remaining
  available onto slots, ready to be activated by incoming data elements, or
\item their tasks may be deployed individually when new input data becomes
  available and activates them.  In this case the same task is deployed
  multiple times, once for each activation: systems that follow this strategy
  minimize the overhead of deployment by accumulating input data into batches
  and deploying a task only once for the entire batch, as well exemplified by
  the micro-batch approach of Spark
  Streaming~\cite{zaharia:SOSP:2013:Discretized_Streams}.
\end{inparaenum}
As we discuss in \s{model:data}, task-level deployment requires a persistent
data bus, to decouple task execution in time.  If the data bus is not
persistent, all tasks in the execution plan need to be simultaneously deployed
to enable the exchange of data.

The deployment process always exploits \emph{static} information about the
resources available in the computing infrastructure, like the address of
workers and their number of slots.  Some systems also exploit \emph{dynamic}
information, such as the current load of workers and the location of data.
This is typically associated to task-level scheduling on activation, where
tasks are deployed when their input data is available and they are ready for
execution.
Finally, the deployment process may have a \emph{system-only} or \emph{shared}
\emph{management of resources}.  System-only management only considers the
resources occupied by the data-intensive system.  Shared management takes
global decisions in the case multiple software systems share the same
distributed computing infrastructure.  For instance, it is common to use
external resource managers such as
Yarn\extendedonly{~\cite{vavilapalli:SOCC:2013:YARN}} for task deployment in
cluster environments.


\subsection{Data management}
\label{sec:model:data}

\begin{table}[hptb]
  \parbox[t]{0.32\textwidth}{
    \centering
    \tiny
    \rowcolors{1}{gray!25}{white}
    \renewcommand{\arraystretch}{0.94}
    \begin{tabular}[t]{|L|C|}
      \hline
      Elements structure & no / general / domain specific \\
      Temporal elements & no / yes \\
      Bus connection type & direct / mediated \\
      Bus implementation &  \\
      Bus persistency & persistent / ephemeral \\
      Bus partitioning & no / yes \\
      Bus replication & no / yes \\
      Bus interaction model & push / pull / hybrid \\
      \hline
    \end{tabular}
    \caption{Classification criteria: data management.}
    \label{tab:criteria:data}
  }
  \hfill
  \parbox[t]{0.32\textwidth}{
    \centering
    \tiny
    \rowcolors{1}{gray!25}{white}
    \renewcommand{\arraystretch}{0.94}
    \begin{tabular}[t]{|L|C|}
      \hline
      Elements structure & no / general / domain specific \\
      Storage medium & memory / disk / hybrid / service \\
      Storage structure & \\
      Task state & no / yes \\
      Shared state & no / yes \\
      Partitioned & no / yes \\
      Replication & no / backup-only / yes \\
      Replication consist. & weak / strong \\
      Replication protocol & leader / consensus / conflict resolution \\
      Update propagation & state / operations \\
      \hline
    \end{tabular}
    \caption{Classification criteria: state management.}
    \label{tab:criteria:state}
  }
  \hfill
  \parbox[t]{0.32\textwidth}{
    \centering
    \tiny
    \rowcolors{1}{gray!25}{white}
    \renewcommand{\arraystretch}{0.94}
    \begin{tabular}[t]{|L|C|}
      \hline
      \rowcolor{gray!50}
      \multicolumn{2}{|c|}{\bf Group atomicity} \\ \hline
      Causes for aborts & system / job \\
      Protocol & blocking / coord. free \\
      Assumptions & \\
      \hline
      \rowcolor{gray!50}
      \multicolumn{2}{|c|}{\bf Group isolation} \\ \hline
      Level & blocking / coord. free \\
      Implementation & lock / timestamp \\
      Assumptions & \\
      \hline
    \end{tabular}
    \caption{Classification criteria: tasks grouping.}
    \label{tab:criteria:grouping}
  }
\end{table}

This section studies the characteristics of data elements and the data bus
used to distribute them.  The resulting classification criteria are in
\tab{criteria:data}.
Recall that in our model data elements are immutable, meaning that once they
are delivered through the data bus they cannot be later updated. Also, they
are used to represent both data and invocations, as they carry some payload
and may trigger the activation of tasks.
Data elements may be \emph{structured}, if they have an associated schema
determining the number and type of fields they contain, or
\emph{unstructured}, otherwise.  Structured data is commonly found in DPSs,
when input datasets or data streams are composed of tuples with a fixed
structure.
The structure of elements may reflect on the data bus, with assumptions of
homogeneous data elements (same schema) in some communication channels.  For
instance, DPSs typically assume homogeneous input and homogeneous output data
for each task.
We further distinguish between systems that accept \emph{general} structured
data, when the developers are free to define their custom data model, and
systems that assume a \emph{domain-specific} structure, when developers are
constrained to a specific data model, as in the case of relational data, time
series, or graph-shaped data.
Finally, data may or may not have a \emph{temporal} dimension: this is
particularly relevant for stream DPSs, where it is used for time-based
analysis.  \s{model:delivery_order} will detail how the temporal dimension
influences the order in which tasks analyze data elements.

The data bus can either consist of \emph{direct} connections between the
communicating parties or it can be \emph{mediated} by some middleware service.
Accordingly, the actual \emph{bus implementation} may range from TCP links
(direct connection) to various types of middleware systems (mediated
connection), like message queuing or distributed storage
services\footnote{When the values associated to a classification criterion are
  specific to individual systems, as in the case of \emph{bus implementation}
  (\tab{criteria:data}), we do not report a list of values in the
  corresponding table.}.
While direct connections are always \emph{ephemeral}, various mediated
connections are \emph{persistent}.  In the first case, receivers need to be
active when the data is transmitted over the bus and they cannot retrieve the
same elements later in time.  In the second case, elements are preserved
inside the bus and receivers can access them multiple times and at different
points in time.
For instance, DPSs that implement job-level deployment usually adopt a direct
(and ephemeral) data bus made of TCP connections among tasks.  Conversely,
DPSs that deploy tasks independently (task-level deployment) require a
persistent and mediated data bus (for instance, a distributed filesystem or a
persistent messaging middleware) where intermediate tasks can store their
results for downstream tasks.

In many systems, the data bus provides communication channels where data
elements are logically \emph{partitioned} based on some criterion, for
instance, based on the value of a given field.  The use of a partitioned data
bus is common in DPSs, where it is associated to data-parallel operators: the
programmer specifies the operator for the data elements in a single partition,
but the operator is concretely implemented by multiple (identical) tasks that
work independently on different partitions.
%
A persistent data bus may also be \emph{replicated}, meaning that the same
data elements are stored in multiple copies.  Replication may serve two
purposes: improving performance, for instance by enabling different tasks to
consume the same data simultaneously from different replicas, or tolerating
faults, to avoid losing data in the case one replica becomes unavailable.  We
will discuss fault tolerance in greater details in \s{model:fault-tolerance}.
Among the systems we analyze in this paper, all those that replicate the data
bus implement a single-leader replication schema, where one \emph{leader}
replica is responsible for receiving all input data and for updating the other
$f$ (\emph{follower}) replicas.  The update is synchronous (or
semi-synchronous), meaning that the addition of an input data element to the
data bus completes when the data element has been successfully applied to all
$f$ follower replicas (or to $r < f$ replicas, if the update is
semi-synchronous).
The data bus offers an \emph{interaction model} that is \emph{push} if the
sender delivers data to recipients, or \emph{pull} if receivers demand data to
senders.  \emph{Hybrid} approaches are possible in the presence of a mediated
bus, a common case being a push approach between the sender and the data bus
and a pull approach between the data bus and the recipients.

\subsection{State management}
\label{sec:model:state}

After discussing data we focus on state, deriving the classification criteria
listed in \tab{criteria:state}.
In absence of state, tasks results (data written on the bus) only depend on
their input (data read from the bus), but many systems support stateful tasks
whose results also depend on some mutable state that they can read and modify
during execution.
This marks a difference between data and state elements: the former are
immutable while the latter may change over time.  As for their
\emph{structure}, state elements resemble data elements: they may be
\emph{unstructured} or \emph{structured}, and in the second case they may rely
on \emph{domain-specific} data models.

When present, state may be stored on different \emph{storage media}:
\begin{inparaenum}
\item many systems store the entire state \emph{in-memory}, and replicate it
  to disk only for fault tolerance;
\item other systems use \emph{disk} storage, or
\item \emph{hybrid} solutions, where state is partially stored in memory for
  improved performance and flushed to disk to scale in size;
\item some systems rely on a storage \emph{service}, as common in cloud-based
  systems that split their core functionalities into independently deployed
  services.
\end{inparaenum}
Some recent work investigates the use of persistent
memory~\cite{liu:VLDB:2021:Zen}, but these solutions are not employed in
currently available systems.

The \emph{storage structure} indicates the data structure used to represent
state on the storage media.  This structure is heavily influenced by the
expected access pattern: for instance, relational state may be stored
row-wise, to optimize access element by element (common in data management
workloads) or column-wise, to optimize access attribute by attribute (common
in data analysis workloads, for instance to compute an aggregation function
over an attribute).  Many DMSs use indexed structures such as B-trees or
log-structured merge (LSM) trees to rapidly identify individual elements.

Data-intensive systems may support two types of state: \emph{task state},
which is private to a single task, and \emph{shared state}, which can be
accessed by multiple tasks.
%
%
The availability of these types of state deeply affects the design and
implementation of the system.
Shared state is central in DMSs, where two tasks (for example, an insert and a
select query) can write and read simultaneously from the same state (for
example, a relational table).
%
%
Conversely, most DPSs avoid shared state to simplify parallel execution.
Frequently, batch DPSs do not offer any type of state, while stream DPSs only
offer task state, which does not require any concurrency control mechanism, as
it is accessed only by one task (sequential unit of execution).
Notice that task state is only relevant in continuous jobs, where it can
survive across multiple activations of the same task.  Indeed, it is used in
DSPs to implement stateful operators like windows.

In our model, workers are responsible for storing separate portions of the
shared state.  Tasks have local access to elements stored (in memory or on
disk) on the shared state portion of the worker they are deployed on.  They
can communicate with remote tasks over the data bus to access shared state
portions deployed on other workers.  For systems that rely on a storage
service, we model the service as a set of workers that are only responsible
for storing shared state portions and offer remote access through the data
bus.
Splitting of the shared state among workers may respond to a criterion of
\emph{partitioning}. For instance, partitioning enables DMSs to scale beyond
the memory capacity of a single node, but also to run tasks belonging
to the same or different jobs (queries) in parallel on different partitions.
Besides partitioning, many data-intensive systems adopt replication.  As in
the case of data bus replication, state replication may also serve two
purposes:
\begin{inparaenum}[(i)]
\item reduce read access latency, by allowing multiple workers to store a copy
  (replica) of the same state elements locally;
\item provide durability and fault tolerance, avoiding potential loss in the
  case of failures. We return on the specific use of replication for fault
  tolerance in \s{model:fault-tolerance}.
\end{inparaenum}
Here, we consider if the replication is \emph{backup-only}, meaning that
replicas are only used for fault tolerance and cannot be accessed by tasks
during execution, or not.

If tasks can access state elements from multiple replicas, different
replication \emph{consistency} models are possible, which define which state
values may be observed by tasks when accessing multiple replicas.  Replication
models have been widely explored in database and distributed systems
theory\extendedonly{~\cite{viotti:CSUR:2016:Consistency}}.  For the goal of
our analysis, we only distinguish between \emph{strong} and \emph{weak}
consistency models, where the former require synchronous coordination among
the replicas while the latter do not.  This classification approach is also in
line with the recent literature that denotes models that do not require
synchronous coordination as being highly
available~\cite{bailis:2013:VLDB:HAT}.
%
Intuitively, strong consistency models are more restrictive and use
coordination to avoid anomalies that may arise when tasks access elements
simultaneously from different replicas.  In practice, most data-intensive
systems that adopt a strong consistency model provide \emph{sequential}
consistency, a model that ensures that accesses to replicated state are the
same as if they were executed in some serial order.  This simplifies reasoning
on the state of the system, as it hides concurrency by mimicking the behavior
of a sequential execution.
In terms of implementation, we distinguish two main classes of mechanisms to
achieve strong consistency: in \emph{leader}-based algorithms, all state
updates are delivered to a single replica (leader) that decides their order;
in \emph{consensus}-based algorithms, replicas use quorum-based or distributed
consensus protocols to agree on the order of state accesses.
Systems that adopt a weak consistency model typically provide \emph{eventual}
consistency, where updates to state elements are propagated asynchronously,
which may lead to (temporary) inconsistencies between replicas.
For this reason, weak consistency is typically coupled with automated
\emph{conflict resolution} algorithms, which guarantee that all replicas solve
conflicts in the same way and eventually converge to the same state.  A
popular approach to conflict resolution are conflict-free replicated
datatypes, which expose only operations that guarantee deterministic conflict
resolution in the presence of simultaneous
updates~\cite{shapiro:SSS:2011:CRDT}.

Finally, replication protocols may employ two approaches to propagate updates:
\emph{state-based} or \emph{operation-based} (also known as active
replication).  In state-based replication, when a task updates a value in a
replica, the new state is propagated to the other replicas.  In
operation-based replication, the operation causing the update is propagated
and re-executed at each replica: this approach may save bandwidth but it may
spend more computational resources to re-execute operations at each replica.

\subsection{Tasks grouping}
\label{sec:model:grouping}

Several systems offer primitives to identify groups of tasks and provide
additional guarantees for such groups, which we classify as \emph{group
  atomicity} (\s{model:grouping:atomicity}) and \emph{group isolation}
(\s{model:grouping:isolation}).
The resulting classification criteria are in \tab{criteria:grouping}.
Atomicity ensures no partial failures for a group of tasks: they either all
fail or all complete successfully.  Isolation limits the ways in which running
tasks can interact and interleave with each other.
In DMSs, these concerns are considered part of transactional management,
together with consistency and durability properties.  In our model, we discuss
consistency constraints as part of group atomicity in the next section, while
we integrate durability with fault tolerance and discuss it in
\s{model:fault-tolerance}.

\subsubsection{Group atomicity}
\label{sec:model:grouping:atomicity}

Atomicity ensures that a group of tasks either entirely succeeds or entirely
fails.  We use the established jargon of database transactions and we say that
a task (or group of tasks) either \emph{commits} or \emph{aborts}.  If the
tasks commit, all the effects of their execution, and in particular all their
changes to the shared state, become visible to other tasks.  If the tasks
abort, none of the effects of their execution becomes visible to other tasks.
We classify group atomicity along two dimensions.  First, we consider the
possible \emph{causes for aborts} and we distinguish between
\emph{system-driven} or \emph{job-driven}.
System-driven aborts (such as a worker running out of memory) derive from
non-deterministic hardware or software failures, while job-driven aborts (such
as database integrity constraints) are part of a job definition and are
triggered if job completion may lead to a logic error.
%
%
Second, we consider how systems implement group atomicity.  Atomicity is
essentially a consensus problem~\cite{maiyya:VLDB:19:unifying}, where tasks
need to agree on a common outcome: commit or abort.
The established protocol to implement atomicity is \emph{two-phase commit}.
In this protocol, one of the participants (tasks) takes the role of a
coordinator, collects all votes (commit or abort) from participants (phase
one), and distributes the common decision to all of them (phase two).
Notice that this protocol is not robust against failures of the coordinator,
however, data-intensive systems typically adopt orthogonal mechanisms to deal
with failures, as discussed in \s{model:fault-tolerance}.
Most importantly, two-phase commit is a \emph{blocking} protocol as
participants cannot make progress before receiving the global outcome from the
coordinator.
For these reasons, some systems adopt simplified, \emph{coordination free}
protocols, which reduce or avoid coordination under certain assumptions.
Being specific to individual systems, we discuss such protocols in
\s{systems_db}.


\subsubsection{Group isolation}
\label{sec:model:grouping:isolation}

Group isolation constrains how tasks belonging to different groups can
interleave with each other, and is classically organized into
\emph{levels}~\cite{Adya:ICDE:2000:Generalized_isolation}.  The stronger,
serializable isolation, requires the effects of execution to be the same as if
all groups were executed in some serial order, with no interleaving of tasks,
while weaker levels enable some disciplined form of concurrency that may lead
to anomalies in the results that clients observe.
In line with the approach adopted for replication consistency and for
atomicity, in this work we consider only two broad classes of isolation
levels: those that require \emph{blocking coordination} between tasks and
those that are \emph{coordination free} (referred to as being highly-available
in the literature~\cite{bailis:2013:VLDB:HAT}).
This is also motivated by the systems under analysis, which either provide
strong isolation levels (typically, serializable) or do not provide isolation
at all.

Implementation-wise, strong isolation is traditionally achieved with two
classes of coordination protocols: \emph{lock-based} and
\emph{timestamp-based}\extendedonly{~\cite{bernstein:CSUR:81:concurrency}}.
With lock-based protocols, tasks acquire non-exclusive or exclusive locks to
access shared resources (shared state in our model) in read-only or read-write
mode.  Lock-based protocols may incur distributed deadlocks: to avoid them,
protocols implement detection or prevention schemes that abort and restart
groups in the case of deadlock.
Timestamp-based protocols generate a serialization order for groups before
execution, and then the tasks need to enforce that order.  Pessimistic
timestamp protocols abort and re-execute groups when they try to access shared
resources out of order.  Multi-version concurrency control protocols reduce
the probability of aborts by storing multiple versions of shared state
elements, and allowing tasks to read old versions when executed out of order.
Optimistic concurrency control protocols allow out of order execution of
tasks, but check for conflicts before making the effects of a group of tasks
visible to other tasks.  Finally, a few systems adopt special protocols that
reduce or avoid coordination under certain assumptions: as in the case of
group atomicity, we discuss these protocols in \s{systems_db}.


\subsection{Delivery and order guarantees}
\label{sec:model:delivery_order}

\begin{table}[hptb]
  \parbox[t]{0.32\textwidth}{
    \centering
    \tiny
    \rowcolors{1}{gray!25}{white}
    \renewcommand{\arraystretch}{0.94}
    \begin{tabular}[t]{|L|C|}
      \hline
      Delivery guarantees & at most / at least / exactly once \\
      Nature of timestamps & n.a. / ingestion / event \\
      Order guarantees & n.a. / eventually / always \\
      \hline
    \end{tabular}
    \caption{Classification criteria: delivery and order.}
    \label{tab:criteria:delivery-order}
  }
  \hfill
  \parbox[t]{0.32\textwidth}{
    \centering
    \tiny
    \rowcolors{1}{gray!25}{white}
    \renewcommand{\arraystretch}{0.94}
    \begin{tabular}[t]{|L|C|}
      \hline
      Detection & leader-worker / distributed \\
      Scope & comput - task state - shared state \\
      Computation recovery & absent / job / task \\
      State recovery & checkp. (indep./coord./per-activ.) / log (WAL/CL) / repl. \\
      Guarantees for state & none / valid / same \\
      Assumptions & \\
      \hline
    \end{tabular}
    \caption{Classification criteria: fault tolerance.}
    \label{tab:criteria:fault-tolerance}
  }
  \hfill
  \parbox[t]{0.32\textwidth}{
    \centering
    \tiny
    \rowcolors{1}{gray!25}{white}
    \renewcommand{\arraystretch}{0.94}
    \begin{tabular}[t]{|L|C|}
      \hline
      Goal & \\
      Automated & no / yes \\
      Mechanisms & state migr. / task migr. \\
      Restart & no / yes \\
      \hline
    \end{tabular}
    \caption{Classification criteria: dynamic reconfiguration.}
    \label{tab:criteria:reconf}
  }
\end{table}

\emph{Delivery} and \emph{order} guarantees define how external actors
(driver programs, sources, and sinks) observe the effects of their actions
(submitting input data and invocations).
%
%
Both topics are crucial for distributed systems and have been widely explored
in the literature.  Here, we focus on the key concepts that characterize the
behavior of the systems we analyzed, and we offer a description that embraces
different styles of interaction, from invocation-based (as in DMSs queries) to
data-driven (as in stream DPSs).
The resulting classification criteria are in \tab{criteria:delivery-order}.

Delivery focuses on the effects of a single input $I$ (data element or
invocation).
Under \emph{at most once} delivery, the system behaves as if $I$ was either
received and processed once or never.
Under \emph{at least once} delivery, the system behaves as if $I$ was received
and processed once or more than once.
Under \emph{exactly once} delivery, the system behaves as if $I$ was received
and processed once and only once.
A well known theoretical result in the area of distributed systems states the
impossibility to deliver an input exactly once in a distributed environment
where components can fail.
Nevertheless, a system can behave as if the input was processed exactly once
under some assumptions: the most common are that driver programs and sources
can resubmit the input upon request (to avoid loss of data), while sinks can
detect duplicate output results and discard them (to avoid duplicate
processing and output).
To exemplify, DMSs offer exactly once delivery when they guarantee group
atomicity through transactions: in this case, a job entirely succeeds or
entirely fails, and in the case of a failure the system either notifies the
driver program (that may retry until success) or internally retries, allowing
the jobs to be executed exactly once.
DPSs offer exactly once delivery by replaying data from sources (or from
intermediate results stored in a persistent data bus) in the case of a
failure.  In presence of continuous jobs (stream processing), systems also
need to avoid duplicating the effects of processing on task state when
replaying data: to do so, they often discard the effects of previous
executions by reloading an old task state from a checkpoint (see also the role
of checkpoints on fault tolerance in \s{model:fault-tolerance}).

Order focuses on multiple data elements or invocations and defines in which
order their effects become visible.
Order ultimately depends on the \emph{nature of timestamps} physically or
logically attached to data elements.  In some systems, no timestamp is
associated to data elements and in these cases no ordering guarantees are
provided.  Conversely, when data elements represent occurrences of events in
the application domain, they have an associated timestamp that can be set by
the original source or by the system when it first receives the element.  We
rely on established terminology, and we refer to the former case as
\emph{event time} and to the latter case as \emph{ingestion
  time}~\cite{akidau:VLDB:2015:dataflow}.
When a timestamp is provided, systems may ensure that the associated order is
guaranteed \emph{always} or \emph{eventually}.
Systems in the first class wait until all data elements before a given
timestamp become available and then process them in order.  To do so, they
typically rely on a contract between the sender components and the data bus,
where sender components use special elements (denoted as \emph{watermarks}) to
indicate that all elements up to a given time $t$ have been produced, and the
data bus delivers elements up to time $t$ in the correct order.
Systems in the second class execute elements out-of-order, but \emph{retract}
previously issued results and correct them when they receive new data with an
older timestamp.  Thus, upon receiving all input data up to time $t$, the
system eventually returns the correct results.  Notice that this mechanism
requires the elements receiving output data to tolerate temporarily incorrect
results.
According to our definitions above, retraction is not compatible with exactly
once delivery, as it changes the results provided to sinks after they have
already been delivered, thus breaking the illusion that they have been
produced once and only once.

\subsection{Fault tolerance}
\label{sec:model:fault-tolerance}

Fault tolerance is the ability of a system to tolerate failures that may occur
during its execution.  We consider hardware failures, such as a disk failing
or a node crashing or becoming unreachable, and non-deterministic software
failures, such as a worker exhausting node's memory. We assume that the logic
of jobs is correct, which guarantees that the re-execution of a failed job
does not deterministically lead to the same failure.
Our minimal unit of fault is the worker and we assume the typical approach to
tolerate failures that involves first detecting the failure and then
recovering from it.
The classification criteria for fault tolerance are in
\tab{criteria:fault-tolerance}.

\emph{Fault detection} is usually addressed as a problem of group membership:
given a group of workers, determine the (sub)set of those active and available
to execute jobs.  Systems address this problem either using a
\emph{leader-worker} approach, which assumes one entity with a special role
(leader) that cannot fail and can supervise normal workers, or using a
\emph{distributed} protocol, like gossip-based protocols.

After a failure is detected, \emph{fault recovery} brings the system into a
state from which it can resume with the intended semantics.  We describe the
recovery process by focusing on five aspects: scope, computation recovery,
state recovery, guarantees, and assumptions.
Depending if tasks are stateless or stateful and if they can share state or
not, the \emph{scope} of recovery may involve recovering the
\emph{computation} of failing tasks, the \emph{task state} of failing tasks,
and/or the \emph{shared state} portions held by failing workers.

\emph{Computation recovery} may be \emph{absent}, in which case failing jobs
are simply discarded and the system offers at most once delivery (see
\s{model:delivery_order}).  Otherwise, the system recovers the computation by
restarting it: we distinguish between systems that need to restart an entire
\emph{job} and systems that can restart individual \emph{tasks}.
DMSs typically restart entire jobs to satisfy transactional (atomicity)
guarantees that require a job to either entirely succeed or entirely fail.
Some DPSs (those using a persistent data bus to save the intermediate results
of tasks) may restart only failed tasks.
Restarting a computation requires that input data and invocations are
persisted and replayable, and that duplicate output data can be detected and
discarded by sinks (if the system wants to ensure exactly once delivery, see
\s{model:delivery_order}).
To replay input data and invocations, systems either rely on replayable
sources, such as persistent message services, or keep a log internally (see
the discussion on logging below).

To recover state, systems may rely on \emph{checkpointing}, \emph{logging},
and \emph{replication}.  Frequently, they combine these mechanisms.
Checkpointing involves saving a copy of the entire state to durable storage.
When a failure is detected, the last available checkpoint can be reloaded and
the execution of jobs may restart from that state.
Different workers may take \emph{independent} checkpoints or they may
\emph{coordinate}, for instance by using the distributed snapshot
protocol~\cite{chandi:TOCS:1985:distributed} to periodically save a consistent
view of the entire system state.
A third alternative (\emph{per-activation} checkpoint) is sometimes used for
continuous jobs to save task state: at each activation, a task stores its task
state together with the output data.  This approach essentially transforms a
stateful task into a stateless task, where state is encoded as a special data
element that the system receives in input at the next activation.  In
practice, per-activation checkpoint is used in presence of a persistent data
bus that stores checkpoints.
Frequent checkpointing may be resource consuming and affect the response time
of the system.  \emph{Logging} is an alternative approach that saves either
individual operations or state changes rather than the entire content of the
state.  These two forms of logging are known in the database literature as:
\begin{inparaenum}[(i)]
\item command logging (CL) persists input data and invocations, and in the
  case of failure re-processes the same input to restore
  state~\cite{malviya:ICDE:14:rethinking};
\item write ahead log (WAL) persists state changes coming from tasks to
  durable storage before they are applied, and in the case of a failure
  re-applies the changes in the log to restore the
  state\extendedonly{~\cite{mohan:TODS:92:aries}}.
\end{inparaenum}
As logs may grow unbounded with new invocations entering the system, they are
always complemented with (infrequent) checkpoints.  In the case of failure,
the state is restored from the last checkpoint and then the log is replayed
from the checkpoint on.
Finally, systems may \emph{replicate} state portions in multiple workers.  In
this case, a state change performed by a task succeeds only after the change
has been applied to a given number of replicas $r$.  This means that the
system can tolerate the failure of $r-1$ replicas without losing the state and
without the need to restore it from a checkpoint.
As already discussed in \s{model:state}, the same replicas used for fault
tolerance may also be used by tasks during normal processing to improve state
access latency.

The recovery mechanisms above provide different \emph{guarantees} on the state
of the system after recovery.  It can be \emph{any} state, a \emph{valid}
system state, or the \emph{same} state the system was before failing.
In \emph{any} state recovery, the presence of a failure may bring the system
to a state that violates some of the invariants for data and state management
that hold for normal (non failing) executions: for instance, the system may
drop some input data.
A \emph{valid} state recovery mechanism brings the system to a state that
satisfies all invariants, but it may differ from those states traversed before
the failure: for instance, a system that provides serializability for groups
of tasks may recover by re-executing groups of tasks in a different (but still
serial) order.  Depending on the system, clients may be able or not to observe
the differences between the two states (before and after the failure): for
instance, in a DMS, two read queries before and after the
failure may observe different states.
A \emph{same} state recovery mechanism brings the system in the same state it
was before the failure.
Replication, write-ahead logging, and per-activation checkpointing bring the
system to the same state it was prior to fail, while independent checkpointing
only guarantees to bring the system back to a valid state, as the latest
checkpoints of each task may not represent a consistent cut of the system.
The same happens for command logging, due to different interleavings in the
original execution and in the recovery phase that may lead to different
(albeit valid) states.

A final aspect related to recovery are the \emph{assumptions} under which it
operate, like assuming no more than $k$ nodes can fail simultaneously as in
the case of replication.


\subsection{Dynamic reconfiguration}
\label{sec:model:reconf}

With dynamic reconfiguration we refer to the ability of a system to modify the
deployment and execution of jobs on the distributed computing infrastructure
at runtime.
The corresponding classification criteria are in \tab{criteria:reconf}.
Reconfiguration may be driven by different \emph{goals}, which may involve
providing some minimum quality of service, for instance in terms of throughput
or response time and/or minimizing the use of resources to cut down operation
costs.
It may be activated manually or be \emph{automated}, if the system can monitor
the use of resources and make changes that improve the state of the system
with respect to the intended goals.
The reconfiguration process may involve different mechanisms: \emph{state
  migration} across workers, for instance to rebalance shared state portions
if they become unbalanced; \emph{task migration}, to change the association of
tasks to slots, including the addition or removal of slots, to add
computational resources if the load increases or release them when they are
not necessary.
State migration is common in DMSs, where the distribution of shared state
across workers may affect performance.  Task migration is instead used in DPSs
in presence of continuous jobs, where tasks are migrated across invocations.
In both cases, the migration may adapt the system to the addition or removal
of slots.
Some systems can continue operating during a reconfiguration process, while
other systems need to temporarily stop and \emph{restart} the jobs they are
running: this approach appears in some system that adopt job-level deployment;
in this case, reconfiguration takes place by saving the current state,
restarting the whole system, and restoring the last recorded state.


\section{Survey of Data-Intensive Systems}
\label{sec:systems}

To keep our survey compact and general we decided not to survey data-intensive
systems one by one (this is provided in Appendix), but to group them into the
taxonomy of \fig{survey_org} and to discuss systems class by class in
\s{systems_db}, \s{systems_proc}, and \s{systems_other}.  The caption of
\fig{survey_org} lists all systems we grouped in each class.


\begin{figure}[hptb]
  \begin{center}
    \begin{small}
      \begin{forest}
        for tree={
          anchor=west,
          calign=center,
          grow'=east,
          scale=0.6,
          parent anchor=east,
          child anchor=west,
          s sep=-1.0mm,
          l sep+=0mm,
          edge path={
            \noexpand\path[\forestoption{edge}](!u.parent anchor) -- (.child anchor)\forestoption{edge label};},        
          for root={
            parent anchor=east,
          },
        },
        [\makecell{Data intensive}, tier=root
          [\makecell{Data management\\systems (DMSs) \secref{systems_db}}, tier=sec
            [\makecell{NoSQL DMSs \secref{systems_db:nosql}}, tier=subsec
              [\makecell{Key-value \secref{systems_db:nosql:kv}} (1), tier=subsubsec]
              [\makecell{Wide-column \secref{systems_db:nosql:wide_column}} (2), tier=subsubsec]
              [\makecell{Document \secref{systems_db:nosql:document}} (3), tier=subsubsec]
              [\makecell{Time-series \secref{systems_db:nosql:ts}} (4), tier=subsubsec]
              [\makecell{Graph \secref{systems_db:nosql:graph}} (5), tier=subsubsec]
            ]
            [\makecell{NewSQL DMSs \secref{systems_db:newsql}}, tier=subsec
              [\makecell{Key-value \secref{systems_db:newsql:kv}} (6), tier=subsubsec]
              [\makecell{Structured / relational \secref{systems_db:newsql:relational}}, tier=subsubsec
                [\makecell{Time-based protocols} (7), tier=par]
                [\makecell{Deterministic execution} (8), tier=par]
                [\makecell{Explicit partition./repl.} (9), tier=par]
                [\makecell{Primary-based protocols} (10), tier=par]
              ]
              [\makecell{Object \secref{systems_db:newsql:object}} (11), tier=subsubsec]
              [\makecell{Graph \secref{systems_db:newsql:graph}} (12), tier=subsubsec]
            ]
          ]
          [\makecell{Data processing\\systems (DPSs) \secref{systems_proc}}, tier=sec
            [\makecell{Dataflow with task depl. \secref{systems_proc:task_depl}} (13), tier=subsec]
            [\makecell{Dataflow with job depl. \secref{systems_proc:job_depl}} (14), tier=subsec]
            [\makecell{Graph processing \secref{systems_proc:graph}} (15), tier=subsec]
          ]
          [\makecell{Other systems \secref{systems_other}}, tier=sec
            [\makecell{Computations on data management \secref{systems_other:computations_db}}, tier=subsec
              [\makecell{Incremental computations \secref{systems_other:computations_db:incremental}} (16), tier=subsubsec]
              [\makecell{Long-running jobs \secref{systems_other:computations_db:long}} (17), tier=subsubsec]
              [\makecell{Graph processing \secref{systems_other:computations_db:graph}} (18), tier=subsubsec]
            ]
            [\makecell{New programming models \secref{systems_other:programming_models}}, tier=subsec
              [\makecell{Stateful dataflow \secref{systems_other:programming_models:stateful_dataflow}} (19), tier=subsubsec]
              [\makecell{Reactors \secref{systems_other:programming_models:reactors}} (20), tier=subsubsec]
            ]
            [\makecell{Hybrid systems \secref{systems_other:hybrid_systems}} (21), tier=subsec]
          ]
        ]    
      \end{forest}
    \end{small}
  \end{center}
  \caption{A taxonomy of data intensive systems.\\{\scriptsize List of systems:
    (1) Dynamo, DynamoDB, Redis, Voldemort, Riak KV, Aerospike, PNUTS, Memcached, 
    (2) BigTable, Cassandra,
    (3) MongoDB, CouchDB, AsterixDB,
    (4) InfluxDB, Gorilla, Monarch, Peregreen,
    (5) TAO, Unicorn,
    (6) Deuteronomy, FoundationDB, Solar,
    (7) Spanner, CockroachDB,
    (8) Calvin,
    (9) VoltDB,
    (10) Aurora, Socrates,
    (11) Tango,
    (12) A1,
    (13) MapReduce, Dryad, HaLoop, CIEL, Spark, Spark Streaming,
    (14) MillWheel, Flink, Storm, Kafka Streams, Samza / Liquid, Timely dataflow,
    (15) Pregel, GraphLab, PowerGraph, Arabesque, G-Miner,
    (16) Percolator,
    (17) F1,
    (18) Trinity,
    (19) SDG, TensorFlow, Tangram,
    (20) ReactDB,
    (21) S-Store, SnappyData, StreamDB, Tspoon, Hologres.}}
  \label{fig:survey_org}
\end{figure}

Our taxonomy groups existing systems in a way that emphasizes their
commonalities with respect to our classification criteria
(Tables~\ref{tab:criteria:functional}--\ref{tab:criteria:reconf}), while also
capturing pre-existing classifications widely adopted by experts.
The top-level distinction between data management systems (DMSs,
\s{systems_db}) and data processing systems (DPSs, \s{systems_proc}) is well
known to researchers and practitioners and it is also well captured by our
classification criteria, with all DMSs providing shared state but not task
state and DPSs having opposite characteristics: this distinction impacts on
the value of most fields of \tab{criteria:state}.

\begin{table}[tb]
  \centering
  \tiny
  \renewcommand{\arraystretch}{0.94}
  \begin{tabular}[t]{|l|c|c|c|c|c|c|c|c|}
    \rowcolor{gray!50}
    \hline
    & \multicolumn{2}{|c|}{\textbf{Data management}} & \multicolumn{3}{|c|}{\textbf{Data processing}} & \multicolumn{3}{|c|}{\textbf{Other}} \\ \hline

    \rowcolor{gray!50}
    & \textbf{NoSQL} & \textbf{NewSQL} & \textbf{\Gape[0pt]{\makecell{Dataflow\\task depl}}} & \textbf{\Gape[0pt]{\makecell{Dataflow\\jobs depl}}} & \textbf{Graph} & \textbf{\Gape[0pt]{\makecell{Comput on\\data manag}}} & \textbf{\Gape[0pt]{\makecell{New prog\\models}}} & \textbf{Hybrid} \\ \hline

    \rowcolor{gray!25}
    \multicolumn{9}{|c|}{\textbf{Functional components}} \\ \hline
    
    \textbf{\makecell[l]{Driver exec}}         & \makecell{client+sys\stproc}   & \makecell{client+sys\stproc}   & \makecell{sys/conf}               & \sysdep                           & sys      & client\exc  & sys\exc & sys\exc      \\ \hline
    \textbf{\makecell[l]{Driver exec time}}    & \makecell{reg+start\stproc}    & \makecell{reg+start\stproc}    & reg                               & reg                               & reg      & reg\exc     & reg\exc & reg\exc      \\ \hline
    \textbf{\makecell[l]{Invocation of jobs}}  & \makecell{sync (+async)}       & \makecell{sync (+async)}       & \sysdep                           & async\exc                         & \sysdep  & \sysdep     & \sysdep & \sysdep      \\ \hline
    \textbf{Sources}                           & \sysdep                        & no                             & \makecell{pass\batch/act\stream}  & \makecell{pass\batch/act\stream}  & pass     & \sysdep     & \sysdep & \sysdep      \\ \hline
    \textbf{Sinks}                             & \sysdep                        & no\exc                         & yes                               & yes                               & yes      & yes\exc     & yes\exc & yes          \\ \hline
    \textbf{State}                             & yes                            & yes                            & \makecell{no\batch/yes\stream}    & \makecell{no\batch/yes\stream}    & yes      & yes         & yes     & yes          \\ \hline
    \textbf{Deployment}                        & \sysdep                        & \sysdep                        & cluster                           & cluster                           & cluster  & cluster\exc & cluster & cluster      \\ \hline

    \rowcolor{gray!25}
    \multicolumn{9}{|c|}{\textbf{Jobs definition}} \\ \hline

    \textbf{Jobs def API}      & \classdep  & \classdep  & \makecell{lib (+DSL)}                  & \makecell{lib (+DSL)}                  & lib    & \sysdep                              & lib           & \makecell{lib (+DSL)}                  \\ \hline
    \textbf{Exec plan def}     & impl\exc   & impl\exc   & \makecell{expl (+impl)}                & \makecell{expl (+impl)}                & expl   & \makecell{impl (+expl)}              & \sysdep       & \makecell{expl (+impl)}                \\ \hline
    \textbf{Task comm}         & impl       & impl\exc   & impl                                   & impl                                   & expl   & \makecell{impl (+expl)}              & impl\exc      & impl                                   \\ \hline
    \textbf{Exec plan struct}  & worfkl\exc & worfkl\exc & dataflow                               & dataflow                               & graph  & \sysdep                              & \sysdep       & \sysdep                                \\ \hline
    \textbf{Iterations}        & no\exc     & no\exc     & \sysdep                                & \sysdep                                & yes    & no\exc                               & yes           & no\exc                                 \\ \hline
    \textbf{Dyn creation}      & no\exc     & no\exc     & no\exc                                 & no                                     & no\exc & no                                   & no\exc        & no                                     \\ \hline
    \textbf{Nature of jobs}    & one-shot   & one-shot   & \makecell{one-shot\batch\\cont\stream} & \makecell{one-shot\batch\\cont\stream} & cont   & one-shot                             & one-shot\exc  & \makecell{one-shot\batch\\cont\stream} \\ \hline
    \textbf{State man.}        & expl       & expl       & \makecell{absent\batch/impl\stream}    & \makecell{absent\batch/impl\stream}    & expl   & expl                                 & expl          & expl\exc                               \\ \hline
    \textbf{Data par. API}     & no         & no\exc     & yes                                    & yes                                    & yes    & no\exc                               & yes\exc       & yes\exc                                \\ \hline
    \textbf{Placem-aware API}  & no         & no\exc     & no                                     & no                                     & yes    & no                                   & no\exc        & \sysdep                                \\ \hline

    \rowcolor{gray!25}
    \multicolumn{9}{|c|}{\textbf{Jobs compilation}} \\ \hline
    
    \textbf{Jobs compil time}    & on exec\exc & on exec\exc & on exec & on exec & on exec & on exec\exc & on exec\exc & on exec\exc \\ \hline
    \textbf{Use resources info}  & static      & \sysdep     & dynamic & static  & static  & static      & static\exc  & static      \\ \hline

    \rowcolor{gray!25}
    \multicolumn{9}{|c|}{\textbf{Jobs deployment and execution}} \\ \hline
    
    \textbf{Granular of depl}   & job\exc    & job     & task       & job         & \sysdep & job      & \sysdep     & \sysdep \\ \hline
    \textbf{Depl time}          & compil\exc & compil  & activ      & compil      & \sysdep & compil   & \sysdep     & \sysdep \\ \hline
    \textbf{Use resources info} & static     & static  & dynamic    & static\exc  & \sysdep & static   & static\exc  & \sysdep \\ \hline
    \textbf{Managem of res}     & sys\exc    & sys\exc & shared\exc & sys\exc     & sys\exc & sys\exc  & sys\exc     & sys     \\ \hline

    \rowcolor{gray!25}
    \multicolumn{9}{|c|}{\textbf{Data management}} \\ \hline
    
    \textbf{Elem struct}   & \classdep & \classdep    & \makecell{gen (+spec)}          & \makecell{gen (+spec)}              & graph                       & \sysdep  & \sysdep      & spec\exc     \\ \hline
    \textbf{Temp elem}     & \sysdep   & no           & \makecell{no\batch\\yes\stream} & \makecell{no\batch\\yes\stream}     & no                          & no\exc   & no           & \sysdep      \\ \hline
    \textbf{Bus conn type} & direct    & direct       & mediated\exc                    & direct\exc                          & \sysdep                     & direct   & direct\exc   & direct\exc   \\ \hline
    \textbf{Bus impl}      & net chan  & net chan\exc & fs (+RAM)                       & \makecell{net chan/\\msg service}   & \makecell{net chan/\\mem}   & net chan & net chan\exc & net chan\exc \\ \hline
    \textbf{Bus persist}   & ephem     & ephem        & pers\exc                        & ephem\exc                           & \sysdep                     & ephem    & ephem\exc    & ephem\exc    \\ \hline
    \textbf{Bus partition} & yes       & yes          & yes                             & yes                                 & yes                         & yes      & yes          & yes          \\ \hline
    \textbf{Bus repl}      & no        & no           & no                              & no\exc                              & no                          & no       & no\exc       & no\exc       \\ \hline
    \textbf{Bus interact}  & pull      & pull\exc     & hybrid\exc                      & push\exc                            & \sysdep                     & push\exc & push\exc     & push\exc     \\ \hline
    
  \end{tabular}
  \caption{Survey of systems: functional components; jobs definition,
    compilation, deployment, and execution; data management.  Legend: \sysdep
    = system dependent; \classdep = differences captured by the specific
    sub-classes in our taxonomy (see \fig{survey_org}); \exc = with few
    system-specific exceptions; \stproc = in the case of stored procedures;
    \batch = in the case of batch processing; \stream = in the case of stream
    processing.}
  \label{tab:classes:functional_jobs_data_state}
\end{table}
    
\begin{table}[tb]
  \centering
  \tiny
  \renewcommand{\arraystretch}{0.94}
  \begin{tabular}[t]{|l|c|c|c|c|c|c|c|c|}
    \rowcolor{gray!50}
    \hline
    & \multicolumn{2}{|c|}{\textbf{Data management}} & \multicolumn{3}{|c|}{\textbf{Data processing}} & \multicolumn{3}{|c|}{\textbf{Other}} \\ \hline
    \rowcolor{gray!50}
    & \textbf{NoSQL} & \textbf{NewSQL} & \textbf{\Gape[0pt]{\makecell{Dataflow\\task depl}}} & \textbf{\Gape[0pt]{\makecell{Dataflow\\jobs depl}}} & \textbf{Graph} & \textbf{\Gape[0pt]{\makecell{Comput on\\data manag}}} & \textbf{\Gape[0pt]{\makecell{New prog\\models}}} & \textbf{Hybrid} \\ \hline

    \rowcolor{gray!25}
    \multicolumn{9}{|c|}{\textbf{State management}} \\ \hline

    \textbf{Elem struct}   & \classdep                        & \classdep                            & --                              & --                              & graph    & \sysdep                   & gen\exc & spec\exc  \\ \hline
    \textbf{Stor medium}   & \sysdep                          & \sysdep                              & --                              & --                              & mem      & \sysdep                   & mem     & mem\exc   \\ \hline
    \textbf{Stor struct}   & \sysdep                          & \sysdep                              & --                              & --                              & user-def & \sysdep                   & \sysdep & \sysdep   \\ \hline
    \textbf{Task state}    & no                               & no                                   & \makecell{no\batch/yes\stream}  & \makecell{no\batch/yes\stream}  & yes      & no                        & no      & \sysdep   \\ \hline
    \textbf{Shared state}  & yes                              & yes                                  & no                              & no                              & no       & yes                       & yes     & yes\exc   \\ \hline
    \textbf{Partitioned}   & yes                              & yes                                  & --                              & --                              & --       & yes                       & yes     & yes\exc   \\ \hline
    \textbf{Replication}   & \makecell{yes/backup}            & \makecell{yes/backup}                & --                              & --                              & --       & \makecell{yes/backup}   & no      & \sysdep   \\ \hline
    \textbf{Repl consist}  & \makecell{weak/conf}             & strong                               & --                              & --                              & --       & \sysdep                   & --      & --\exc    \\ \hline
    \textbf{Repl protocol} & \makecell{lead/cons\\+confl res} & \makecell{lead/cons}                 & --                              & --                              & --       & lead.\exc                 & --      & --        \\ \hline
    \textbf{Upd propag}    & op\exc                           & op\exc                               & --                              & --                              & --       & op\exc                    & --      & --\exc    \\ \hline

    \rowcolor{gray!25}
    \multicolumn{9}{|c|}{\textbf{Group atomicity}} \\ \hline

    \textbf{Aborts}      & --\exc   & \makecell{job+sys\exc}           & --   & --   & --   & \makecell{job+sys\exc} & --\exc  & \makecell{job+sys\exc}    \\ \hline
    \textbf{Protocol}    & --\exc   & blocking\exc                     & --   & --   & --   & blocking\exc           & --\exc  & blocking\exc              \\ \hline
    \textbf{Assumptions} & --       & \makecell{--/\detjob/\onewrite}  & --   & --   & --   & no                     & --      & no/\detjob\exc         \\ \hline

    \rowcolor{gray!25}
    \multicolumn{9}{|c|}{\textbf{Group isolation}} \\ \hline
    
    \textbf{Level}       & \makecell{--/coord free\exc} & \makecell{blocking\exc}         & --   & --   & --   & blocking                   & \sysdep & \sysdep           \\ \hline
    \textbf{Implement}   & --/\seq                      & \makecell{ts/lock}              & --   & --   & --   & ts/lock\exc                & \sysdep & \sysdep           \\ \hline
    \textbf{Assumptions} & \makecell{--/\oneportion}    & \makecell{no/\detjob/\onewrite} & --   & --   & --   & \makecell{no/\oneportion}  & no      & no/\detjob\exc    \\ \hline

    \rowcolor{gray!25}
    \multicolumn{9}{|c|}{\textbf{Delivery and order}} \\ \hline
    
    \textbf{Delivery guar} & \makecell{most/exact\exc} & exact\exc & exact                              & exact/least                 & exact & exact\exc & exact & exact        \\ \hline
    \textbf{Nature of ts}  & \makecell{no/event}       & no        & \makecell{no\batch/event\stream}   & \makecell{no/event}         & no    & no\exc    & no    & \sysdep      \\ \hline
    \textbf{Order guar}    & --                        & --        & \makecell{--\batch/alw\stream}     & \makecell{alw/event\exc}    & --    & --        & --    & --/alw       \\ \hline

    \rowcolor{gray!25}
    \multicolumn{9}{|c|}{\textbf{Fault tolerance}} \\ \hline

    \textbf{Detection}      & \sysdep                           & lead-work\exc                                & lead-work                                 & lead-work                        & lead-work                & lead-work\exc                 & lead-work\exc                 & lead-work\exc   \\ \hline
    \textbf{Scope}          & shared st                         & shared st                                    & \makecell{comp\\+task st\stream}          & \makecell{comp\\+task st\stream} & \makecell{comp+task st}  & \makecell{shared st\\(+comp)} & \makecell{shared st\\(+comp)} & \sysdep         \\ \hline
    \textbf{Comput recov}   & --                                & --                                           & task                                      & job\exc                          & job\exc                  & \sysdep                       & \sysdep                       & \sysdep         \\ \hline
    \textbf{State recov}    & \makecell{log (+repl)}            & \makecell{log (+repl)}                       & \makecell{--\batch/checkp\stream}         & \makecell{checkp\exc}            & checkp                   & \sysdep                       & checkp\exc                    & log+checkp      \\ \hline
    \textbf{Guar for state} & \makecell{none/conf\exc}          & \makecell{same/conf}                         & \makecell{--\batch\\valid/same\stream}    & \makecell{valid/same}            & \makecell{valid/same}    & same\exc                      & valid                         & valid/same\exc  \\ \hline
    \textbf{Assumptions}    & \makecell{\durstore/\\\durrepl}   & \makecell{\durstore/\\\durrepl (+\detjob)}   & \makecell{\replaysource}                  & \makecell{\replaysource}         & --                       & \makecell{\durstore}          & \sysdep                       & \sysdep         \\ \hline

    \rowcolor{gray!25}
    \multicolumn{9}{|c|}{\textbf{Dynamic reconfiguration}} \\ \hline

    \textbf{Goal}          & \makecell{avail\\+load bal+elast} & \makecell{change schema\\+load bal+elast} & \makecell{--/elast}              & \makecell{--/\\load bal+elast} & \makecell{load bal} & \sysdep & --/elast & \makecell{--/\\load bal+elast} \\ \hline
    \textbf{Automated}     & \sysdep                           & \sysdep                                   & \makecell{--/yes}                & \sysdep                        & yes                 & yes     & --/yes   & \sysdep                        \\ \hline
    \textbf{State migr.}   & yes                               & yes                                       & \makecell{--\batch/yes\stream}   & yes                            & yes                 & yes     & --/yes   & --/yes                         \\ \hline
    \textbf{Task migr.}    & --                                & --                                        & \makecell{--\batch/yes\stream}   & yes                            & yes                 & \sysdep & --/yes   & --/yes                         \\ \hline
    \textbf{Add/rem slots} & yes                               & yes                                       & \makecell{--/yes}                & \makecell{--/yes}              & no                  & yes     & --/yes   & \sysdep                        \\ \hline
    \textbf{Restart}       & no                                & no\exc                                    & \makecell{--/no}                 & \sysdep                        & no                  & no      & --/no    & \sysdep                        \\ \hline

  \end{tabular}
  \caption{Survey of systems: state management; group atomicity and isolation;
    delivery and order; fault tolerance; dynamic reconfiguration. Legend:
    \sysdep = system dependent; \classdep = differences captured by the
    specific sub-classes in our taxonomy (see \fig{survey_org}); \exc = with
    few system-specific exceptions; \batch = in the case of batch processing;
    \stream = in the case of stream processing; \detjob = jobs are
    deterministic; \seq = jobs are executed sequentially, with no
    interleaving; \onewrite = a single worker handles all writes; \oneportion
    = jobs access a single state portion; \durstore = storage layer is
    durable; \durrepl = replicated data is durable; \replaysource = sources
    are replayable.}
  \label{tab:classes:guarantees_fault_reconf}
\end{table}


Within the class of DMSs, the ability to offer strong guarantees in terms of
consistency (\tab{criteria:state}), group atomicity, and group isolation
(\tab{criteria:grouping}) draws a sharp distinction between those systems
usually known as ``No SQL'' and those known as ``New SQL''.
NoSQL and NewSQL systems can be further classified looking at the data model
they offer (field ``elements structure'' in \tab{criteria:data}).

Within the class of DPSs, criterion ``execution plan structure''
(\tab{criteria:jobs}) differentiates dataflow and graph processing systems,
while criterion ``granularity of deployment'' (\tab{criteria:jobs}) further
separates dataflow systems into those offering task-level deployment and those
offering job-level deployment.

Finally, other systems (\s{systems_other}) that do not clearly fall into the two
main categories of DMSs and DPSs include those that implement data processing
(as DPSs) but on top of shared state abstractions usually offered by DMSs
only, those that aim to offer new programming models, and hybrid systems that
explicitly try to integrate data processing and management capabilities within
a unified solution.


\tab{classes:functional_jobs_data_state} and
\tab{classes:guarantees_fault_reconf} show how the classes of systems at the
second level of our taxonomy map on the classification criteria of our model.
Next sections describe each class in details, explaining the values in these
tables and making practical examples that refer to specific systems.

\section{Data Management Systems}
\label{sec:systems_db}

Data management systems (DMSs) offer the abstraction of a mutable state store
that many jobs can access simultaneously to query, retrieve, insert, and
update elements.
Differently from data processing systems, they mostly target lightweight jobs,
which do not involve computationally expensive data transformations and are
short-lived.
Since their advent in the seventies, relational databases represented the
standard approach to data management, offering a uniform data model
(relational), query language (SQL), and execution semantics (transactional).
Over the last two decades, new requirements and operational conditions brought
the idea of a unified approach to data management to an
end~\cite{stonebraker:ICDE:2005:one_size_fits_all,
  stonebraker:VLDB:2007:hstore}: new applications emerged with different needs
in terms of data and processing models, for instance, to store and retrieve
unstructured data; scalability concerns related to data volume, number of
simultaneous users, and geographical distribution pointed up the cost of
transactional semantics.
This state of things fostered the development of the DMSs presented in this
section.  \s{systems_db:overview} discusses the aspects in our model that are
common to all such systems.  Then, following an established terminology, we
organize them in two broad classes: NoSQL
databases~\cite{davoudian:CSur:2018:NoSQL} (\s{systems_db:nosql}) emerged
since the early 2000s, providing simple and flexible data models such as
key-value pairs, and trading consistency guarantees and strong (transactional)
semantics for horizontal scalability, high availability, and low response
time; NewSQL databases~\cite{stonebraker:CACM:2012:NewSQL}
(\s{systems_db:newsql}) emerged in the late 2000s and take an opposite
approach: they aim to preserve the traditional relational model and
transactional semantics by introducing new design and implementation
strategies that reduce the cost of coordination.

\subsection{Overview}
\label{sec:systems_db:overview}

\paragraph{Functional model}

All DMSs provide a global shared state that applications can simultaneously
access and modify.
In the more traditional systems there is a sharp distinction between the
application logic (the driver, executed client-side on registration) and the
queries (the jobs, executed by the DMS).  Recent systems increasingly allow to
move the part of the application logic that orchestrates jobs execution within
the DMS, in the form of stored procedures that run system-side on start.
Stored procedures may bring two advantages:
\begin{inparaenum}[(i)]
\item reducing the interactions with external clients, thus improving latency;
\item moving part of the overhead for compiling jobs from driver execution
  time to driver registration time.
\end{inparaenum}

Being conceived for interactive use, all DMSs offer synchronous APIs to invoke
jobs from the driver program.  Many also offer asynchronous APIs that allow
the driver to invoke multiple jobs and receive notifications of their results
when they terminate.  A common approach to reduce the cost of communication
when starting jobs from a client-side driver is batching multiple invocations
together, which is offered in some NoSQL systems such as
MongoDB~\cite{chodorow:2013:MongoDB} and Redis~\cite{macedo:2011:Redis}.

Several DMSs can interact with active sources and sinks.  Active sources push
new data into the system, leading to insertion or modification of state
elements.  Sinks register to state elements of interest (e.g., by specifying a
key or a range of keys) and are notified upon modification of such elements.

DMSs greatly differ in terms of deployment strategies, which are vastly
influenced by the coordination protocols that govern replication, group
atomicity and isolation.
NoSQL systems do not offer group atomicity and isolation.  Those designed for
cluster deployment typically use blocking (synchronous or semi-synchronous)
replication protocols.  Those that support wide area deployments either use
coordination-free (asynchronous) replication protocols that reduce durability
and consistency guarantees, or employ a hybrid strategy, with synchronous
replication within a data center and asynchronous replication across data
centers.
Many NewSQL systems claim to support wide area deployments while offering
strong consistency, group atomicity and isolation.  We review their
implementation strategies to achieve this result in \s{systems_db:newsql}.

\paragraph{Jobs}

All DMSs implement one-shot jobs with explicit state management primitives to
read and modify a global shared state.
Jobs definition API greatly differ across systems, ranging from pure key-value
stores that offer CRUD (create, read, update, and delete) primitives for
individual state elements to expressive domain specific libraries (e.g., for
graph computation) or languages (e.g., SQL for relational data).
In almost all DMSs, the execution plan and the communication between tasks are
implicit and do not include iterations or dynamic creation of new tasks.  A
notable exception are graph databases, which support iterative algorithms
where developers explicitly define how tasks update portions of the state (the
graph) and exchange data.
The structure of the execution plan also varies across systems: common
structures include the use of a single task that implements CRUD primitives,
workflows orchestrated by a central coordinator task, or hierarchical
structures.
Jobs are compiled on driver execution, except for those systems where part of
the driver program is registered server-side (as stored procedures).  Job
compilation always uses static information about resources, such as the
allocation of shared state portions onto nodes.  Some structured NewSQL DMSs
such as Spanner~\cite{bacon:SIGMOD:2017:Spanner} and
CockroachDB~\cite{taft:SIGMOD:2020:CockroachDB} also exploit dynamic
information about resources, for instance to configure a given task (e.g.,
select a sort-merge or a hash-based strategy for a join task) depending on
some resource utilization or statistics about state (e.g., cardinality of the
tables to join).

All DMSs perform deployment at job level, when the job is compiled, with the
only exception of AsterixDB~\cite{alsubaiee:VLDB:2014:AsterixDB}, which
compiles jobs into a dataflow plan and deploys individual tasks when they are
activated~\cite{borkar:ICDE:2011:Hyracks}.
Deployment is always guided by the location of the state elements to be
accessed, which we consider as a static information.  Indeed, under normal
execution, shared state portions do not move, and our model captures dynamic
relocation of shared state portions (e.g., for load balancing) as a distinct
aspect (\emph{dynamic reconfiguration}).  Also, we keep saying that deployment
is based on static information even for those systems that exploit dynamic
information (e.g., the load of workers) but only to deploy the tasks that
manage the communication between the system and external clients.
Finally, DMSs are not typically designed to operate in scenarios where the
compute infrastructure is shared with other software applications.  This is
probably due to the interactive and short-lived nature of jobs, which would
make it difficult to predict and adapt the demand of resources in those
scenarios.  The only case in which we found explicit mention of a shared
platform is in the description of the Google infrastructure, where DMSs
components (BigTable~\cite{chang:OSDI:06:bigtable},
Percolator~\cite{peng:OSDI:2010:LargeScale},
Spanner~\cite{corbett:TOCS:2013:Spanner}) share the same physical machines and
their scheduling, execution, and monitoring is governed by an external
resource management service.

\paragraph{Data and state management}

The data model (that is, the structure of data and state elements) is a key
distinguishing characteristic of DMSs, which we use to organize our discussion
in \s{systems_db:nosql} and \s{systems_db:newsql}.
Some systems explicitly consider the temporal dimension of data elements: for
instance, some wide columns stores associate timestamps to elements and let
users store multiple versions of the same element, while time series databases
are designed to efficiently store and query sequences of measurements over
time.
DMSs differ in terms of storage medium and structure.  We detail the choices
of the different classes of systems in \s{systems_db:nosql} and
\s{systems_db:newsql}, but we can identify some common concerns and design
strategies.
First, the representation of state on storage is governed by the data model
and the expected access pattern: relational tables are stored by row, whereas
time series are stored by column to facilitate computations on individual
measurements over time (e.g., aggregation, trend analysis).
Second, there is a tension between read and write performance: read can be
facilitated by indexed data structures such as B-trees, but they incur higher
insertion and update costs.  Hierarchical structures such as log-structured
merge (LSM) trees improve write performance by buffering updates in
higher-level logs (frequently in-memory) that are asynchronously merged into
lower-level indexed structured, at the expense of read performance, due to the
need to navigate multiple layers.
Third, most DMSs exploit main memory to reduce data access latency.  For
instance, systems based on LSM-trees store the write buffer in memory.
Similarly, most systems that use disk-based storage frequently adopt some
in-memory caching layer.
Finally, some systems adopt a modular architecture that supports different
storage layers.  This is common in DMSs offered as a service in public cloud
environments (e.g., Amazon Aurora~\cite{verbitski:SIGMOD:2017:Aurora}) or in
private data centers (e.g., Google Spanner~\cite{corbett:TOCS:2013:Spanner}),
where individual system components (including the storage layer) are
themselves services.

Tasks always communicate using direct, ephemeral connections, which implement
a partitioned, non-replicated data bus.
Shared state is always partitioned across workers to enable concurrent
execution of tasks.  Most DMSs also adopt state replication to improve read
access performance, with different guarantees in terms of consistency between
replicas: NoSQL databases provide weak (or configurable) consistency
guarantees to improve availability and reduce response time.  NewSQL databases
provide strong consistency using leader-based or consensus protocols and
restricting the types of transactions (jobs) that can read from non-leader
replicas -- typically snapshot transactions, which are a subset of read-only
transactions that read a consistent version of the state, without guarantees
of it being the most recent one.
Group atomicity and isolation are typically absent or optional in NoSQL
databases, or restricted to very specific cases, such as jobs that operate on
single data elements.  Instead, NewSQL databases provide strong guarantees for
atomicity and isolation, at the cost of blocking coordination.
The transactional semantics of NewSQL systems ensures exactly once delivery.
Indeed, transactions (jobs) either complete successfully (and their effects
become durable) or abort, in which case they are either automatically retried
or the client is notified and can decide to retry them until success.
Conversely, NoSQL systems frequently offer at most once semantics, as they do
not guarantee that the results of job execution are safely stored on
persistent storage or replicated.  In some cases, users can balance durability
and availability by selecting the number of replicas that are updated
synchronously.
Finally, systems that support timestamps use event time semantics, where
timestamps are associated to state elements by clients, while none of the
systems provides order guarantees.  Even in the presence of timestamps, DMSs
do not implement mechanisms to account for elements produced or received out
of timestamp order.

\paragraph{Fault tolerance}

Frequently, DMSs offer multiple mechanisms for fault tolerance and durability
that administrators can enable and combine depending on their needs.  Fault
detection can be centralized (leader-worker) or distributed, depending on the
specific system.
Fault recovery mostly targets the durability of shared state.  Since jobs are
lightweight, DMSs simply abort them in the case of failure and do not attempt
to recover (partial) computations.
Transactional systems guarantee group atomicity: in the case of failure, none
of the effects of a job become visible.  To enable recovery of failed jobs,
they either notify the clients about an abort, allowing them to restart the
failed job, or restart the job automatically.
Almost all DMSs adopt logging mechanisms to ensure that the effects of jobs
execution on shared state are durable.  Logging enables changes to be recorded
on some durable append-only storage before being applied to shared state: most
systems adopt a write-ahead log that records changes to individual data
elements, while few others adopt a command log that stores the operations
(commands) that perform the change.
Individual systems make different assumptions on what they consider as durable
storage: in some cases logs are stored on a single disk, but more frequently
they are replicated (or saved on third party log services that are internally
replicated).
Logging is frequently used in combination with replication of shared state
portions on multiple workers: in these cases, each worker stores its updates
on a persistent log to recover from software crashes, while replication on
other workers may avoid unavailability in the case of hardware crashes or
network disconnections.
In addition, most systems offer geo-replication for disaster recovery, where
the entire shared state is replicated in a different data center and
periodically synchronized with the working copy.
Similarly, many systems provide periodic or manual checkpointing to store a
copy of the entire database at a given point in time.
Depending on the specific mechanisms adopted, DMSs provide either no
guarantees for state, for instance in the case of asynchronous replication, or
same state guarantees, for instance in the case of persistent log or
consistent replication.

\paragraph{Dynamic reconfiguration}

Most DMSs support adding and removing workers at runtime, and migrating shared
state portions across workers, which enable dynamic load balancing and scaling
without restarting the system.
All systems support dynamic reconfiguration as a manual administrative
procedure, and some can also automatically migrate state portions for load
balancing.
A special case of reconfiguration for NewSQL systems that rely on structured
state (in particular, relational systems) involves changing the state schema:
many systems support arbitrary schema changes as long-running procedures that
validate state against the new schema and migrate it while still serving
clients using the previous schema.  Instead, systems such as
VoltDB~\cite{stonebraker:IEEEB:2013:voltdb} rely on a given state partitioning
scheme and prevent changes that violate such scheme.

\subsection{NoSQL systems}
\label{sec:systems_db:nosql}

Using an established terminology, we classify as NoSQL all those DMSs that aim
to offer high availability and low response time by relinquishing features and
guarantees that require blocking coordination between workers.
In particular, they typically:
\begin{inparaenum}[(i)]
\item avoid expressive job definition API that may lead to complex execution
  plans (as in the case of SQL, hence the
  name)~\cite{stonebraker:CACM:2010:SQLvsNoSQL}.  In fact, the majority of the
  systems we analyzed focuses on jobs comprising a single task that operates
  on an individual element of the shared state.
\item Use asynchronous replication protocols: depending on the specific
  protocol, this may affect consistency, durability (if replication is used
  for fault tolerance) and may generate conflicts (when clients are allowed to
  write to multiple replicas).
\item Abandon or restrict group guarantees (atomicity and isolation), when
  jobs with multiple tasks are supported.
\end{inparaenum}
In our discussion, we classify systems by the data model they offer.

\subsubsection{Key-value stores}
\label{sec:systems_db:nosql:kv}

Key-value stores offer a very basic API for managing shared state:
\begin{inparaenum}[(i)]
\item shared state elements are represented by a key and a value;
\item elements are schema-less, meaning that different elements can have
  different formats, such as free text or JSON objects with heterogeneous
  attributes;
\item the key-space is partitioned across workers;
\item jobs consist of a single task that retrieves the value of an element
  given a key (\texttt{get}) or insert/update an element given its key
  (\texttt{put}).
\end{inparaenum}

Individual systems differ in the way they physically store elements.  For
instance, Dynamo~\cite{deCandia:SOSP:2007:Dynamo} supports various types of
physical storage,
DynamoDB\extendedonly{\footnote{\url{http://aws.amazon.com/dynamodb/}}} uses
B-Trees on disk but buffers incoming updates in main memory to improve write
performance, Redis~\cite{macedo:2011:Redis} stores elements in memory only.
Keys are typically partitioned across workers based on their hash (hash
partitioning), but some systems also support range partitioning, where each
worker stores a sequential range of keys, as in the case of Redis.
Given the focus on latency, some systems cache the association of keys to
workers client-side, allowing clients to directly forward requests to workers
responsible for the key they are interested in.

Some systems provide richer API to simplify the interaction with the store:
\begin{inparaenum}[(i)]
\item Keys can be organized into tables, mimicking the concept of a table in a
  relational database, for instance, DynamoDB and
  PNUTS~\cite{cooper:VLDB:2008:PNUTS} let developers split state elements into
  tables;
\item Elements may have an associated structure, for instance, PNUTS lets
  developers optionally define a schema for each table, DynamoDB specifies a
  set of attributes but does not constrain their internal structure, Redis
  provides built-in datatypes to define values and represent them efficiently
  in memory;
\item Most systems provide functions to iterate (\texttt{scan}) on the
  key-space or on individual tables (range-based partitioning may be used to
  speedup such range-based iterations), as it happens for DynamoDB, PNUTS, and
  Redis;
\item Finally, some systems provide query operations (\texttt{select}) to
  retrieve elements by value and in some cases these operations are supported
  by secondary indexes that are automatically updated when the value of an
  element changes, as in DynamoDB.
\end{inparaenum}

All key-value stores replicate shared state to increase availability, but use
different replication protocols.
Dynamo,
Voldemort\extendedonly{\footnote{\url{https://www.project-voldemort.com/}}},
and Riak KV\extendedonly{\footnote{\url{https://riak.com/products/riak-kv/}}}
use a quorum approach, where read and write operations for a given key need to
be processed by a given number of replica workers responsible for that key.
After a write quorum is reached, updates are propagated to remaining replicas
asynchronously.  A larger number of replicas and a larger write quorum better
guarantee durability and consistency at the cost of latency and availability.
However, being designed for availability, these systems adopt mechanisms to
avoid blocking on write when some workers are not responsive: for instance,
other workers can supersede and store written values on their behalf.  In some
cases, these mechanisms can lead to conflicting simultaneous updates: Dynamo
tracks causal dependencies between writes to solve conflicts automatically
whenever possible, and stores conflicting versions otherwise, leaving manual
reconciliation to the users.
DynamoDB, Redis, and Aerospike~\cite{srinivasan:VLDB:2016:Aerospike} use
single-leader protocols where all writes are processed by one worker and
propagated synchronously to some replicas and asynchronously to others,
depending on the configuration.  DynamoDB also supports consistent reads that
are always processed by the leader at a per-operation granularity.  Redis
supports multi-leader replication in the case of wide area scenarios, using
conflict free replicated datatypes for automated conflict resolution.

In summary, key-value stores represent the core building block of a DMS.  They
expose a low-level but flexible API to balance availability, consistency, and
durability, and to adapt to different deployment scenarios.
Systems that adopt a modular implementation can use key-value stores as a
storage layer or a caching layer~\cite{nishtala:NSDI:2013:Memcached} and build
richer job definition API, job execution engines, protocols for group
atomicity, group isolation, and consistent replication on top.

\subsubsection{Wide-column stores}
\label{sec:systems_db:nosql:wide_column}

Wide-column stores organize shared state into tables (multi-dimensional maps),
where each row associates a unique key to a fixed number of column families,
and each column family contains a value, possibly organized into more
columns (attributes).
State is physically stored per column family and keys need not have a value
for each column family (the table is typically sparse).
One could define the wide-column data model as middle ground between the
key-value and the relational model: it is similar to the key-value model, but
associates a key to multiple values (column families); it defines tables as
the relational model, but tables are sparse and lack referential integrity.
The main representative systems of this class are Google
BigTable~\cite{chang:OSDI:06:bigtable}, with its open-source implementation
HBase\extendedonly{\footnote{\url{https://hbase.apache.org}}}, and Apache
Cassandra~\cite{lakshman:SIGOPS:2010:cassandra}.
As the official documentation of Cassandra
explains\extendedonly{\footnote{\url{https://cassandra.apache.org/doc/latest/cassandra/data\_modeling/index.html}}},
the typical use of wide-column systems is to compute and store answers to
frequent queries (read-only jobs) for each key, at insertion/update time,
within column families.  In contrast, relational databases normalize tables to
avoid duplicate columns and compute results at query time (rather then
insertion/update time) by joining data from multiple tables.
In fact, wide-column stores offer rich API to scan, select, and update values
by key, but do not offer any join primitive.
To support the above scenario, wide-column systems:
\begin{inparaenum}[(i)]
\item aim to provide efficient write operations to modify several column
  families, for instance both BigTable and Cassandra adopt log-structured
  merge trees for storage and improve write latency by buffering writes in
  memory;
\item provide isolation for operations that involve the same key.
\end{inparaenum}
These two design choices allow users to update all entries for a given key
(answers to queries) efficiently and in isolation.

BigTable and Cassandra have different approaches to replication.  BigTable
uses replication only for fault tolerance and executes all tasks that involve
a single key on the leader worker responsible for that key.  It also supports
wide-area deployment by fully replicating the data store in additional data
centers: replicas in these data centers can be used for fault tolerance but
also to perform jobs, in which case they are synchronized with eventual
consistency.
Cassandra uses quorum replication as in Dynamo, and allows users to configure
the quorum protocols to trade consistency and durability for availability.

\subsubsection{Document stores}
\label{sec:systems_db:nosql:document}

Document stores represent a special type of key-value stores where values are
structured documents, such as XML or JSON objects.  Document stores offer an
API similar to key-value stores but they can exploit the structure of
documents to update only some of their fields.
Physical storage solutions vary across systems, ranging from disk-based, to
memory solutions, to hybrid approaches and storage-agnostic solutions.  In
most cases document stores support secondary indexes to improve retrieval of
state elements using criteria different from the primary key.
Most document stores offer group isolation guarantees for jobs that involve a
single document.  This is the case of MongoDB~\cite{chodorow:2013:MongoDB} and
AsterixDB~\cite{alsubaiee:VLDB:2014:AsterixDB}.  Recent versions of MongoDB
also implement multi-document atomicity and isolation as an option, using
blocking protocols.

MongoDB supports replication for fault tolerance or also to serve read-only
jobs.  It implements a single leader protocol with semi-synchronous
propagation of changes, where clients can configure the number of replicas
that need to synchronously receive an update, thus trading durability and
consistency for availability and response time.
CouchDB~\cite{anderson:2010:CouchDB} offers a quorum-based replication
protocol and allows for conflicts in the case a small write quorum is
selected.  In this case, conflict resolution is manual.
AsterixDB does not currently support replication.

Several document stores support some form of data analytic jobs: MongoDB
offers jobs in the form of a pipeline of data transformations that can be
applied in parallel to a set of documents.  CouchDB focuses on Web
applications and can start registered jobs when documents are added or
modified to update some views.  AsterixDB provides a declarative language that
integrates operators for individual and for multiple documents (like joins,
group by), and compiles jobs into a dataflow execution plan.

\subsubsection{Time-series stores}
\label{sec:systems_db:nosql:ts}

Time-series stores are a special form of wide-column stores dedicated to store
sequences of values over time, for instance measurements of a numeric metric
such as the CPU utilization of a computer over time.
Given the specific application scenario, this class of systems stores data by
column, which brings several advantages:
\begin{inparaenum}[(i)]
\item together with the use of an in-memory or hybrid storage layer, it
  improves the performance of write operations, which typically append new
  values (measurements) to individual columns;
\item it offers faster sequential access to columns, which is common in
  read-only jobs that perform aggregations or look for temporal patterns over
  individual series;
\item it enables a higher degree of data compression, for instance by storing
  only the difference between adjacent numerical values (delta compression),
  which is small if measurements change slowly.
\end{inparaenum}

Among the time-series stores we analyzed,
InfluxDB\extendedonly{\footnote{\url{https://www.influxdata.com}}} is the most
general one.  It provides a declarative job definition language that supports
computations on individual columns (measurements).
Gorilla~\cite{pelkonen:VLDB:2015:Gorilla} is used as an in-memory cache to
store monitoring metrics at Facebook.  Given the volume and rate at which
metrics are produced, Facebook keeps most recent data at a very fine
granularity within the Gorilla cache and stores historical data at a coarser
granularity in HBase.
Peregreen~\cite{visheratin:ATC:2020:Peregreen} follows a similar approach and
optimizes retrieval of data through indexing.  It uses a three-tier data
indexing, where each tier pre-computes aggregated statistics (minimum,
maximum, average, etc.) for the data it references.  This allows to quickly
identify chunks of data that satisfy some conditions based on the pre-computed
statistics and to minimize the number of interactions with the storage layer.
Monarch~\cite{adams:VLDB:2020:Monarch} is used to store monitoring data at
Google.  It has a hierarchical architecture: data is stored in the zone (data
center) in which it is generated and sharded (by key ranges,
lexicographically) across nodes called leaves.  Jobs are evaluated
hierarchically: nodes are organized in three layers (global, zone level,
leaves) and the job plan pushes tasks as close as possible to the data they
need to consume.
All time-series stores we analyzed replicate shared state to improve
availability and performance of read operations.  To avoid blocking write
operations, they adopt asynchronous or semi-synchronous replication, thus
reducing durability guarantees.  This is motivated by the specific application
scenarios, where losing individual measurements may be tolerated.

\subsubsection{Graph stores}
\label{sec:systems_db:nosql:graph}

Graph stores are a special form of key-value stores specialized in
graph-shaped data, meaning that shared state elements represent entities
(vertices of a graph) and their relations (edges of the graph).
Despite researchers widely recognized the importance of large scale graph data
structures\extendedonly{~\cite{sakr:CACM:2021:Big_Graphs}}, several graph data
stores do not scale horizontally~\cite{shao:SIGMOD:2013:Trinity}.
A prominent example of distributed graph store is
TAO~\cite{bronson:ATC:2013:TAO}, used at Facebook to manage the social graph
that interconnects users and other entities such as posts, locations, and
actions.  It builds on top of key-value stores with hybrid storage (persisted
on disk and cached in memory), asynchronously replicated with no consistency
or grouping guarantees.

A key distinguishing factor in graph stores is the type of queries (read-only
jobs) they support.  Indeed, a common use of graph stores is to retrieve
sub-graphs that exhibit certain patterns of relations: for instance, in a
social graph, one may want to retrieve people (vertices) that are direct
friends or have friends in common (friendship relation edges) and like the
same posts.
This problem is denoted as \textit{graph pattern matching} and its general
form can only be solved by systems that can express iterative or recursive
jobs, as it needs to traverse the graph following its edges.
These types of vertex-centric computations have been first introduced in the
Pregel data processing system~\cite{malewicz:SIGMOD:2010:Pregel}, also
discussed in \s{systems_proc}.
%

Efficient query of graph stores can also be supported by external systems.
For instance, Facebook developed the Unicorn~\cite{curtiss:VLDB:2013:Unicorn}
system to store indexes that allow to quickly navigate and retrieve data from
a large graph.  Indexes are updated periodically using an external compute
engine.
Unicorn adopts a hierarchical architecture, where indexes (the shared state of
the system) are partitioned across servers and the results of index lookups
(read jobs) are aggregated first at the level of individual racks and then
globally to obtain the complete query results.  This approach aggregates
results as close as possible to the servers producing them to reduce network
traffic.  Unicorn supports graph patterns queries by providing an
\texttt{apply} function that can dynamically start new lookups based on the
results of previous ones: our model captures this feature by saying that jobs
can dynamically start new tasks.

\subsection{NewSQL systems}
\label{sec:systems_db:newsql}

NewSQL systems aim to provide transactional semantics (group atomicity and
isolation), durability (fault tolerance), and strong replication consistency
while preserving horizontal scalability.
Following the same approach we adopted in \s{systems_db:nosql}, we organize
them according to their data model.

\subsubsection{Key-value stores}
\label{sec:systems_db:newsql:kv}

NewSQL key-value stores are conceived as part of a modular system, where the
store offers transactional guarantees to read and update a group of elements
with atomicity and isolation guarantees, and it is used by a job manager that
compiles and optimizes jobs written in some high-level declarative language.
A common design principle of these systems is to separate the layer that
manages the transactional semantics from the actual storage layer, thus
enabling independent scaling based on the application requirements. 
Deuteronomy~\cite{levandoski:CIDR:2011:Deuteronomy} implements transactional
semantics using a locking protocol and is storage agnostic.
FoundationDB~\cite{zhou:SIGMOD:2021:FoundationDB} uses optimistic concurrency
control with a storage layer based on B-Trees.
Solar~\cite{zhu:ToS:2019:Solar} also uses optimistic concurrently control with
log structured merge trees.

\subsubsection{Structured and relational stores}
\label{sec:systems_db:newsql:relational}

Stores for structured and relational data provide the same data model, job
model, and job execution semantics as classic non-distributed relational
databases.
As we clarify in the classification below, they differ in their protocols for
implementing group atomicity, group isolation, and replication consistency,
which reflects on their architectures.

\paragraph{Time-based protocols}

Some systems exploit physical (wall-clock) time to synchronize nodes.
This approach was pioneered by Google's
Spanner~\cite{corbett:TOCS:2013:Spanner}.  It adopts standard database
techniques: two-phase commit for atomicity, two-phase locking and
multi-version concurrency control for isolation, and single-leader synchronous
replication of state portions.  Paxos consensus is used to elect a leader for
each state portion and to keep replicas consistent.
The key distinguishing characteristic of Spanner is the use TrueTime, a clock
abstraction that uses atomic clocks and GPS to return physical time within a
known precision bound.
In Spanner, each job is managed by a transaction coordinator, which assigns
jobs with a timestamp at the end of the TrueTime clock uncertainty range and
waits until this timestamp is passed for all nodes in the system.  This
ensures that jobs are globally ordered by timestamp, thus offering the illusion
of a centralized system with a single clock (external consistency).
Spanner is highly optimized for workloads with many read-only jobs.  Indeed,
multi-version concurrency control combined with TrueTime allows read-only jobs
to access a consistent snapshot of the shared state without locking and
without conflicting with in-progress read-write jobs, as they will be
certainly be assigned a later timestamp.
More recently, Spanner has been extended with support for distributed SQL
query execution~\cite{bacon:SIGMOD:2017:Spanner}.
CockroachDB~\cite{taft:SIGMOD:2020:CockroachDB} is similar to Spanner but uses
an optimistic concurrency control protocol that, in the case of conflicts,
attempts to modify the timestamp of a job to a valid one rather than
re-executing the entire job.
As Spanner, CockroachDB supports distributed execution plans.
It supports wide-area deployment and allows users to define how data is
partitioned across regions, to promote locality of data access or to enforce
privacy regulations.

\paragraph{Deterministic execution}

Calvin~\cite{thomson:SIGMOD:2012:Calvin} builds on the assumption that jobs
are deterministic and achieves atomicity, isolation, and consistency by
ensuring that jobs are executed in the same order in all replicas.
Determinism ensures that jobs either succeed or fail in any replica
(atomicity), interleave in the same way (global order ensures isolation),
leading to the same results (consistency).
Workers are organized into three layers:
\begin{inparaenum}[(i)]
\item A sequencing layer receives jobs invocations from clients, organizes
  them into batches, and orders them consistently across replicas.  Ordering
  of jobs is the only operation that requires coordination and takes place
  before jobs execution.  Calvin provides both synchronous (Paxos) and
  asynchronous protocols for ordering jobs, which bring different tradeoffs
  between latency and cost of recovery in the case of failures.
\item A scheduler layer that executes tasks onto workers in the defined global
  order.  In cases where it is not possible to statically determine which
  shared state portions will be involved in the execution of a job (for
  instance in the case of state-dependent control flow), Calvin uses an
  optimistic protocol and aborts jobs if some of their tasks are received by
  workers out of order.
\item A storage layer that stores the actual data.  In fact, Calvin supports
  any storage engine providing a key-value interface.
\end{inparaenum}

\paragraph{Explicit partitioning and replication strategies}

VoltDB~\cite{stonebraker:VLDB:2007:hstore, stonebraker:IEEEB:2013:voltdb} lets
users control partitioning and replication of shared state, so they can
optimize most frequently executed jobs.  For instance, users can specify that
a \texttt{Customer} and a \texttt{Payment} tables are both partitioned by the
attribute (column) \texttt{customerId}.
Jobs that are guaranteed to access only a shared state portion within a given
worker are executed sequentially and atomically on that worker.  For instance,
a job that accesses tables \texttt{Customer} and \texttt{Payment} to retrieve
information for a given \texttt{customerId} can be fully executed on the
worker with the state portion that includes that customer.
Every table that is not partitioned is replicated in every worker, which
optimizes read access from any worker at the cost of replicating state
changes.
In the case jobs need to access state portions at different workers, VoltDB
resorts to standard two-phase commit and timestamp-based concurrency control
protocols.
Differently from Spanner and Calvin, VoltDB provides strong consistency only
for cluster deployment: geographical replication is supported, but only
implemented with asynchronous and weakly consistent protocols.

\paragraph{Primary-based protocols}

Primary-based protocols are a standard approach to replication used in
traditional transactional databases.  They elect one primary worker that
handles all read-write jobs and acts as a coordinator to ensure transactional
semantics.  Other (secondary) workers only handle read-only jobs and can be
used to fail over if the primary crashes.
Recently, the approach has been revamped by DMSs offered as services on the
cloud.  These systems adopt a layered architecture that decouples jobs
execution functionalities (e.g., scheduling, managing atomicity and isolation)
from storage functionalities (durability): the two layers are implemented as
services that can scale independently from each other.
The execution layer still consists of one primary worker and an arbitrary
number of secondary workers, which access shared state through the storage
service (although they typically implement a local cache to improve
performance).
Amazon Aurora~\cite{verbitski:SIGMOD:2017:Aurora} implements the storage layer
as a sequential log (replicated for availability and durability), which offers
better performance for write operations.  Indexed data structures that improve
read performance are materialized asynchronously without affecting write
latency.
The storage layer uses a quorum approach to guarantee replication consistency
across workers.
Microsoft Socrates~\cite{antonopoulos:SIGMOD:2019:Socrates} adopts a similar
approach but further separates storage into a log layer (that stores write
requests with low latency), durable storage layer (that stores a copy of the
shared state), and a backup layer (that periodically copies the entire state).

\subsubsection{Objects stores}
\label{sec:systems_db:newsql:object}

Object stores\extendedonly{~\cite{banciihon:PODS:1988:OO}} became popular in
the early Nineties, inheriting the same data model as object-oriented
programming languages.
We found one recent example of a DMS that uses this data model, namely
Tango~\cite{balakrishnan:SOSP:2013:Tango}.
In Tango, clients store their view of objects locally, in-memory, and this
view is kept up to date with respect to a distributed (partitioned) and
durable (replicated) log of updates.
The log represents the primary replica of the shared state that all clients
refer to.  All updates to objects are globally ordered on the log through
sequence numbers that are obtained through a centralized sequencer.
Total order guarantees isolation for operations on individual objects: Tango
also offers group atomicity and isolation across objects using the log to
store information for an optimistic concurrency control protocol.

\subsubsection{Graph stores}
\label{sec:systems_db:newsql:graph}

We found one example of a NewSQL graph store, named
A1~\cite{buragohain:SIGMOD:2020:A1}, which provides strong consistency,
atomicity, and isolation using timestamp-based concurrency control.
Its data model is similar to that of NoSQL distributed graph stores, and jobs
can traverse the graph and read and modify its associated data during
execution.
The key distinguishing characteristic of A1 is that it builds on a distributed
shared memory abstraction that uses RDMA (remote direct memory access)
implemented within network interface cards~\cite{dragojevic:NSDI:2014:FaRM}.


\section{Data Processing Systems}
\label{sec:systems_proc}

Data processing systems (DPSs) aim to perform complex computations (long
lasting jobs) on large volumes of data.
Most of today's DPSs inherit from the seminal MapReduce
system~\cite{dean:CACM:2008:mapreduce}: to avoid the hassle of concurrent
programming and to simplify scalability, they organize each job into a
dataflow graph where vertices are functional operators that transform data and
edges are the flows of data across operators.  Each operator is applied in
parallel to independent partitions of its input data, and the system
automatically handles data partitioning and data transfer across workers.
Following an established terminology, we denote as \emph{batch processing}
systems those that take in input static (finite) datasets, and \emph{stream
  processing} systems those that take in input streaming (potentially
unbounded) datasets.
In practice, many systems support both types of input and we do not use the
distinction between batch and stream processing as the main factor to organize
our discussion.
Instead, after discussing the aspects in our model that are common to all DPSs
(\s{systems_proc:overview}), we classify dataflow systems based on the key
aspect that impacts their implementation: if they deploy individual tasks on
activation (\s{systems_proc:task_depl}) or entire jobs on registration
(\s{systems_proc:job_depl}).
Finally, we present systems designed to support computations on large graph
data structures.  They evolved in parallel with respect to dataflow systems,
which originally were not suited for iterative computations that are typical
in graph algorithms (\s{systems_proc:graph}).

\subsection{Overview}
\label{sec:systems_proc:overview}

\paragraph{Functional model}

Most DPSs use a leader-workers architecture, where one of the processes that
compose the system (denoted the leader) has the special role of coordinating
other workers.  Such systems always allow submitting the driver program to the
leader for system-side execution.  Some of them also allow client-side driver
execution, such as Apache Spark~\cite{zaharia:CACM:2016:spark} and Apache
Flink~\cite{carbone:IEEEB:2015:flink}.
Other systems, such as Kafka Streams~\cite{bejeck:2018:KafkaStreams} and
Timely Dataflow~\cite{murray:SOSP:2013:Naiad}, are implemented as libraries
where client processes also act as workers.  Developers start one or more
client processes and the library handles the distributed execution of jobs
onto them.
Stream processing systems support asynchronous invocation of (continuous)
jobs, whereas batch processing systems may offer synchronous or asynchronous
job invocation API, or both.
All DPSs support sources and sinks, as they are typically used to read data
from external systems (sources), perform some complex data analysis and
transformation (jobs), and store the results into external systems (sinks).
Sources are passive in the case of batch processing systems and active in the
case of stream processing systems.
Most batch processing systems are stateless: output data is the result of
functional transformations of input data.  Stream processing systems can
persist a (task) state across multiple activations of a continuous job.
We model iterative graph algorithms as continuous jobs where tasks (associated
to vertices, edges, or sub-graphs) are activated at each iteration and store
their partial results (values associated to vertices, edges, or sub-graphs) in
task state.
DPSs assume a cluster deployment, as job execution typically involves
exchanging large volumes of data (input, intermediate results, and final
results) across workers.

\paragraph{Jobs}

All dataflow systems provide libraries to explicitly define the execution plan
of jobs.  Increasingly often, they also offer higher-level abstractions for
specific domains, such as relational data
processing~\cite{abouzeid:VLDB:2009:HadoopDB,
  camacho-rodriguez:SIGMOD:2019:ApacheHive, armbrust:SIGMOD:2015:SparkSQL},
graph computations~\cite{gonzalez:OSDI:2014:GraphX}, or machine
learning~\cite{meng:JMLR:2016:MLlib}.  Some of these APIs are declarative in
nature and make the definition of the execution plan implicit.
Task communication is always implicitly defined and controlled by the system's
runtime.
Concerning jobs, dataflow systems differ with respect to the following
aspects:
\begin{inparaenum}[(i)]
\item Generality.  MapReduce~\cite{dean:CACM:2008:mapreduce} and some early
  systems derived from it only support two processing stages with fixed
  operators, while later systems like Spark support any number of processing
  stages and a vast library of operators;
\item Support for iterations.  Systems like HaLoop~\cite{bu:VLDB:2010:HaLoop}
  extended MapReduce to efficiently support iterations by caching data
  accessed across iterations in workers.  Spark inherits the same approach
  and, together with Flink, supports some form of iterative computations for
  streaming data.  Timely Dataflow~\cite{murray:SOSP:2013:Naiad} generalizes
  the approach to nested iterations;
\item Dynamic creation of tasks.  Among the systems we analyzed, only
  CIEL~\cite{murray:NSDI:2011:CIEL} enables dynamic creation of tasks
  depending on the results of processing.
\end{inparaenum}

Data parallelism is key to dataflow systems, and all operators in their jobs
definition API are data parallel.  Jobs are one-shot in the case of batch
processing and continuous in the case of stream processing.  In the latter
case, jobs may implicitly define some task state, for instance by expressing
computations that operate on a recent portion (window) of data rather than on
individual data elements.
Jobs cannot control task placement explicitly, but many systems provide
configuration parameters to guide placement decisions, for instance to force
or inhibit the colocation of certain tasks.
An exception to the above rules is represented by graph processing systems,
which are based on a programming model where developers define the behavior of
individual vertices~\cite{mcCune:CSur:2015:TLaV}: the model provides explicit
primitives to access the state of a vertex and to send messages between
vertices (explicit communication).

All DPSs compile jobs on driver execution.  For other characteristics related
to jobs compilation, deployment, and execution, we distinguish between systems
that perform task level deployment (discussed in \s{systems_proc:task_depl})
and systems that perform job level deployment (discussed in
\s{systems_proc:job_depl}).

\paragraph{Data and state management}

Deployment and execution strategies affect the implementation of the data bus.
In the case of job-level deployment, the data bus is implemented using
ephemeral, push-based communication channels between tasks (e.g., direct TCP
connections).  In the case of task-level deployment, the data bus is mediated
and implemented by a persistent service (e.g., a distributed filesystem or a
persistent message queuing system) where upstream tasks push the results of
their computation and downstream tasks pull them when activated.
A persistent data bus can be replicated for fault tolerance, as in the case of
Kafka Streams, which builds on replicated Kafka topics.
CIEL~\cite{murray:NSDI:2011:CIEL} and Dryad~\cite{isard:EuroSys:2007:Dryad}
support hybrid bus implementations, where some connections may be direct while
others may be mediated.
Data elements may range from arbitrary strings (unstructured data) to specific
schemas (structured data).  The latter offer opportunities for optimizations
in the serialization process, for instance allowing for better compression or
for selective deserialization of only the fields that are accessed by a given
task.
In general, DPSs do not provide shared state.  Stream processing and graph
processing systems include a task state to persist information across multiple
activations of a continuous job (for instance, windows).
In absence of shared state, DPSs do not provide group atomicity or isolation
properties.
Almost all systems provide exactly once delivery, under the assumption that
sources can persist and replay data in the case of failure and sinks can
distinguish duplicates.  The concrete approaches to provide such guarantee
depend on the type of deployment (task-level or job-level) and are discussed
later in \s{systems_proc:task_depl} and \s{systems_proc:job_depl}.
Order is relevant for stream processing systems: with the exception of
Storm~\cite{toshniwal:SIGMOD:2014:Storm}, all stream processing systems
support timestamped data (event or ingestion time semantics).
Most systems deliver events in order, under the assumptions that sources
either produce data with a predefined maximum delay or inform the system about
the progress of time using special metadata denoted as watermark.
Kafka Streams takes a different approach: it does not wait for out of order
elements and immediately produces results.  In the case new elements arrive
out of order, it retracts updates the previous results.

\paragraph{Fault tolerance}

All DPSs detect faults using a leader-worker architecture and, in absence of a
shared state, they recover from failures through the mechanisms that guarantee
exactly once delivery.

\paragraph{Dynamic reconfiguration}

DPSs use dynamic reconfiguration to adapt to the workload by adding or
removing slots.
Systems that adopt task-level deployment can decide how to allocate resources
to individual tasks when they are activated, while systems that adopt
job-level deployment need to suspend and resume the entire job, which
increases the overhead for performing a reconfiguration.
The mechanisms that dynamically modify the resources (slots) available to a
DPS can be activated either manually or by an automated
service that monitors the utilization of resources and implements the
allocation and deallocation policies.
All commercial systems implement automated reconfiguration, frequently by
relying on external platforms for containerization, such as
Kubernetes
, or for cluster resources management, such as
YARN. 
The only exceptions for which we could not find official support for automated
reconfiguration are Storm~\cite{toshniwal:SIGMOD:2014:Storm} and Kafka
Streams~\cite{bejeck:2018:KafkaStreams}.

\subsection{Dataflow with task-level deployment}
\label{sec:systems_proc:task_depl}

Systems that belong to this class deploy tasks on activation, when their input
data becomes available.  Tasks store intermediate results on a persistent data
bus, which enables to selectively restart them in the case of failure.  This
approach is best suited for long running batch jobs and was pioneered in the
MapReduce batch processing system~\cite{dean:CACM:2008:mapreduce}.  It has
been widely adopted in various extensions and generalizations.  HaLoop
optimizes iterative computations by caching loop-invariant data and by
co-locating tasks that reuse the same data across
iterations~\cite{bu:VLDB:2010:HaLoop}.  Dryad~\cite{isard:EuroSys:2007:Dryad}
generalizes the programming model to express arbitrary dataflow plans and
enables developers to flexibly select the concrete channels (data bus in our
model) that implement the communication between tasks.
CIEL~\cite{murray:NSDI:2011:CIEL} extends the dataflow model of Dryad by
allowing tasks to create other tasks dynamically, based on the results of
their computation.
Spark is the most popular system of this
class~\cite{zaharia:CACM:2016:spark}: it inherits the dataflow model of Dryad
and supports iterative execution and data caching like HaLoop.
Spark Streaming~\cite{zaharia:SOSP:2013:Discretized_Streams} implements
streaming computations on top of Spark by splitting the input stream into
small batches and by running the same job for each batch.  It implements task
state using native Spark features: the state of a task after a given
invocation is implicitly stored as a special data item that the task receives
as input in the subsequent invocation.

In systems with task-level deployment, job compilation considers dynamic
information to create tasks: for instance, the number of tasks instantiated to
perform a data-parallel operation depends on how the input data is
partitioned.
Similarly, the deployment phase uses dynamic information to submit tasks to
workers running as close as possible to the their input data.  Hadoop
(the open source implementation of MapReduce) and Spark adopt a delay
scheduling~\cite{zaharia:EuroSys:2010:Delay}.  They put jobs (and their tasks)
in a FIFO queue.  When slots become available, the first task in the queue is
selected: if the slot is located near to the input data for the task, then the
task is immediately deployed, otherwise each task can be postponed for some
time to wait for available slots closer to their input data.
Task-level deployment enables sharing of compute resources with other
applications: in fact, most of the systems that use this approach can be
integrated with cluster management systems.

Task-level deployment also influences how systems implement fault tolerance
and ensure exactly once delivery of results.
Batch processing systems simply re-execute the tasks involved in the
failure.  In absence of state, the results of a task depend only on input data
and can be recomputed at need.  Intermediate results may be persisted on
durable storage and retrieved in the case of a failure or recomputed from the
original input data.
Spark Streaming~\cite{zaharia:SOSP:2013:Discretized_Streams} adopts the same
fault tolerance mechanism for streaming computations.  It segments a stream
into a sequence of so-called micro-batches and executes them in order.  Task
state is treated as a special form of data, it is periodically persisted to
durable storage and is retrieved in the case of a failure.  Failure recovery
may require activating failed tasks more than once, to recompute the task
state from the last state persisted before the failure.

Dynamic reconfiguration is available in all systems that adopt task-level
deployment.  Systems that do not provide any state abstraction can simply
exploit new slots to schedule tasks when they become available, and remove
workers when idle.  In the presence of task state, migrating a task involves
migrating its state across activations: as in the case of fault tolerance,
this is done by storing task state on persistent storage.

\subsection{Dataflow with job-level deployment}
\label{sec:systems_proc:job_depl}

In the case of job-level deployment, all tasks of a job are deployed onto the
slots of the computing infrastructure on job registration.  As a result, this
class of systems is better suited for streaming computations that require low
latency: indeed, no scheduling decision is taken at runtime and tasks are
always ready to receive and process new data.
Storm~\cite{toshniwal:SIGMOD:2014:Storm} and its successor
Heron~\cite{kulkarni:SIGMOD:2015:Twitter_Heron} are stream processing systems
developed at Twitter.  They offer lower-level programming API than previously
discussed dataflow systems, asking developers to fully implement the logic of
each processing step using a standard programming language.
Flink~\cite{carbone:IEEEB:2015:flink} is a unified execution engine for batch
and stream processing.  In terms of programming model, it strongly resembles
Spark, with a core API to explicitly define job plans as a dataflow of
functional operators, and domain specific libraries for structural
(relational) data, graph processing, and machine learning.  One notable
difference involves iterative computations: Flink supports them with native
operators (within jobs) rather than controlling them from the driver program.
Timely dataflow~\cite{murray:SOSP:2013:Naiad} offers a lower-level and more
general dataflow model than Flink, where jobs are expressed as a graph of
(data parallel) operators and data elements carry a logical timestamp that
tracks global progress.  Management of timestamps is explicit, and developers
control how operators handle and propagate them, which enables various
execution strategies: for instance, developers may choose to complete a given
computation step before letting the subsequent one start (mimicking a batch
processing strategy as implemented in MapReduce or Spark), or they may allow
overlapping of steps (as it happens in Storm or Flink).  The flexibility of
the model allows for complex workflows, including streaming computations with
nested iterations, which are hard or even impossible to express in other
systems.
The above systems rely on direct and ephemeral channels (typically, TCP
connections) to implement the data bus.  Kafka
Streams~\cite{bejeck:2018:KafkaStreams} and
Samza~\cite{noghabi:VLDB:2017:Samza}, instead, build a dataflow processing
layer on top of Kafka~\cite{kreps:NetDB:2011:kafka} durable channels.
In systems that adopt job-level deployment, job compilation and deployment
only depend on static information about the computing infrastructure: for
instance, the number of tasks for data-parallel operations only depends on the
total number of slots made available in workers.
As a result, this class of systems does not support sharing resources with
other applications: all resources need to be acquired at job compilation,
which prevents scheduling decisions across applications at runtime.

We observed three approaches to implement fault tolerance and delivery
guarantees:
\begin{inparaenum}
\item Systems such as Flink and MillWheel~\cite{akidau:VLDB:2013:MillWheel}
  periodically take a consistent snapshot of the state.  The command to
  initiate a snapshot starts from sources and completes when it reaches the
  sinks.  In the case of failure, the last completed snapshot is restored and
  sources replay data that was produced after that snapshot, in the original
  order.  If sinks can detect and discard duplicate results, this approach
  guarantees exactly once delivery;
\item Storm acknowledges each data element delivered between two tasks:
  developers decide whether to use acknowledgements (and retransmit data if an
  acknowledgement is lost), providing at least once delivery, or not,
  providing at most once delivery;
\item Kafka Streams relies on the persistency of the data bus (Kafka): it
  stores the task state in special Kafka topics and relies on two-phase commit
  to ensure that upon activation a task consumes its input, updates its state,
  and produces results for downstream tasks atomically.  In the case of
  failure, a task can resume from the input elements that were not
  successfully processed, providing exactly once delivery (unless data
  elements are received out-of-order, in which case it retracts and updates
  previous results leading to at least once delivery).
\end{inparaenum}
These three mechanisms are also used for dynamic reconfiguration, as they
allow a system to resume processing after a new deployment.

\subsection{Graph processing}
\label{sec:systems_proc:graph}

Early dataflow systems were not well suited for iterative computations, which
are common in graph processing algorithms.  To overcome this limitation, an
alternative computational model was developed for graph processing, known as
vertex-centric~\cite{mcCune:CSur:2015:TLaV}.
In this model, pioneered by the Google Pregel
system~\cite{malewicz:SIGMOD:2010:Pregel}, jobs are iterative: developers
provide a single function that encodes the behavior of each vertex $v$ at each
iteration.  The function takes in input the current (local) state of $v$ and
the set of messages produced for $v$ during the previous iteration; it outputs
the new state of $v$ and a set of messages to be delivered to connected
vertices, which will be evaluated during the next iteration.  The job
terminates when vertices do not produce any message at a given iteration.
Vertices are partitioned across workers and each task is responsible for a
given partition.  Jobs are continuous, as tasks are activated multiple times
(once for each iteration) and store the vertex state across activations (in
their task state).  Tasks only communicate by exchanging data (messages
between vertices) over the data bus, which is implemented as direct channels.
One worker acts as a leader and is responsible for coordinating the iterations
within the job and for detecting possible failures of other workers.  Workers
persist their state (task state and input messages) at each iteration: in the
case of a failure, the computation restarts from the last completed iteration.
Several systems inherit and improve the original Pregel model in various ways:
\begin{inparaenum}[(i)]
\item By using a persistent data bus, where vertices can pull data when
  executed, to reduce the overhead for broadcasting state updates to many
  neighbor vertices~\cite{low:VLDB:2012:GraphLab};
\item By decoupling communication and processing in each superstep, to combine
  messagess and reduce the communication
  costs~\cite{gonzalez:OSDI:2012:PowerGraph};
\item By allowing asynchronous execution of supersteps, to reduce
  synchronization overhead and inactive time~\cite{low:VLDB:2012:GraphLab};
\item By optimizing the allocation of vertices to tasks based on topological
  information, to reduce the communication overhead;
\item By dynamically migrating vertices between tasks (dynamic
  reconfiguration) across iterations, to keep the load balanced or to place
  frequently communicating vertices on the same
  worker~\cite{chen:EuroSys:2018:G-Miner};
\item By offering sub-graph centric abstractions, suitable to express graph
  mining problems that aim to find sub-graphs with given
  characteristics~\cite{teixeira:SOSP:2015:Arabesque}.
\end{inparaenum}
For space sake, we do not discuss all systems that derived from Pregel here,
but the interested reader can refer to the detailed survey by McCune et
al~\cite{mcCune:CSur:2015:TLaV}.


\section{Other Systems}
\label{sec:systems_other}

This section includes all systems that do not clearly fall in either of the
two classes identified above.
Due to their heterogeneity, we do not provide a common overview, but we
organize and we discuss them within three main classes:
\begin{inparaenum}[(i)]
\item systems that support analytical jobs on top of shared state
  abstractions;
\item systems that propose new programming models;
\item systems that integrate concepts from both DMSs and DPSs in an attempt to
  provide a unifying solution.
\end{inparaenum}

\subsection{Computations on data management systems}
\label{sec:systems_other:computations_db}

DMSs are designed to execute lightweight jobs that read and modify a shared
state.  We identified a few systems that support some form of heavy-weight
job.

\subsubsection{Incremental computations}
\label{sec:systems_other:computations_db:incremental}

Percolator~\cite{peng:OSDI:2010:LargeScale} builds on top of the BigTable
column store and incrementally updates its shared state.
It adopts observer processes that periodically scan the shared state: when they
detect changes, they start a computation that may update other tables with its
results.
In Percolator, computations are broken down into a set of small updates to the
current shared state.  This differentiates it from DPSs, which are not
designed to be incremental.  For instance, Percolator can incrementally update
Web search indexes as new information about Web pages and links become
available.
Percolator jobs may involve multiple shared-state elements, and the system
ensures group atomicity using using two-phase commit and group isolation using
a timestamp-based protocol.

\subsubsection{Long-running jobs}
\label{sec:systems_other:computations_db:long}

F1~\cite{shute:VLDB:2013:F1} implements a SQL query executor on top of
Spanner.
It supports long-running jobs, which are compiled to execution plans where the
tasks (or at least part of them) are organized into a dataflow to enable
distributed execution as in DPSs.
F1 also introduces optimistic transactions (jobs), which consist of a read
phase to retrieve all the data needed for the computation and a write phase to
store the results.  The read phase does not block other concurrent jobs, so
they can run for long time (as in the case of analytical jobs).  The write
phase completes only if no conflicting updates from other jobs occurred during
the read phase.

\subsubsection{Graph processing}
\label{sec:systems_other:computations_db:graph}

In graph data stores, long-running jobs appear as computations that traverse
multiple hops of the graph (for instance, jobs that search for paths or
patterns in the graph) or as iterative analytical jobs (such as vertex-centric
computations).
Trinity~\cite{shao:SIGMOD:2013:Trinity} inherits the same model of NoSQL graph
stores such as TAO, but implements features designed specifically to support
long-running jobs.  It lets users define the communication protocols that
govern the exchange of data over the data bus, to optimize them for each
specific job: for instance, data may be buffered and aggregated at the sender
or at the receiver.  It checkpoints the intermediate state of a long-running
jobs to resume it in the case of failure.

\subsection{New programming models}
\label{sec:systems_other:programming_models}

\subsubsection{Stateful dataflow}
\label{sec:systems_other:programming_models:stateful_dataflow}

The absence of shared mutable state in the dataflow model forces developers to
encode all information as data that flows between tasks.  However, some
algorithms would benefit from the availability of state that can be modified
in-place, for instance machine learning algorithms that iteratively refine a
set of parameters.
Thus, several systems propose extensions to the dataflow programming model
that accommodate shared mutable state.
In stateful dataflow graphs
(SDG)~\cite{fernandez:ATC:2014:MakingStateExplicit}, developers write a driver
program using imperative (Java) code that includes state and methods to access
and modify it.  Code annotations are used to specify state access patterns
within methods.
The resulting jobs are compiled into a dataflow graph where operators access
the shared state. If possible, state elements are partitioned across workers,
otherwise they are replicated in each worker and the programming model
supports user-defined functions to merge changes applied to different
replicas.
Deployment and exectution rely on a DPS with job-level
deployment~\cite{castro:SIGMOD:2013:SEEP}.

Tangram~\cite{huang:ATC:2019:Tangram} implements task-based deployment and
allows tasks to access and update an in-memory key-value store as part of
their execution.
By analyzing the execution plan, Tangram can understand which parts of the
computation depend on mutable state and which parts do not, and optimizes
fault tolerance for the job at hand.

TensorFlow~\cite{abadi:ATC:2016:TensorFlow} is a library to define machine
learning models.  Jobs represent models with transformations (tasks) and
variables (shared state elements).  As strong consistency is not required for
the application scenario, tasks can execute and update variables
asynchronously, with only barrier synchronization at each step of an iterative
algorithm.  While TensorFlow was conceived for distributed execution, other
machine learning libraries such as
PyTorch\extendedonly{\footnote{\url{https://pytorch.org}}} were initially
designed for single-machine execution and later implemented distributed
training using the same approach as TensorFlow.

\subsubsection{Relational actors}
\label{sec:systems_other:programming_models:reactors}

ReactDB~\cite{shah:SIGMOD:2018:reactors} extends the actor-based programming
model\extendedonly{~\cite{agha:90:actors}} with data management concepts such
as relational tables, declarative queries, and transactional semantics.
It builds on logical actors that embed state in the form of relational tables.
Actors can query their internal state using a declarative language and
asynchronously send messages to other actors.
ReactDB lets developers explicitly control how the shared state is partitioned
across actors.  Jobs are submitted to a coordinator actor that governs their
execution.  The system guarantees transactional semantics for the entire
duration of the job, across all actors that are directly or indirectly invoked
by the coordinator.

\subsection{Hybrid systems}
\label{sec:systems_other:hybrid_systems}

Several works aim to integrate data management and processing within a unified
solution.
S-Store~\cite{cetintemel:VLDB:2014:sstore} integrates stream processing
capabilities within a transactional database.  It uses an in-memory store to
implement the shared state (visible to all tasks), the task state (visible
only to individual tasks of stream processing jobs), and the data bus (where
data flowing from task to task of stream processing jobs is temporarily
stored).  S-Store uses the same concepts as
VoltDB~\cite{stonebraker:IEEEB:2013:voltdb} to offer transactional guarantees
with low overhead.  Data management and stream processing tasks are scheduled
on the same engine in an order that preserves transactional semantics and is
consistent with the dataflow.
S-Store unifies input data (for streaming jobs) and invocations (of data
management jobs, in the form of stored procedures): this is in line with the
conceptual view we provide in \s{model}.

SnappyData~\cite{barzan:CIDR:17:SnappyData} has a similar goal to S-Store but
a different programming and execution model.  It builds on Spark and
Spark Streaming, and augments them with the ability to access a key-value
store (shared state) during their execution.
In the attempt to efficiently support heterogeneous types of jobs, SnappyData
lets developers select how to organize the shared state, for instance in terms
or format (row oriented or column oriented), partitioning, and replication.
It supports group atomicity and isolation using two-phase commit and
multi-version concurrency control, and integrates fault detection and recovery
mechanisms for Spark tasks and their effects on the shared state.

StreamDB~\cite{chen:SCC:18:streamDB} and
TSpoon~\cite{affetti:JPDC:2020:tspoon} take the opposite approach with respect
to S-Store by integrating data management capabilities within a stream
processor.
StreamDB models database queries as stream processing jobs that receive
updates from external sources and output new results to sinks.  Stream
processing tasks can read and modify portions of a shared state: all database
queries that need to access a given portion will include the task responsible
for that portion.
StreamDB ensures group atomicity and isolation without explicit locks:
invocation of jobs are timestamped when first received by the system and each
worker executes tasks from different jobs in timestamp order.
TSpoon does not provide a shared state, but enriches the dataflow model with:
\begin{inparaenum}[(i)]
\item the ability to read (query) task state on demand;
\item transactional guarantees in the access to task state.
\end{inparaenum}
Developers can identify portions of the dataflow graph (denoted as
transactional sub-graphs) that need to be read and modified with group
atomicity and isolation.
TSpoon implements atomicity and isolation by decorating the dataflow graph
with additional operators that act as transaction managers.  It supports
different levels of isolation (from read committed to serializable) with
different tradeoffs between guarantees and overhead.

Hologres~\cite{jiang:VLDB:2020:Hologres} is used within Alibaba to execute
both analytical jobs and interactive jobs.
The system is designed to support high volume ingestion data from external
sources and continuous jobs that derive information to be stored in the shared
state or to be presented to external sinks.
The shared state that is partitioned across workers.  A worker stores an
in-memory representation of the partition it is responsible for and delegates
durability to an external storage service.
The distinctive features of the system are:
\begin{inparaenum}[(i)]
\item a structured data model where state is represented as tables that can be
  physically stored row-wise or column-wise depending on the access pattern;
\item a scheduling mechanism where tasks are deployed and executed onto
  workers based on load-balancing and prioritization of jobs that require low
  latency.
\end{inparaenum}


\section{Discussion}
\label{sec:discussion}

In building and discussing our model and taxonomy, we derived several
observations.  We report them in this section, pointing out ideas for
future research.

\fakeparagraph{State and data management}
The dichotomy between DMSs and DPSs is frequently adopted in the literature,
but not defined in precise terms.  Our model makes the characteristics that
contribute to this dichotomy clear and explicit, introducing the state
component and a sharp distinction between shared and task state.
DMSs offer primitives to read and modify a mutable shared state, while DPSs
target computationally expensive data transformations and do not support state
at all, or support it only within individual tasks.
This distinction brings together other differences (which we made explicit
with the classification criteria in \tab{criteria:state}) such that the two
classes of systems complement each other and are often used in combination to
support heterogeneous workloads.

This complementarity pushed researchers to extend their DMSs or DPSs to break
the dichotomy, adding features typical of the other class.  For instance, some
DMSs such as AsterixDB support long-lasting queries using the dataflow
processing model typical of DPSs, while recent versions of stream DPSs such as
Flink and Kafka started to offer primitives to access their task state with
read-only queries (one-shot jobs).
This triggered interesting research on declarative APIs that integrate
streaming data and state changes into a unifying
abstraction~\cite{sax:BIRTE:2018:StreamsTables}.
Few systems, such as S-Store and TSpoon, pushed this effort even further,
integrating transactional semantics within stream DPSs.

Future research efforts could continue to explore approaches that extend the
capabilities of individual systems, with the goal of better supporting hybrid
workloads that demand for both state management and data processing
capabilities, reducing the need to deploy many different systems, thus
simplifying the overall architecture of data-intensive applications.

\fakeparagraph{Coordination avoidance}
In distributed scenarios, the coordination between workers may easily become a
bottleneck.  Avoiding or reducing coordination is a recurring principle we
observed in all data-intensive systems.
Most DPSs circumvent this problem by forcing developers to think in terms of
functional and data-parallel transformations.  As state is absent or local to
tasks, tasks may freely proceed in parallel.  Coordination, if present, is
limited to barrier synchronization in systems that support iterative jobs
(e.g., iterative dataflow systems and graph processing systems).

Conversely, DMSs require coordination to control concurrent access to shared
state from multiple jobs.  Indeed, the approach to coordination is the main
criterion we used to classify them in \s{systems_db}.
NoSQL systems partition state by key: they either only support jobs that
operate on individual keys or relinquish group guarantees for jobs that span
multiple keys, effectively treating accesses to different keys as if they came
from independent jobs that do not coordinate with each other.
NewSQL systems do not entirely avoid coordination, but try to limit the
situations in which it is required or its cost. In our analysis we identified
four main approaches to reach this goal:
\begin{inparaenum}
\item use of precise clocks~\cite{corbett:TOCS:2013:Spanner},
\item pre-ordering of jobs and deterministic
  execution~\cite{thomson:SIGMOD:2012:Calvin},
\item explicit partitioning strategies to maximize jobs executed
  (sequentially) in a single slot~\cite{stonebraker:IEEEB:2013:voltdb},
\item primary-based protocols that delegate the scheduling of all read-write
  jobs to a single worker~\cite{verbitski:SIGMOD:2017:Aurora}.
\end{inparaenum}
In addition, all DMSs adopt strategies that optimize the execution of
read-only jobs and minimize their impact on read-write jobs.  They include the
use of replicas to serve read-only jobs and multi-version concurrency control
to let read-only jobs access a consistent view of the state without
conflicting with read-write jobs.

An open area of investigation for future research is a detailed study of the
assumptions and performance implications of coordination avoidance strategies
under different workloads. This study could guide the selection of the best
strategies for the scenario at hand and open the room for dynamic adaptation
mechanisms.

\fakeparagraph{Architectures for data-intensive applications}
Data-intensive applications typically rely on complex software architectures
that integrate different data-intensive systems to harness their complementary
capabilities~\cite{davoudian:CSur:2020:BigData_SE}.
For instance, many scenarios require integrating OLTP (on line transaction
processing) workloads, which consist of read-write jobs that mutate the state
of an application (e.g., user requests in an e-commerce portal), and OLAP (on
line analytical processing) workloads, which consist of read-only analytic
jobs (e.g., analysis of sales segmented by time, product, and region).
To support these scenarios, software architectures typically delegate OLTP
jobs to DMSs that efficiently support concurrent read-only and read-write
queries (e.g., relational databases), and use DMSs optimized for read-only
queries (e.g., wide-column stores) for OLAP jobs.
The process of extracting data from OLTP systems and loading it into OLAP
systems is denoted as ETL (extract, transform, load), and is handled by DPSs
that pre-compute and materialize views to speedup read queries in OLAP systems
(e.g., by executing expensive grouping, joins, aggregates, as well as building
secondary indexes).
Traditionally, ETL was executed periodically by batch DPSs, with the downside
that analytical jobs do not always access the latest available data, while
recent architectural patterns (e.g., Lambda and Kappa
architectures~\cite{lin:InternetComputing:2017:LambdaKappa}) advocate the use
of stream DPSs for this task.

In general, the architectural patterns of data-intensive applications are in
continuous evolution~\cite{davoudian:CSur:2020:BigData_SE}, and our study
highlights a vast choice of diverse data-intensive systems, with partially
overlapping features.
Future research could build on our model and classification to simplify the
design of applications.
Indeed, while the primary goal of our model was to present the key
characteristics of data-intensive systems to researchers and practitioners
with diverse backgrounds, it may inspire high-level modeling frameworks to
capture the requirements of data-intensive applications and guide the design
of a suitable software architecture for the specific scenario at hand.
Recent work already explored similar model-driven development in the context
of stream processing applications~\cite{guerriero:TOSEM:2021:StreamGen}.

\fakeparagraph{Modular implementations}
Several data-intensive systems have a modular design, where the
functionalities of the system are implemented by distinct components that can
be developed and deployed independently.
This approach is well suited for cloud environments where individual
components are offered as services and can be scaled independently depending
on the workload.  Also, the same service can be used in multiple products: for
example, storage services, log services, lock services, key-value stores may
be used as stand-alone products or adopted as building block of a relational
DMS.
We observed this strategy in systems developed at
Google~\cite{chang:OSDI:06:bigtable, corbett:TOCS:2013:Spanner},
Microsoft~\cite{antonopoulos:SIGMOD:2019:Socrates}, and
Amazon~\cite{verbitski:SIGMOD:2017:Aurora}.

Future research could bring this idea forward, proposing more general
component models that promote re-usability and adaptation to heterogeneous
scenarios.  Our work may guide the identification of the abstract components
that build data-intensive systems, the interfaces they offer, the assumptions
they rely on, and the functionalities they provide.
These research efforts may complement the aforementioned study of
architectural patterns, promoting the definition of complex architectures from
predefined components.

\fakeparagraph{Wide area deployment}
The systems we analyzed are primarily designed for cluster deployment.
In DPSs, tasks exchange large volumes of data over the data bus and the
limited bandwidth of wide-area deployment may easily become a bottleneck.
Some DMSs support wide-area deployment through replication, but in doing so
they either drop consistency guarantees or implement mechanisms to reduce the
cost for updating remote replicas. For instance, deterministic
databases~\cite{thomson:SIGMOD:2012:Calvin} define an order for jobs and force
all replicas to follow this order, with no need to explicitly synchronize job
execution.

On the other hand, increasingly many applications work at a geographical scale
and the edge computing paradigm~\cite{shi:IoT:2016:Edge} is emerging to
exploit processing and storage resources at the edge of the network, close to
the end users.  Designing data-intensive systems that embrace this paradigm is
an important topic of investigation.

\fakeparagraph{New and specialized hardware}
The use of specialized hardware to improve the performance of data-intensive
systems is an active area of research.
Recent works study hardware acceleration for
DPSs~\cite{hoozemans:CSM:2021:FPGA} and DMSs~\cite{lee:VLDB:2021:RateupDB,
  fang:VLDB:2020:in-memory} using GPUs or FPGAs.  Offloading of tasks to GPUs
is also supported in recent versions of DPSs such as Spark and is a key
feature for systems that target machine learning problems, such as
TensorFlow~\cite{abadi:ATC:2016:TensorFlow}.

Open research problems in the area include: devising suitable programming
abstractions to simplify the deployment of tasks onto hardware accelerators,
building new libraries of tasks that may run onto hardware accelerators,
explore new types of accelerators.
More in general, the availability of new hardware solutions stimulates the
definition of design choices that better exploit the characteristics of those
solutions.
In the context of DMSs, non-volatile memory offers durability at nearly the
same performance as main-memory.  The interested reader can refer to the work
by Arulraj and Pavlo~\cite{arulraj:SIGMOD:2017:HowTo} that discusses the use
of non-volatile memory to implement a database system.
As pointed out in recent studies, another area of investigation consists in
using remote memory access to better exploit data locality and reduce data
access latency.  The potential of remote memory access has been pointed out in
recent studies both in the domain of
DMSs~\cite{ziegler:SIGMOD:2022:ScaleStore} and in the domain of
DPSs~\cite{delMonte:SIGMOD:2022:rethinking}.

\fakeparagraph{Dynamic adaptation}
Our model captures the ability of some systems to adapt to mutating workload
conditions (\tab{criteria:reconf}).  Many works use this feature to implement
automated control systems for DPSs that monitor the use of resources and adapt
the deployment to meet the quality of service specified by the users, while
using the minimum amount of resources.
The interested reader can refer to recent work on dynamic adaptation for
batch~\cite{baresi:TSE:2021:allocation} and
stream~\cite{cardellini:CSur:2022:adaptation} DPSs.

Future studies in the area of dynamic adaptation could intersect with topics
already presented in this section: in particular they may consider the
availability of geographically distributed processing, memory, and storage
resources, as well as heterogeneous and specialized hardware platforms.


\section{Conclusion}
\label{sec:conclusion}

This paper presented a unifying model for distributed data-intensive systems,
which defines a system in terms of abstract components that cooperate to offer
the system functionalities.  The model precisely captures the possible design
and implementation strategies for each component, with the assumptions they
rely on and the guarantees they provide.
From the model, we derive a list of classification criteria that we use to
organize state-of-the-art systems into a taxonomy and survey them,
highlighting their commonalities and distinctive features.
Our work can be useful for engineers that need to deeply understand the range
of possibilities to select the best systems for their application, but also to
researchers and practitioners that work on data-intensive systems, to acquire
a wide yet precise view of the field.


\newpage
\begin{appendix}
  \section{Summary of Terms and Conceptual Map}
\label{sec:terms}

\begin{figure}[hptb]
  \centering
  \includegraphics[width=\textwidth]{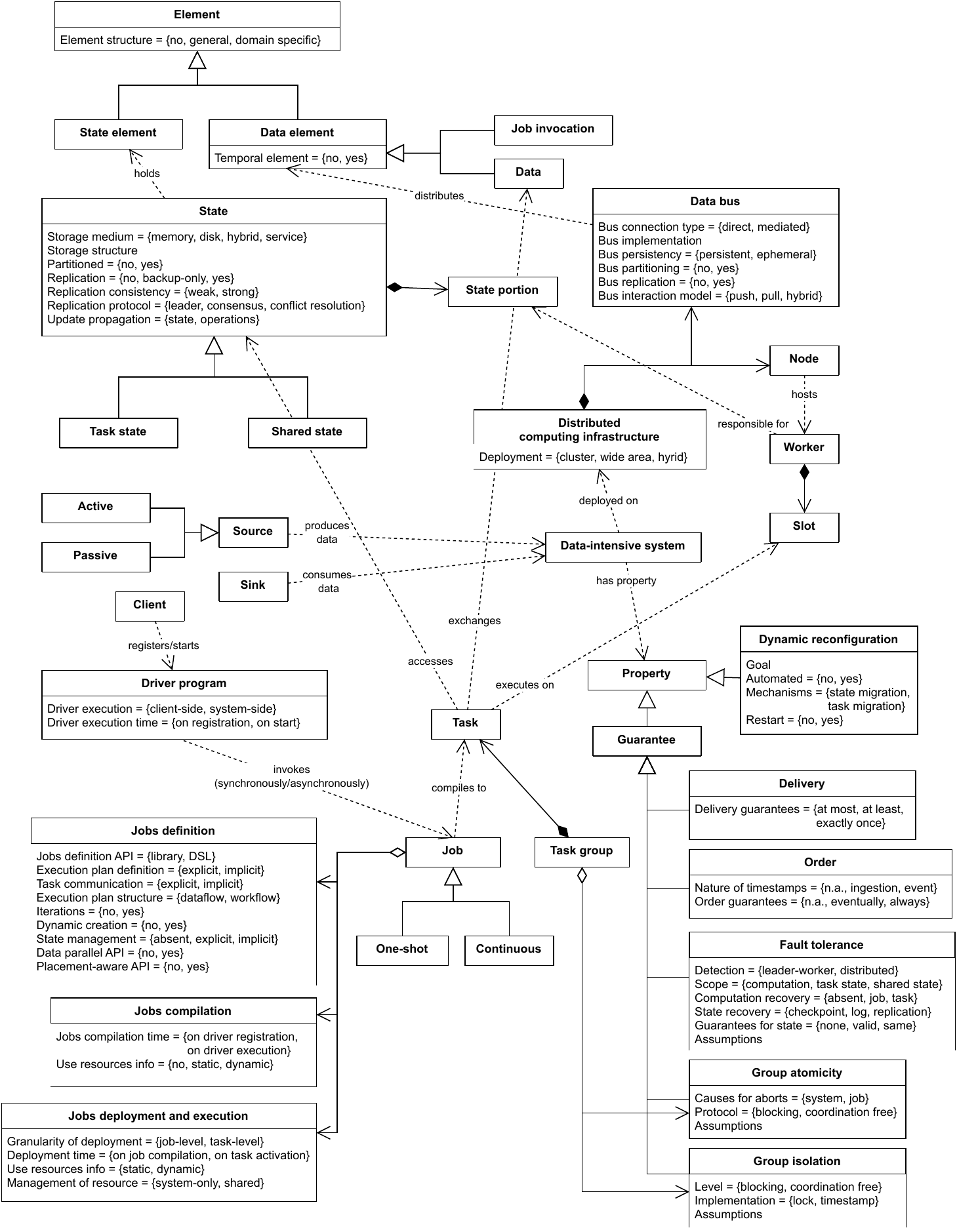}
  \caption{Conceptual map of all the concepts introduced in the unifying model
    for data-intensive systems (\s{model}) and their relations.}
  \label{fig:conceptual_map}
\end{figure}

This section presents additional resources to help the reader navigate through
the concepts discussed in the paper.

\begin{table}[hptb]
  \centering
  \tiny
  \rowcolors{2}{gray!25}{white}
  \renewcommand{\arraystretch}{0.94}
  \begin{tabular}[t]{|l|m{10cm}|}
    \rowcolor{gray!50}
    \hline
    \textbf{Term} & \textbf{Definition} \\
    \hline
    
    Client & Software component that exploits the functionalities offered by a
             data-intensive system by registering and starting driver
             programs. \\
    Driver program & Part of the application logic that interacts with the
                     data-intensive system and exploits its functionalities by
                     invoking one or more jobs. \\ 
    Node & Physical or virtual machine. \\
    Worker & Process running on a node. \\
    Slot & Processing resource unit offered by a worker. \\
    Distributed computing infrastructure & Set of nodes on top of which a
                                           distributed data-intensive system
                                           runs. \\
    Job & Largest unit of execution that can be offloaded onto the distributed
          computing infrastructure. \\
    One-shot job & Job that is executed once and terminates. Invoking the same
                   jobs multiple times leads to separate executions of the
                   same code. \\
    Continuous job & Job that persists across invocations (in this case, we
                     call them \emph{activations} of the same job). It may
                     persist some state across activations. \\ 
    Task & Elementary unit of execution.  Tasks derive from the compilation of
           a job.  They are executed sequentially on a slot. \\
    Data (elements) & Immutable units of information. Delivered through the
                      data bus. \\ 
    State (elements) & Mutable units of information.  Split into state
                       portions.  Stored within workers. \\
    Task state & State that is private to / accessible from a single task. \\
    Shared state & State that can be accessed simultaneously from multiple
                   tasks, belonging to the same or different jobs. \\
    Data bus & Communication channel that distributed data elements and jobs
               invocations. \\ 
    Source & Component that provides input data for the data-intensive
             system. \\
    Passive source & Source that provides a static input dataset. \\
    Active source & Source that provides a dynamic input dataset, that is,
                    continuously produces new data. \\
    Sink & Component that consumes output data from the data-intensive
           system. \\
    Execution plan & A workflow of tasks: it is the result of the compilation
                     of a job. \\
    Data parallel API & API where a computation is defined for a single data
                        element but executed in parallel on multiple
                        elements. \\
    Placement-aware API & API where developers can control or influence the
                          placement of tasks onto slots. \\
    Resources information & Information about the resources of the distributed
                            computing infrastructure and their use. \\
    Static resources information & Resources information that only considers
                                   the resources available in the distributed
                                   computing infrastructure. \\
    Dynamic resources information & Resources information that considers the
                                    use of the resources available in the
                                    distributed computing infrastructure. \\
    Group atomicity & Property of a group of tasks: ensures that either all of
                      the tasks complete successfully or none of them (and
                      none of their effects become visible). \\
    Group isolation & Property that constrains how tasks belonging to different
                      groups can interfere with each other. \\
    Delivery guarantees & Define how external components (driver programs,
                          sources, sinks) observe the effects of a single
                          input data element (or invocation). \\
    Order guarantees & Define the order in which external components (driver
                       programs, sources, sinks) observe the effects of multiple
                       data element (or invocation). \\
    Event time & Timestamp associated to data element by the original source
    \\
    Ingestion time & Timestamp associated to data elements when they first
                     enter the data-intensive system \\ 
    Watermark & Special input element containing a timestamp $t$. It is
                delivered by input components (e.g., sources) and indicates
                that the components will not produce any further data element
                with timestamp lower than $t$. \\
    Checkpointing & Process that stores a copy of state on durable storage. \\
    Logging & Process that stores individual operations or state changes on
              durable storage. \\
    Dynamic reconfiguration & Ability of a system to modify the deployment and
                              execution of jobs at runtime. \\
    \hline
  \end{tabular}
  \caption{Summary of terms introduced in the unifying model for
    data-intensive systems (\s{model}).}
  \label{tab:terms}
\end{table}

\fig{conceptual_map} organizes the entities introduced in our model into a
map, which highlights their relations.
The map adopts a UML-like notation, where entities are characterized by a
(possibly empty) list of attributes, and may be connected to each other
through different types of relations:
\begin{inparaenum}[(1)]
\item \emph{specialization} (empty arrow), when an entity is a specialization
  of a more general entity,
\item \emph{aggregation} (empty diamond), when an entity is included into a
  more comprehensive entity,
\item \emph{composition} (filled diamond), when an entity is constituting part
  of another entity,
\item \emph{dependency} (dashed arrow), when an entity depends or uses another
  entity, in which case we also denote the type of use as a label of the
  arrow.
\end{inparaenum}

\tab{terms} complements this map by reporting the definition of all the terms
we introduced in our unifying model for data-intensive systems (\s{model}) and
we used throughput the paper to classify and describe individual systems.


  \section{Data Management Systems}
\label{sec:systems_db_list}

\begin{table}[htpb]
  \centering
  \tiny
  \rowcolors{2}{gray!25}{white}
\begin{tabular}[t]{|l|c|c|c|c|c|c|c|}
    \hline
    \rowcolor{gray!50}
             & Driver exec       & Driver exec time   & Invoc of jobs & Sources & Sinks & State & Deployment   \\ \hline

\rowcolor{gray!50}
\multicolumn{8}{|c|}{NoSQL systems} \\ \hline

Dynamo       & client            & reg                & sync          & no      & no    & yes   & wide         \\
DynamoDB     & client            & reg                & sync          & no      & yes   & yes   & wide         \\
Redis        & client+sys\stproc & reg+start\stproc   & sync          & no      & yes   & yes   & hybrid       \\
BigTable     & client+sys\stproc & reg+start\stproc   & sync          & no      & no    & yes   & hybrid       \\
Cassandra    & client            & reg                & sync+async    & no      & yes   & yes   & wide         \\
MongoDB      & client            & reg                & sync          & active  & yes   & yes   & cluster      \\
\hline    
  \end{tabular}
  \caption{Data management systems: functional model. Legend: \stproc = stored procedures.}
  \label{tab:db:functional}
\end{table}

\begin{table}[tpb]
  \centering
  \tiny
  \rowcolors{2}{gray!25}{white}
  \begin{tabular}[t]{|l|c|c|c|c|c|c|c|c|c|c|}
    \hline
    \rowcolor{gray!50}
             & Jobs def API                                   & \Gape{\makecell{Exec plan\\def}} & \Gape{\makecell{Task\\comm}} & \Gape{\makecell{Exec plan\\struct}} & Iter & \Gape{\makecell{Dyn\\creat.}} & \Gape{\makecell{Nature\\of jobs}} & State man. & \Gape{\makecell{Data par\\API}} & \Gape{\makecell{Placem.\\-aware API}} \\ \hline

\rowcolor{gray!50}
\multicolumn{11}{|c|}{NoSQL systems} \\ \hline

Dynamo       & crud                                           & impl                             & impl                         & 1 task (put/get)                    & no   & no                            & one-shot                          & expl       & no                              & no                  \\
DynamoDB     & crud+scan                                      & impl                             & impl                         & 1 task (put/get)                    & no   & no                            & one-shot                          & expl       & no                              & no                  \\
Redis        & \Gape{\makecell{crud+scan\\+ops on vals}}      & impl                             & impl                         & 1 task (put/get)                    & no   & no                            & one-shot                          & expl       & no                              & no                  \\
BigTable     & crud+scan                                      & impl                             & impl                         & 1 task (put/get)                    & no   & no                            & one-shot                          & expl       & no                              & no                  \\
\hline
  \end{tabular}
  \caption{Data management systems: jobs definition.}
  \label{tab:db:job_definition}
\end{table}

\begin{table}[tpb]
  \centering
  \tiny
  \rowcolors{2}{gray!25}{white}
  \begin{tabular}[t]{|l|c|c|c|c|c|c|}
    \hline
    \rowcolor{gray!50}
             & Jobs comp time & Use resources info (comp) & Granul of depl & Depl time & Use resources info (depl) & Manag of res \\ \hline

\rowcolor{gray!50}
\multicolumn{7}{|c|}{NoSQL systems} \\ \hline

Dynamo       & exec            & static               & job            & job comp   & static               & sys-only             \\
DynamoDB     & exec            & static               & job            & job comp   & static               & sys-only             \\
Redis        & exec            & static               & job            & job comp   & static               & sys-only             \\
BigTable     & exec            & static               & job            & job comp   & static               & shared               \\
Cassandra    & exec            & static               & job            & job comp   & static               & sys-only             \\
MongoDB      & exec            & static               & job            & job comp   & static               & sys-only             \\
\hline
  \end{tabular}
  \caption{Data management systems: jobs compilation and execution.}
  \label{tab:db:jobs_compilation_execution}
\end{table}

\begin{table}[tpb]
  \centering
  \tiny
  \rowcolors{2}{gray!25}{white}
  \begin{tabular}[t]{|l|c|c|c|c|c|c|c|c|}
    \hline
    \rowcolor{gray!50}
             & Elem struct            & Temp elem & Bus conn & Bus impl   & Bus persist & Bus partition & Bus repl & Bus inter \\ \hline

\rowcolor{gray!50}
\multicolumn{9}{|c|}{NoSQL systems} \\ \hline

Dynamo       & key-value              & no        & direct   & net chan   & ephemeral   & yes           & no       & pull      \\
DynamoDB     & key-value              & no        & direct   & net chan   & ephemeral   & yes           & no       & pull      \\
Redis        & key-value              & no        & direct   & net chan   & ephemeral   & yes           & no       & pull      \\
BigTable     & wide-col               & yes       & direct   & net chan   & ephemeral   & yes           & no       & pull      \\
Cassandra    & wide-col               & no        & direct   & net chan   & ephemeral   & yes           & no       & pull      \\
MongoDB      & document               & no        & direct   & net chan   & ephemeral   & yes           & no       & pull      \\
\hline
  \end{tabular}
  \caption{Data management systems: data management.}
  \label{tab:db:data}
\end{table}

\begin{table}[tpb]
  \centering
  \tiny
  \rowcolors{2}{gray!25}{white}
  \begin{tabular}[t]{|l|c|c|c|c|c|c|c|c|c|c|c|}
    \hline
    \rowcolor{gray!50}
             & Elem struct            & \Gape{\makecell{Stor\\medium}} & \Gape{\makecell{Stor\\struct}}  & Task st & Shared st & St part & Repl                        & Repl consist                                  & Repl prot                                           & \Gape{\makecell{Update\\propag}} \\ \hline

\rowcolor{gray!50}
\multicolumn{11}{|c|}{NoSQL systems} \\ \hline

Dynamo       & key-value              & agnostic    & agnostic                           & no      & yes       & yes     & yes                                          & conf                                          & \Gape{\makecell{quorum\\+confl}}                    & op                               \\
DynamoDB     & key-value              & disk        & B-trees                            & no      & yes       & yes     & yes                                          & conf                                          & leader                                              & unknown                          \\
Redis        & key-value              & mem         & user-def                           & no      & yes       & yes     & yes                                          & weak                                          & \Gape{\makecell{leader(clus)\\confl(wide)}}         & op                               \\
BigTable     & wide-col               & hybrid      & LSM trees                          & no      & yes       & yes     & \Gape{\makecell{backup(clus)\\yes(wide)}}    & weak                                          & confl                                               & op                               \\
Cassandra    & wide-col               & hybrid      & LSM trees                          & no      & yes       & yes     & yes                                          & conf                                          & \Gape{\makecell{quorum\\+confl}}                    & op                               \\
\hline
  \end{tabular}
  \caption{Data management systems: state management.}
  \label{tab:db:state}
\end{table}

\begin{table}[tpb]
  \centering
  \tiny
  \rowcolors{2}{gray!25}{white}
  \begin{tabular}[t]{|l|c|c|c|c|c|c|c|c|c|}
    \hline
    \rowcolor{gray!50}
             & Aborts  & Protocol   & Assumptions        & Level      & Impl     & Assumptions  & Delivery          & Nature of ts & Order \\ \hline

\rowcolor{gray!50}
\multicolumn{10}{|c|}{NoSQL systems} \\ \hline

Dynamo       & n.a.    & n.a.       & n.a.               & n.a.       & n.a.     & n.a.         & conf (most/exact) & no           & n.a.  \\
DynamoDB     & sys     & unknown    & cluster deployment & conf       & unknown  & cluster depl & conf (most/exact) & no           & n.a.  \\
Redis        & n.a.    & n.a.       & n.a.               & n.a.       & n.a.     & n.a.         & most              & no           & n.a.  \\
BigTable     & n.a.    & n.a.       & n.a.               & coord free & \seq     & \oneportion  & exact             & event        & n.a.  \\
Cassandra    & n.a.    & n.a.       & n.a.               & coord free & \seq     & \oneportion  & conf (most/exact) & no           & n.a.  \\
MongoDB      & sys+job & blocking   & n.a.               & blocking   & ts       & none         & conf (most/exact) & no           & n.a.  \\
\hline          
  \end{tabular}
  \caption{Data management systems: group atomicity, group isolation,
    delivery, order.  Legend: \detjob = jobs are deterministic; \seq = jobs
    are executed sequentially, with no interleaving; \onewrite = a single
    worker handles all writes; \oneportion = jobs access a single state
    portion; OCC = optimistic concurrency control.}
  \label{tab:db:guarantees}
\end{table}

\begin{table}[tpb]
  \centering
  \tiny
  \rowcolors{2}{gray!25}{white}
  \begin{tabular}[t]{|l|c|c|c|c|c|c|}
    \hline
    \rowcolor{gray!50}
             & Detection & Scope     & Comput recov & State recov     & Guarantees for state       & Assumptions \\ \hline

\rowcolor{gray!50}
\multicolumn{7}{|c|}{NoSQL systems} \\ \hline

Dynamo       & p2p       & shared st & n.a.         & repl            & none                       & none        \\
DynamoDB     & lead-work & shared st & n.a.         & log+checkp+repl & same                       & \durrepl    \\
Redis        & p2p       & shared st & n.a.         & log+checkp+repl & conf (none or same)        & \durrepl    \\
BigTable     & lead-work & shared st & n.a.         & log+repl        & same                       & \durstore   \\
Cassandra    & p2p       & shared st & n.a.         & log+repl        & none                       & none        \\
MongoDB      & lead-work & shared st & n.a.         & log+repl        & conf (none or same)        & \durrepl    \\
\hline
  \end{tabular}
  \caption{Data management systems: fault tolerance.  Legend: \durstore =
    storage layer is durable; \durrepl = replicated data is durable.}
  \label{tab:db:fault}
\end{table}

\begin{table}[tpb]
  \centering
  \tiny
  \rowcolors{2}{gray!25}{white}
  \begin{tabular}[t]{|l|c|c|c|c|c|c|}
    \hline
    \rowcolor{gray!50}
             & Goal                     & Automated & State migration & Task migration & Add/del slots & Restart \\ \hline

\rowcolor{gray!50}
\multicolumn{7}{|c|}{NoSQL systems} \\ \hline
    
Dynamo       & load balan+elast         & yes       & yes             & n.a.           & yes           & no      \\
DynamoDB     & load balan+elast         & yes       & yes             & n.a.           & yes           & no      \\
Redis        & load balan               & no        & yes             & n.a.           & yes           & no      \\
BigTable     & load balan               & yes       & yes             & n.a.           & yes           & no      \\
Cassandra    & load balan               & yes       & yes             & n.a.           & yes           & no      \\
MongoDB      & load balan               & yes       & yes             & n.a.           & yes           & no      \\
\hline
  \end{tabular}
  \caption{Data management systems: dynamic reconfiguration.}
  \label{tab:db:reconf}
\end{table}


\subsection{NoSQL systems}

\subsubsection{Key-value stores}

\fakeparagraph{Dynamo} Dynamo~\cite{deCandia:SOSP:2007:Dynamo} is a NoSQL
key-value store used by Amazon to save the state of its services.
State elements are arbitrary values (typically, binary objects) identified by
a unique key.  Jobs consist of operations on individual state elements:
retrieve the value associated to a key (get) or insert/update a value with a
given key (put).
Dynamo builds a distributed hash table: workers have an associated unique
identifier, which determines the portion of shared state (set of keys) they
are responsible for.  Shared state is replicated, so multiple workers are
responsible for the same key.
When a worker receives a client request for a key, it forwards the request to
one of the workers responsible for that key.  Clients may also be aware of the
distribution of shared state portions across workers, and use this information
to better route their requests.
Dynamo uses a quorum-based approach where read and write operations on a key
need to be processed by a quorum of the replicas responsible for that key.
Users can set the number of replicas for each key, the read quorum, and the
write quorum to trade consistency and durability for performance and
availability.  After a write quorum is reached, updates are asynchronously
propagated to remaining replicas.
In the case of transient unavailability of one replica, another node can
temporarily store writes on its behalf, which avoids blocking write operations
while preserving the desired degree of replication for durability.
When Dynamo is configured to trade consistency for availability, replicas may
diverge due to concurrent writes: to resolve conflicts, Dynamo adopts a
versioning system, where multiple versions of a given key may exist at
different replicas.  Upon write, clients specify which version they want to
overwrite: Dynamo uses this information to trace causality between updates and
automatically resolves conflicts when it can determine a unique causal order
of updates.  In the case of concurrent updates, Dynamo preserves conflicting
versions and presents them to clients (upon a read) for semantic
reconciliation.
Dynamic reconfiguration is a key feature of Dynamo.  Nodes can be added and
removed dynamically, and shared state portions automatically migrate.  Dynamo
adopts a distributed (gossip based) failure detection model, while failure
recovery takes place by simply redistributing state portions over remaining
nodes.

\fakeparagraph{DynamoDB}
DynamoDB\footnote{\url{http://aws.amazon.com/dynamodb/}} is an evolution of
Dynamo that aims to simplify operational concerns: it is fully managed and
offered as a service.  It exposes the same data model as Dynamo, but uses a
single-leader replication protocol, where all writes for a key are handled by
the leader for that key, which propagates them synchronously to one replica
(for durability) and asynchronously to another replica.  Read operations can
be either strongly or weakly consistent: the former are always processed by
the leader, while the latter may be processed by any replica, even if the
replica lags behind of updates.  Replicas can be located in different data
centers to support wide area deployments.
DynamoDB stores data on disk using B-trees and buffers incoming write requests
in a write-ahead log.  It supports secondary indexes that are asynchronously
updated from the log upon write.
Recent versions of DynamoDB enable automated propagation of changes to
external sinks.  They also offer an API to group individual operations in
transactions that provide group atomicity and configurable group isolation.
Transactions can be configured to be idempotent, thus offering exactly once
delivery.
To detect failures, the leader of each partition periodically sends heartbeats
to all replicas.  After some heartbeats are lost, the remaining nodes use the
Paxos consensus algorithm to elect a new leader.  As an additional fault
tolerance mechanism, B-trees and logs are periodically checkpointed to durable
storage.
Amazon offers automatic scaling of DynamoDB as a service: users can select the
expected read and write throughput for a given table (blocks of key-value
pairs) and DynamoDB automatically increases the amount of resources dedicated
to that table to meet the requirements.

\fakeparagraph{Redis} Redis~\cite{macedo:2011:Redis} is a single-node
in-memory key-value store.  Since version 3.0, Redis Cluster provides a
distributed implementation of the store.  In terms of data model, Redis
differs from other key-value stores in that it provides typed values and
optimized operations for those types: for instance, a value may be declared as
a list, which supports appending new elements without overwriting the entire
list.
Redis provides a scripting language to express driver programs (stored
procedures) that run system-side.
Redis supports data dispatching to sinks in the form of a publish-subscribe
service: clients can subscribe to a given key and be notified about the
changes to that key.
Users can configure their desired level of durability (for fault tolerance),
with options ranging from no persistence to using periodic checkpointing to
command logging.
Redis Cluster partitions data by key and uses single leader asynchronous
replication.  Alternative solutions are available for wide area deployments,
where redirecting all writes to a single leader may be unfeasible: for
instance, Redis CRDTs offer multi-leader replication with automated conflict
resolution based on conflict-free replicated data types.
Dynamic reconfiguration with migration of shared state portions is supported
but not automated.

\fakeparagraph{Other key-value stores}  Several other stores implement the
key-value model with design and implementation strategies that are similar to
those presented above.  For the sake of space, we only discuss their key
distinguishing factors.
Voldemort\footnote{\url{https://www.project-voldemort.com/}} and Riak
KV\footnote{\url{https://riak.com/products/riak-kv/}} follow the same design
as Dynamo.
Like Redis, Aerospike~\cite{srinivasan:VLDB:2016:Aerospike} adopts single
leader replication within a single data center and multi-leader replication
for wide area deployments.  It focuses on both horizontal scalability (with
workers on multiple nodes) and vertical scalability (with multiple workers on
the same node).  It adopts a hybrid storage model where key indexes are kept
in memory but concrete values can be persisted on disk, and a threading model
that reduces locking and contention by assigning independent shared state
portions to threads.
PNUTS~\cite{cooper:VLDB:2008:PNUTS} provides single leader replication in wide
area deployments: it uses an intermediate router component to dispatch jobs
invocations to responsible workers, and relies on an external messaging system
to implement the data bus that propagates updates to replicas.  The same
service can also notify external sinks.
Thanks to their simple interface, key-value stores are sometimes used as
building blocks in distributed software architectures: Memcached is used as a
memory-based distributed cache to reduce data access latency at
Facebook~\cite{nishtala:NSDI:2013:Memcached}.
RocksDB\footnote{\url{https://rocksdb.org}} is used to persist task state in
the Flink stream processing system (also discussed later).
Other data stores offering a key-value model are presented below as part of
NewSQL databases.

\subsubsection{Wide-column stores}

\fakeparagraph{BigTable} BigTable~\cite{chang:OSDI:06:bigtable} and its
open-source implementation HBase\footnote{\url{https://hbase.apache.org}} use
a wide-column data model: shared state is organized in tables, where each row
is associated to a fixed number of column families.  A column family typically
contains multiple columns that are frequently accessed together.  Tables
associate a value (binary object) to a row and a column (within a column
family), they are range-partitioned across workers by row and physically
stored (compressed) per column family.
Jobs can read and update individual values and perform table scans.  Rows are
units of isolation: accesses to columns of the same row from different jobs
are serialized, and this is the only task grouping guarantee that BigTable
offers.
BigTable adopts a leader-worker deployment, where a leader is responsible for
assigning shared state portions to workers.  Initially, each table is
associated to a single worker, but it is automatically split when it increases
in size.  Clients retrieve and cache information about state distribution and
submit their requests (tasks) involving a given state portion to the worker
responsible for that state portion.
BigTable can store multiple versions for each value.  Versions are also
visible to clients, which can retrieve old versions and control deletion
policies.  Writes always append new versions of a value, which improves write
throughput to support frequent updates.  Workers store recent versions in
memory in LSM trees and use an external storage service (the GFS distributed
filesystem) for durability.  Background compaction procedures prune old
versions from memory and from the storage.
BigTable uses replication only for fault tolerance.  Specifically, it relies
on the replication of the GFS storage layer, where it also saves a command
log.  In the case of failure of a worker, the latest snapshot of its shared
state portion is restored from GFS and from the commands in the log that were
not part of the snapshot.
BigTable supports wide area deployments by fully replicating the data store in
each data center.  Replicas in other data centers may be only used for fault
tolerance or they can serve client invocations, in which case they provide
eventual consistency.
Similar to key-value stores, BigTable provides dynamic reconfiguration by
migrating state across available workers.

\fakeparagraph{Cassandra} Cassandra~\cite{lakshman:SIGOPS:2010:cassandra}
combines the wide-column data model of BigTable and the distributed
architecture of Dynamo.  Like BigTable, Cassandra uses LSM trees with
versioning and background compaction tasks to improve the performance of
write-intensive workloads.  It offers a richer job definition language that
includes predefined and user-defined types and operations.  It supports task
grouping only for compare-and-swap operations within a single partition.
Like Dynamo, it uses a distributed hash table to associate keys to workers,
and provides replication both in cluster and wide-area deployments, using a
quorum-based approach for consistency.  The quorum protocol can be configured
to trade consistency and durability for performance, setting the number of
local replicas (within a data center) and global replicas (in the case of wide
area deployments) that need to receive and approve a task before the system
returns to the client.  In the case of weak (eventual) consistency, Cassandra
uses an anti-entropy protocol to periodically and asynchronously keep replicas
up-to-date, using automated conflict resolution (last write wins).

\subsubsection{Document stores}

\fakeparagraph{MongoDB} MongoDB~\cite{chodorow:2013:MongoDB} is representative
of document stores, an extension of key-value stores where values are
semistructured documents, such as binary JSON in the case of MongoDB.  MongoDB
jobs can insert, update, and delete entire documents, but also scan, retrieve,
and update individual values within documents.  Recent versions of MongoDB
support simple data analytic jobs expressed as pipelines of data
transformations.  External systems can register as sinks to be notified about
changes to documents.
Shared state partitioning can be either hash-based or range-based.  Clients
are oblivious of the location of state portions and interact with the
data store through a special worker component that acts as a router.
Shared state portions can be replicated for fault tolerance only or also to
serve read queries.  Replication is implemented using a single leader protocol
with semi-synchronous propagation of changes, where clients can configure the
number of replicas that need to synchronously receive an update, thus trading
durability and consistency for availability and response time.
By default, only single-document jobs are atomic.  Recent versions also
support distributed transactions using two-phase commit for atomicity and
multi-version concurrency control for snapshot isolation.

\fakeparagraph{CouchDB} CouchDB~\cite{anderson:2010:CouchDB} adopts the same
document data model as MongoDB.  Early versions only support complete
replication of shared state, allowing clients to read and write from any
replica to improve availability.  Replicas are periodically synchronized and
conflicts are handled by storing multiple versions of conflicting documents or
fields, delegating resolution to users.
Since version 2.0, CouchDB provides a cluster mode with support for shared
state partitioning and quorum-based replication, where users can configure the
number of replicas for shared state portions, and quorum for read and write
operations, thus balancing availability and consistency.
CouchDB is conceived for Web applications and provides a synchronous HTTP API
for job invocation.  It makes the list of changes (change feed) to a document
available for external components, which can consume it either in pull mode or
in push mode (thus representing the sink components in out model).
CouchDB lets users define multiple views for each document.  In our model, we
can see views as the results of registered jobs that are triggered by changes
to documents.  Computations that create views are restricted to execute on
individual documents, but their results can then be aggregated by key
(mimicking the MapReduce programming model).
CouchDB supports dynamic reconfiguration with addition and removal of nodes,
and redistribution of shared state portions across nodes.  However,
reconfiguration is a manual procedure.

\fakeparagraph{AsterixDB}
AsterixDB\footnote{\url{https://asterixdb.apache.org}} is a semi-structured
(document) store born as a research project that integrates ideas from
NoSQL databases and distributed processing platforms.
AsterixDB offers a SQL-like declarative language that integrates operators for
individual documents as well as for multiple document (like joins, group by).
Jobs are converted into a dataflow format and run on the Hyracks data-parallel
platform~\cite{borkar:ICDE:2011:Hyracks}.  Interestingly, the platform deploys
jobs task-by-task, but does not rely on a persistent data bus.  Rather, when
all input data for a task become ready, it dynamically establishes network
connections from upstream tasks that deliver the results of their computation
before terminating.
AsterixDB supports querying shared state as well as external data from active
and passive sources.  Like BigTable and Cassandra, it stores state in LSM
trees, which improve the performance of write
operations~\cite{alsubaiee:VLDB:2014:AsterixDB}.  It supports partitioning but
does not currently support replication.  As part of its shared state,
AsterixDB can store indexes to simplify accessing external data.
It offers isolation only for operations on individual values, using locking to
update indexes.  Its data structure offer fault tolerance through logging, but
it does not currently implement fault detection nor dynamic reconfiguration
mechanisms.

\subsubsection{Time-series stores}

\fakeparagraph{InfluxDB} InfluxDB\footnote{\url{https://www.influxdata.com}}
is a DMS for time series data.  Shared state is organized into measurements,
which are sparse tables resembling wide columns in BigTable and Cassandra.
Each row maps a point in time (primary key) to the values of one or more
columns.  Developers need to explicitly state which columns are indexed and
which are not, thus balancing read and write latency.
InfluxDB uses a storage model (time structured merge -- TSM trees) that
derives from LSM trees.  It stores measurements column-wide and integrates
disk-based storage and an in-memory write cache to improve write performance.
Jobs are defined in the InfluxQL declarative language and are restricted to
single measurements.  InfluxDB supports active sources, and mimics continuous
jobs through periodic execution of one-shot jobs, which write their results
inside the database and/or send them to sinks.
InfluxDB partitions and replicates shared state across workers.  Write
operations are propagated semi-synchronously across replica workers for fault
tolerance.  Read operations can access any of the replica workers that contain
the requested data, leading to weak consistency.
InfluxDB adopts a leader-worker approach, where leader nodes store meta-data
about membership, data partitioning, continuous queries, and access rights,
and worker nodes store the actual data.  Leader nodes are replicated with
strong consistency for fault tolerance, using the Raft consensus protocol.
InfluxDB offers dynamic reconfiguration to migrate data and add new workers,
but the process is manual.

\fakeparagraph{Gorilla} Gorilla~\cite{pelkonen:VLDB:2015:Gorilla} is an
in-memory time series store that Facebook uses as a cache to an HBase data
store.  HBase stores historical data compressed with a coarser time
granularity, while Gorilla persists the most recent data (26 hours) in memory.
Gorilla uses a simple data model where values for each measure always consist
of a 64-bit timestamp and a 64-bit floating point number.  It uses an encoding
scheme based on bit difference that reduces the data size by 12 times on
average.
Jobs in Gorilla only perform simple read, write, and scan operations.  Few
ad-hoc jobs have been implemented to support correlation between time series
and in-memory aggregation, which is used to compress old data before writing
it to HBase.
Gorilla supports geo-replication for disaster recovery, but trades durability
for availability.  Written data is asynchronously replicated and it can be
lost before it is made persistent.
Gorilla supports dynamic adaptation by re-distributing the key space across
workers.

\fakeparagraph{Monarch} Monarch~\cite{adams:VLDB:2020:Monarch} is a
geo-distributed in-memory time series store designed for monitoring
large-scale systems and used within Google.  Its data model stores time series
data as schematized tables.  Each table consists of multiple key columns that
form the time series key, and a value column, which stores one value for each
point in the history of the time series.  Key columns include a target field,
which is the entity that generates the time series, and a metrics field, which
represents the aspect being measured.
Monarch has a hierarchical architecture.  Data is stored in the zone (data
center) in which it is generated and sharded (by key ranges,
lexicographically) across nodes called leaves.  Data is stored in main memory
and asynchronously persisted to logs on disk, trading durability to reduce
write delay.
Monarch offers a declarative, SQL-like language to express jobs, which can be
either one-shot or continuous.  In the latter case, they are evaluated
periodically and store their results in new derived tables (materialized
views).  Jobs are evaluated hierarchically: nodes are organized in three
layers (global, zone level, leaves), and the job plan pushes tasks as close as
possible to the data they need to consume.
Each level also stores an approximate view (compressed index) of what the
nodes in the lower level store.  This enables optimizing communication across
levels by avoiding pushing tasks to nodes that have no data related to that
task.
Monarch supports dynamic reconfiguration: it monitors lower-level nodes and
re-distributes data (key ranges) across nodes to adapt to changes in the load.

\fakeparagraph{Peregreen} The design of the
Peregreen~\cite{visheratin:ATC:2020:Peregreen} time series database aims to
satisfy the following requirements.
\begin{inparaenum}
\item Cloud deployment of large volumes of historical data: as the scale of
  data is prohibitive for in-memory solutions, Peregreen relies on storage
  services such as distributed filesystems or block storage.  It limits raw
  data footprint by supporting only numeric values and by representing them in
  a compressed columnar format: each column is split into chunks and data in
  each chunk is represented using differences between adjacent values (delta
  encoding), further compressed to form a single binary array.
\item Fast retrieval of data through indexing: Peregreen uses a three-tier
  data indexing, where each tier pre-computes aggregated statistics (minimum,
  maximum, average, etc.) for the data it references.  This allows to quickly
  identify chunks of data that satisfy some conditions based on the
  pre-computed statistics and to minimize the number of interactions with the
  storage layer.
\end{inparaenum}
Peregreen jobs can only insert, update, delete, and retrieve data elements.
Retrieval supports limited conditional search (only based on pre-computed
statistics) and data transformation.  Chunks are versioned and a new version
is created in the case of modifications.
Peregreen is designed for cluster deployments and uses the Raft algorithm to
reach consensus on the workers available at any point in time and the state
portions (indexes to the storage layer) they are responsible for.  This
enables dynamic adaptation, with addition and removal of workers for
load-balancing and elasticity.  State portions are replicated at multiple
workers for fault tolerance.

\subsubsection{Graph stores}

\fakeparagraph{TAO} TAO~\cite{bronson:ATC:2013:TAO} is a data store that
Facebook developed to manage its social graph, which contains billions of
entities (such as people and locations) and relations between them (such as
friendship).
TAO offers a simple data model where entities and relations have a type and
may contain data in the form of key-value pairs.  It provides a restricted API
for job definition, to create, delete, and modify entities and relations, and
to query relations for a given entity.  With respect to other graph data
stores, it does not support queries that search for subgraphs that satisfy
specific constraints (path queries or subgraph pattern queries).
TAO is designed to optimize the latency of read jobs, as it needs to
handle a large number of simultaneous user-specific queries.
To do so, TAO implements shared state in two layers: a persistent storage
layer based on a relational database (MySQL) and a key-value in-memory cache
based on memcache.
TAO is designed for wide area deployments.  Within a single data center, both
the persistent layer and the cache are partitioned.  The cache is also
replicated using a single-leader approach: clients always interact with
follower cache servers, which reply to read operations in the case of cache
hit and propagate read operations in the case of cache miss as well as write
operations to the leader cache server, which is responsible for interacting
with the storage layer and for propagating changes.  Across data centers, the
storage layer is fully replicated with a single-leader approach: reads are
served using the data center local cache or storage layer, so they do not
incur latency, while all write operations are propagated to the leader data
center for the storage layer.
Data is replicated between the storage layer and the cache as well as across
data centers asynchronously, which provides weak consistency and durability.
Dynamic reconfiguration is possible in the case of failures: in the case a
leader cache or storage server fails, replicas automatically elect a new
leader.

\fakeparagraph{Unicorn} Unicorn~\cite{curtiss:VLDB:2013:Unicorn} is an
indexing system used at Facebook to search its social graph.
The shared state of Unicorn consists of inverted indexes that enable
retrieving graph entities (vertices) based on their relations (edges) and the
data associated to them: for instance, in a social graph, one could use an
index on relations to retrieve all people (vertices) that are in a friend
relation with a given person, or a string prefix index to retrieve all people
with a name that starts with a given prefix.
Unicorn is optimized for read-only queries (jobs), in the form of index
lookups and set operations (union, intersection, difference) on lookup
results.  Jobs are evaluated by exploiting a hierarchical organization of
workers into three layers:
\begin{inparaenum}[(i)]
\item index servers store the shared state (the indexes) partitioned by
  results, meaning that any index server will store a subset of the results
  for each index lookup;
\item a rack aggregator per rack is responsible for merging the (partial)
  query results coming from individual index servers;
\item a top aggregator is responsible for merging the (partial) query results
  coming from each rack.
\end{inparaenum}
Interestingly, Unicorn jobs can dynamically start new tasks: this feature is
implemented as an \texttt{apply} function that performs new lookups starting
from the results obtained in previous ones.  This feature can be used to
implement iterative computations: for instance, to find the friends of friends
of a given person, Unicorn can first retrieve the direct friends and then
lookup for all their friends, using result set of the first lookup (direct
friends) as a parameter for the second lookup.
Unicorn indexes are kept up-to-date with the content of the social graph using
a periodic procedure that runs on an external computation engine.
Unicorn tolerates failures through replication.  However, it is possible for
some shared state portions to remain temporarily unavailable: this may result
in incomplete results when searching (if some index serves do not reply),
which is acceptable in its specific application domain.

\subsection{NewSQL systems}

\subsubsection{Key-value stores}

\fakeparagraph{Deuteronomy}
Deuteronomy~\cite{levandoski:CIDR:2011:Deuteronomy} is a data store designed
for wide area deployments that decouples storage functionalities from
transactional execution of jobs.
In Deuteronomy, the driver program runs client-side and submits jobs (queries)
to a transaction component.  The component ensures group atomicity and
isolation for all read and write operations performed on the shared state,
using a locking protocol.
The actual storage is implemented in a separate layer.  Deuteronomy supports
any distributed storage component that offers read and write operations for
individual elements (such as a key-value store), and is oblivious of the
actual location of the workers implementing the storage component, which may
be geographically distributed.
Deuteronomy provides fault tolerance through logging and replication, and
enables dynamic reconfiguration by independently scaling both the
transactional and the storage component.

\fakeparagraph{FoundationDB} FoundationDB~\cite{zhou:SIGMOD:2021:FoundationDB}
is a transactional key-value store, which aims to offer the core building
blocks (hence the name) to build scalable distributed data management systems
with heterogeneous requirements.  The use of key-value abstractions provides
flexibility in the data model, on top of which developers can build various
types of abstractions: for instance, each element may encode a relation
indexed by its primary key.
Jobs consist of a group of read and write requests issued by a driver program
that runs client-side.
Like Deuteronomy, FoundationDB is organized into layers, each of them offering
one of the core functionalities of a transactional DMS and each implemented
within a different set of workers, thus enabling independent scaling.
A storage layer persists data and serves read and write requests.  A log layer
manages a write ahead log.  A transaction system handles isolation and
atomicity for multiple read and write requests.
Transactional semantics is enforced by assigning timestamps to operations and
by checking for conflicts between concurrent transactions after they have been
executed: in the case of conflicts, the transaction aborts and the driver
program is notified.
Fault tolerance is implemented by replicating both the log layer
(synchronously) and the storage layer (asynchronously).  Replication of the
storage layer is also used to serve read requests in parallel.
Moving data between workers that implement the storage layer and the log layer
is also used when adding or removing workers for scalability (dynamic
reconfiguration).

\fakeparagraph{Solar} Like Deuteronomy and FoundationDB,
Solar~\cite{zhu:ToS:2019:Solar} is a transactional key-value store that
decouples storage of shared state from transaction processing.  The
transaction layer uses a timestamp-based optimistic concurrency control and
stores a write ahead log in main memory.  The storage layer persists
checkpoints of the shared state.  Together, they form a LSM tree.  The key
distinguishing feature of Solar is that the transaction layer is implemented
as a centralized service, replicated for fault tolerance of the write ahead
log.

\subsubsection{Structured and relational stores}

\paragraph{Time-based protocols}

\fakeparagraph{Spanner} Spanner~\cite{corbett:TOCS:2013:Spanner} is a
semirelational database: shared state is organized into tables, where each
table has an ordered set of one or more primary key columns and defines a
mapping from these key columns to non key columns.
Spanner provides transactional semantics and replication with strong
consistency for cluster and wide area deployments.
At its core, Spanner uses standard database techniques: two-phase locking for
isolation, two-phase commit for atomicity, and synchronous replication of jobs
results using Paxos state machine replication.
Jobs are globally ordered using timestamps, and workers store multiple
versions of each state element (multi-version concurrency control).  This way,
read-only jobs can access a consistent snapshot of the shared state and do not
conflict with read-write jobs.  A consistent snapshot preserves causality,
meaning that if a job reads a version of a state element (cause) and
subsequently updates another state element (effect), the snapshot cannot
contain the effect without the cause.  Traditional databases ensure causality
by acquiring read and write locks to prevent concurrent accesses, but this
could be too expensive in a distributed environment.
The key distinguishing idea of Spanner is to serve a consistent snapshot to
read-only jobs without locking.  To do so, it uses an abstraction called
TrueTime, which returns real (wall-clock) time within a known precision bound
using a combination of GPS and atomic clocks.  Tasks of read-write jobs first
obtain the locks for all the data portions they access: a coordinator for the
job assigns that job with a timestamp at the end of its time uncertainty
range, then it waits until this timestamp is passed for all workers in the
system, releases the locks and writes the results (commits).  The waiting time
ensures that jobs with later timestamps read all writes of jobs with earlier
timestamps without explicit locking.
Using TrueTime, Spanner also supports consistent reconfiguration, for instance
to change the database schema or to move data for load balancing.
More recently, Spanner has been extended with support for distributed SQL
query execution~\cite{bacon:SIGMOD:2017:Spanner}.

\fakeparagraph{CockroachDB} CockroachDB~\cite{taft:SIGMOD:2020:CockroachDB} is
a relational database for wide area deployments.  It shares many similarities
with Spanner and integrates storage and processing capabilities within each
node.
As a storage layer, it relies on RocksDB, a disk-based key-value store that
organizes data in LSM trees.  It replicates data across nodes, ensuring strong
consistency through Raft consensus.
On top of this, it partitions data with transactional semantics: it uses an
isolation mechanism based on hybrid physical and logical clocks (similar to
Spanner) but integrates it with an optimistic protocol that, in the case of
conflicts, attempts to modify the timestamp of a job to a valid one rather
than re-executing the entire job.
CockroachDB compiles SQL queries into a plan of tasks that can be either fully
executed on a single worker or in a distributed dataflow fashion.
Interestingly, CockroachDB also enables users to configure data placement
across data centers.  For instance, a table can be partitioned across a
\emph{Region} column to ensure that all data about one region is stored within
a single data center.  This may improve access time from local client and
enforce privacy regulations.

\paragraph{Deterministic execution}

\fakeparagraph{Calvin} Calvin~\cite{thomson:SIGMOD:2012:Calvin} is a job
scheduling and replication layer to provide transactional semantics and
replication consistency on top of non-transactional distributed data stores
such as Dynamo, Cassandra, and MongoDB.  It is currently implemented within
the Fauna database\footnote{\url{https://fauna.com}}.
The core Calvin layer is actually agnostic with respect to the specific data
and query model.  In fact, Fauna supports document and graph-based models in
addition to the relational model.
Calvin builds on the assumption that jobs are deterministic.  Its core idea is
to avoids as much as possible expensive coordination \emph{during} job
execution by defining a global order for tasks \emph{before} the actual
execution.
In Calvin, workers are organized in regions: each region contains a single
copy of the entire shared state, and each worker in a region is fully
replicated in every other region.  All workers that contain a replica of the
same state portion in different regions are referred to as a replication
group.
Invocations from clients are organized in batches: all workers in a
replication group receive a copy of the batch and they coordinate to agree on
a global order of execution.  Under the assumption of deterministic jobs, this
approach ensures consistent state across all replicas in a replication group.
Replicas are used both to improve access for read-only jobs, which can be
executed on any replica, and for fault tolerance, as in the case of a failure
remaining replicas can continue to operate without interruption.
Invocations are stored in a durable log, and individual workers can be resumed
from a state snapshot by reapplying all invocations that occurred after that
snapshot.
Calvin supports different replication protocols with different tradeoffs
between job response time and complexity in fail over.
Deterministic jobs executed in the same order lead to the same results in all
(non failing) replicas.  Specifically, they either commit or abort in all (non
failing) replicas.  Accordingly, under the assumption that at least one
replica for each shared state portion does not fail, Calvin can provide
atomicity without expensive protocols such as the classic two-phase commit: if
some tasks may abort due to violation of integrity constrains, they simply
inform other tasks of the same transaction with a single (one-phase)
communication.
Global execution order is also used for isolation: Calvin exploits a locking
mechanism where tasks acquire locks on their shared state portion in the
agreed order.  This requires knowing upfront the exact state portions accessed
within each job: when they cannot be statically determined, for instance due
to state-dependent control flow, Calvin runs reconnaissance jobs that perform
all read accesses to determine the state portions of interest.  However,
during the actual execution, shared state may have changed, and the real jobs
may deviate from reconnaissance jobs and try to access different portions, in
which case they are deterministically restarted.
Interestingly, Calvin provides the same strong semantics both for cluster and
for wide area deployments.  The initial coordination increases the latency to
schedule a batch of jobs, affecting the response time of individual jobs in
wide area deployments, but the batching mechanisms can preserve throughput.

\paragraph{Explicit partitioning and replication strategies}

\fakeparagraph{VoltDB} VoltDB~\cite{stonebraker:IEEEB:2013:voltdb} is an
in-memory relational database developed from the HStore research
project~\cite{stonebraker:VLDB:2007:hstore}.
In VoltDB, clients register stored procedures, which are driver programs
written in Java and executed system side.  They include multiple jobs, are
compiled on registration, and executed on invocation.  Jobs can also write
data to sinks.
The key idea of VoltDB is to let users control database partitioning
and replication, so they can optimize most frequently executed jobs.
In particular, VoltDB preserves the same (transactional) execution semantics
as centralized databases while minimizing the overhead of concurrency control.
By default, all tasks that derive from a single driver program represent a
transaction and are guaranteed to execute with group atomicity and isolation.
In the worst case, this is achieved through blocking coordination (two-phase
commit for atomicity and timestamp-based concurrency control for isolation).
However, VoltDB avoids coordination for specific types of jobs by exploiting
user-provided data about data partitioning and replication.  Users can specify
that a relational table is partitioned based on the value of a column, for
instance a \texttt{Customer} table may be partitioned by \texttt{region},
meaning that all customers that belong to the same region (have the same value
for the attribute \texttt{region}) are stored in the same shared state
portion.  Jobs that only refer to a given region can then be executed by the
single worker responsible for that state portion, sequentially, without
incurring expensive concurrency control overhead.  Every table that is not
partitioned is replicated in every worker, which optimizes read access from
any worker at the cost of replicating state changes.  Users need to select the
best partitioning and replication schema to improve performance for the most
frequent jobs.
VoltDB also supports replicating tables (including partitioned ones) for fault
tolerance.  Replicas are kept up-to-date by propagating and executing tasks to
all replicas, under the assumption that tasks are deterministic.
VoltDB is designed for cluster deployment as clock synchronization and low
latency are necessary to guarantee that timestamp-based concurrency control
and replication work well in practice.
However, VoltDB also provides a hybrid deployment model where a database is
fully replicated at multiple geographical regions.  These replicas can be used
only as an additional form of fault tolerance (with asynchronous propagation
of state) or can serve local clients.  In this case, consistency across
regions is not guaranteed, and conflicts get resolved using predefined
automatic rules.
VoltDB supports manual reconfiguration of both partitioning and workers, but
requires stopping and restarting the system.

\paragraph{Primary-based protocols}

\fakeparagraph{Aurora} Aurora~\cite{verbitski:SIGMOD:2017:Aurora} is a
relational database offered as a service by Amazon.
Aurora builds on two key design choices:
\begin{inparaenum}[(i)]
\item decouple the storage layer from the query processing layer;
\item store the log of changes (write ahead log) in the storage layer instead
  of the actual shared state.  Shared state is materialized only to improve
  read performance, and materialization can be performed asynchronously
  without increasing write latency.
\end{inparaenum}
The storage layer (that is, the write log) is replicated both to improve read
performance and for fault tolerance.  To ensure consistency, read and write
operations use a quorum-based approach.
The processing layer accepts jobs from clients in the form of SQL queries,
which are always executed within a single worker.  Specifically, to guarantee
isolation, Aurora assumes that a single worker is responsible for processing
all read-write jobs at any given point in time.  Read-only jobs can be
executed on any worker, which can read a consistent snapshot of the state
without conflicting with concurrent writes.
In Aurora, the storage and processing layers can scale independently: the
processing layer is stateless, while the storage layer only needs to replicate
the log.

\fakeparagraph{Socrates} Socrates~\cite{antonopoulos:SIGMOD:2019:Socrates} is
a relational database offered as a service in the Azure cloud platform (under
the name of SQL DB
Hyperscale\footnote{\url{https://docs.microsoft.com/en-us/azure/azure-sql/database/}}).
Like Aurora, Socrates decomposes the functionality of a DMS and implements
them as independent services.
Its design goals include quick recovery from failure and fast reconfiguration.
To do so, it relies on four layers, each implemented as a service that can
scale out when needed.
\begin{inparaenum}[(1)]
\item Compute nodes handle jobs, including protocols for group atomicity and
  isolation. There is one primary compute node that processes read-write jobs
  and an arbitrary number of secondary nodes that handle read-only jobs and
  may become primary in the case of failure.  Compute nodes cache shared state
  pages in main memory and on SSD.
\item A log service logs write requests with low latency.
\item A storage service periodically applies writes from the log to store the
  shared state durably.
\item A backup service stores copies of the storage layer for fault tolerance.
\end{inparaenum}

\subsubsection{Objects stores}

\fakeparagraph{Tango} Tango~\cite{balakrishnan:SOSP:2013:Tango} is a service
for storing metadata.  Application code (the driver program) executes
client-side and reads and accesses a shared state consisting of Tango objects.
Clients store their view of Tango objects locally in-memory, and this view is
kept up to date with respect to a distributed (partitioned) and durable
(replicated) totally ordered log of updates.  Tango objects can contain
references to other Tango objects, thus enabling the definition of complex
linked data structures such as trees or graphs.
Although the log is physically partitioned across multiple computers, all
operations are globally ordered through sequence numbers, which are obtained
through a centralized sequencer service: the authors demonstrate that this
service does not become a bottleneck for the system when serving hundreds of
thousands of requests per second.
Clients check if views are up-to-date before performing updates, thus ensuring
totally ordered, linearizable updates.  Tango also guarantees group atomicity
and isolation using the log for optimistic concurrency control.

\subsubsection{Graph stores}

\fakeparagraph{A1} A1~\cite{buragohain:SIGMOD:2020:A1} is an in-memory
database that resembles TAO and Trinity in terms of data model (typed graphs)
and jobs (changes to graph entities and graph pattern matching queries).
The distinguishing characteristic of A1 is that it builds on a distributed
shared memory abstraction that uses RDMA (remote direct memory access)
implemented within network interface cards~\cite{dragojevic:NSDI:2014:FaRM}.
A1 stores all the elements of the graph in a key-value store.  Jobs may
traverse the graph and read and modify its associated data during execution.
A1 provides strong consistency, atomicity, and isolation using timestamp-based
concurrency control.
Fault tolerance is implemented using synchronous replication in memory and
asynchronous replication to disk.  In the case of failures, users can decide
whether to recover the last available state or only state that is guaranteed
to be transactionally consistent.


  \section{Data processing systems}
\label{sec:systems_proc_list}

\begin{table}[htpb]
  \centering
  \tiny
  \rowcolors{2}{gray!25}{white}
\begin{tabular}[t]{|l|c|c|c|c|c|c|c|}
    \hline
    \rowcolor{gray!50}
                & Driver exec  & Driver exec time & Invoc of jobs & Sources & Sinks & State & Deployment   \\ \hline

\rowcolor{gray!50}
\multicolumn{8}{|c|}{Dataflow task deployment systems} \\ \hline

MapReduce       & sys          & reg              & sync          & passive & yes   & no    & cluster \\
Dryad           & sys          & reg              & unknown       & passive & yes   & no    & cluster \\
HaLoop          & sys          & reg              & unknown       & passive & yes   & no    & cluster \\
CIEL            & sys          & reg              & async         & passive & yes   & no    & cluster \\
Spark           & configurable & reg              & sync+async    & passive & yes   & no    & cluster \\
Spark Streaming & configurable & reg              & async         & both    & yes   & yes   & cluster \\
\hline    
  \end{tabular}
  \caption{Data processing systems: functional model.}
  \label{tab:proc:functional}
\end{table}

\begin{table}[tpb]
  \centering
  \tiny
  \rowcolors{2}{gray!25}{white}
  \begin{tabular}[t]{|l|c|c|c|c|c|c|c|c|c|c|}
    \hline
    \rowcolor{gray!50}
                & Jobs def API & \Gape{\makecell{Exec plan\\def}} & \Gape{\makecell{Task\\comm}} & \Gape{\makecell{Exec plan\\struct}} & Iter    & \Gape{\makecell{Dyn\\creat.}} & \Gape{\makecell{Nature\\of jobs}}             & State man. & \Gape{\makecell{Data par\\API}} & \Gape{\makecell{Placem.\\-aware API}} \\ \hline

\rowcolor{gray!50}
\multicolumn{11}{|c|}{Dataflow task deployment systems} \\ \hline

MapReduce       & lib          & expl                             & impl                         & dataflow                            & no      & no                            & one-shot                                      & absent     & yes                             & no                        \\
Dryad           & lib          & expl                             & impl                         & dataflow                            & no      & no                            & one-shot                                      & absent     & yes                             & no                        \\
HaLoop          & lib          & expl                             & impl                         & cycl. dataflow                      & yes     & no                            & one-shot                                      & absent     & yes                             & no                        \\
CIEL            & lib          & expl                             & impl                         & dyn. dataflow                       & no      & yes                           & one-shot                                      & absent     & yes                             & no                        \\
Spark           & lib (+DSLs)  & expl (+impl)                     & impl                         & dataflow                            & no      & no                            & one-shot                                      & absent     & yes                             & no                        \\
\hline    
  \end{tabular}
  \caption{Data processing systems: jobs definition.  Legend: \batch = in batch processing; \stream = in stream processing.}
  \label{tab:proc:job_definition}
\end{table}

\begin{table}[tpb]
  \centering
  \tiny
  \rowcolors{2}{gray!25}{white}
  \begin{tabular}[t]{|l|c|c|c|c|c|c|}
    \hline
    \rowcolor{gray!50}
                & Jobs comp time & Use resources info (comp) & Granul of depl & Depl time  & Use resources info (depl) & Manag of res \\ \hline

\rowcolor{gray!50}
\multicolumn{7}{|c|}{Dataflow task deployment systems} \\ \hline

MapReduce       & reg             & dynamic              & task           & task activ & dynamic              & shared       \\
Dryad           & reg             & dynamic              & task           & task activ & dynamic              & sys-only     \\
HaLoop          & reg             & dynamic              & task           & task activ & dynamic              & shared       \\
CIEL            & reg             & dynamic              & task           & task activ & dynamic              & sys-only     \\
Spark           & reg             & dynamic              & task           & task activ & dynamic              & shared       \\
Spark Streaming & reg             & dynamic              & task           & task activ & dynamic              & shared       \\
\hline
  \end{tabular}
  \caption{Data processing systems: jobs compilation and execution.}
  \label{tab:proc:jobs_compilation_execution}
\end{table}

\begin{table}[tpb]
  \centering
  \tiny
  \rowcolors{2}{gray!25}{white}
  \begin{tabular}[t]{|l|c|c|c|c|c|c|c|c|}
    \hline
    \rowcolor{gray!50}
                & Elem struc   & Temp elem & Bus conn                              & Bus impl          & Bus persist                         & Bus partition & Bus repl  & Bus inter                         \\ \hline

\rowcolor{gray!50}
\multicolumn{9}{|c|}{Dataflow task deployment systems} \\ \hline

MapReduce       & general      & no        & mediated                              & distr fs          & persist                             & yes           & no        & hybrid                            \\
Dryad           & general      & no        & \Gape{\makecell{direct\\or mediated}} & configurable      & \Gape{\makecell{persist\\or ephem}} & yes           & no        & \Gape{\makecell{push\\or hybrid}} \\
HaLoop          & general      & no        & mediated                              & distr fs          & persist                             & yes           & no        & hybrid                            \\
CIEL            & general      & no        & \Gape{\makecell{direct\\or mediated}} & configurable      & \Gape{\makecell{persist\\or ephem}} & yes           & no        & \Gape{\makecell{push\\or hybrid}} \\
Spark           & general+spec & no        & mediated                              & distr fs (+cache) & persist                             & yes           & no        & hybrid                            \\
Spark Streaming & general+spec & yes       & mediated                              & distr fs (+cache) & persist                             & yes           & no        & hybrid                            \\
\hline  
  \end{tabular}
  \caption{Data processing systems: data management.}
  \label{tab:proc:data}
\end{table}

\begin{table}[tpb]
  \centering
  \tiny
  \rowcolors{2}{gray!25}{white}
  \begin{tabular}[t]{|l|c|c|c|c|c|c|c|c|c|c|}
    \hline
    \rowcolor{gray!50}
                & Elem struct & \Gape{\makecell{Stor\\medium}} & \Gape{\makecell{Stor\\struct}}  & Task st & Shared st & St part & Repl & Repl consist  & Repl prot & \Gape{\makecell{Update\\propag}} \\ \hline
    
\rowcolor{gray!50}
\multicolumn{11}{|c|}{Dataflow task deployment systems} \\ \hline

MapReduce       & n.a.         & n.a.    & n.a.     & no                  & no        & n.a.    & n.a. & n.a.         & n.a. & n.a.     \\
Dryad           & n.a.         & n.a.    & n.a.     & no                  & no        & n.a.    & n.a. & n.a.         & n.a. & n.a.     \\
HaLoop          & n.a.         & n.a.    & n.a.     & no                  & no        & n.a.    & n.a. & n.a.         & n.a. & n.a.     \\
CIEL            & n.a.         & n.a.    & n.a.     & no                  & no        & n.a.    & n.a. & n.a.         & n.a. & n.a.     \\
Spark           & n.a.         & n.a.    & n.a.     & no                  & no        & n.a.    & n.a. & n.a.         & n.a. & n.a.     \\
Spark Streaming & n.a.         & n.a.    & n.a.     & yes                 & no        & n.a.    & n.a. & n.a.         & n.a. & n.a.     \\
\hline
  \end{tabular}
  \caption{Data processing systems: state management. Legend: \batch = in batch processing; \stream = in stream processing.}
  \label{tab:proc:state}
\end{table}

\begin{table}[tpb]
  \centering
  \tiny
  \rowcolors{2}{gray!25}{white}
  \begin{tabular}[t]{|l|c|c|c|c|c|c|c|c|c|}
    \hline
    \rowcolor{gray!50}
                & Aborts & Protocol & Assumptions & Level & Impl & Assumptions & Delivery         & Nature of ts & Order \\ \hline

\rowcolor{gray!50}
\multicolumn{10}{|c|}{Dataflow task deployment systems} \\ \hline

MapReduce       & n.a.   & n.a.     & n.a.        & n.a.  & n.a. & n.a.        & exact            & no           & n.a.       \\
Dryad           & n.a.   & n.a.     & n.a.        & n.a.  & n.a. & n.a.        & exact            & no           & n.a.       \\
HaLoop          & n.a.   & n.a.     & n.a.        & n.a.  & n.a. & n.a.        & exact            & no           & n.a.       \\
CIEL            & n.a.   & n.a.     & n.a.        & n.a.  & n.a. & n.a.        & exact            & no           & n.a.       \\
Spark           & n.a.   & n.a.     & n.a.        & n.a.  & n.a. & n.a.        & exact            & no           & n.a.       \\
Spark Streaming & n.a.   & n.a.     & n.a.        & n.a.  & n.a. & n.a.        & exact            & no or ingest & always     \\
\hline
  \end{tabular}
  \caption{Data processing systems: group atomicity, group isolation, delivery, order.}
  \label{tab:proc:guarantees}
\end{table}

\begin{table}[tpb]
  \centering
  \tiny
  \rowcolors{2}{gray!25}{white}
  \begin{tabular}[t]{|l|c|c|c|c|c|c|}
    \hline
    \rowcolor{gray!50}
                & Detection & Scope          & Comput recov & State recov & Guarantees for state  & Assumptions \\ \hline

\rowcolor{gray!50}
\multicolumn{7}{|c|}{Dataflow task deployment systems} \\ \hline

MapReduce       & lead-work & comput         & task         & n.a.        & n.a.                  & \replaysource \\
Dryad           & lead-work & comput         & task         & n.a.        & n.a.                  & \replaysource \\
HaLoop          & lead-work & comput         & task         & n.a.        & n.a.                  & \replaysource \\
CIEL            & lead-work & comput         & task         & n.a.        & n.a.                  & \replaysource \\
Spark           & lead-work & comput         & task         & n.a.        & n.a.                  & \replaysource \\
Spark Streaming & lead-work & comput+task st & task         & log+checkp  & valid or same         & \replaysource \\
\hline
  \end{tabular}
  \caption{Data processing systems: fault tolerance.  Legend: \replaysource =
    sources are replayable.}
  \label{tab:proc:fault}
\end{table}

\begin{table}[tpb]
  \centering
  \tiny
  \rowcolors{2}{gray!25}{white}
  \begin{tabular}[t]{|l|c|c|c|c|c|c|}
    \hline
    \rowcolor{gray!50}
                & Goal       & Automated & State migration & Task migration & Add/del slots & Restart \\ \hline

\rowcolor{gray!50}
\multicolumn{7}{|c|}{Dataflow task deployment systems} \\ \hline
    
MapReduce       & elast      & yes       & n.a.            & n.a.           & yes           & no      \\
Dryad           & n.a.       & n.a.      & n.a.            & n.a.           & n.a.          & n.a.    \\
HaLoop          & elast      & yes       & n.a.            & n.a.           & n.a.          & no      \\
CIEL            & n.a.       & n.a.      & n.a.            & n.a.           & n.a.          & n.a.    \\
Spark           & elast      & yes       & n.a.            & n.a.           & yes           & no      \\
Spark Streaming & elast      & yes       & yes             & yes            & yes           & no      \\
\hline
  \end{tabular}
  \caption{Data processing systems: dynamic reconfiguration.}
  \label{tab:proc:reconf}
\end{table}


\subsection{Task-level deployment}

\fakeparagraph{MapReduce} MapReduce~\cite{dean:CACM:2008:mapreduce} is a
distributed processing model and system developed at Google in the early
2000s, and later implemented in open source projects such as Apache
Hadoop\footnote{\url{https://hadoop.apache.org}}.
Its programming and execution models represent a paradigm shift in distributed
data processing that influenced virtually all DPSs discussed in this paper:
developers are forced to write jobs as a sequence of functional
transformations, avoiding by design the complexity and cost associated to
state management.
In the specific case of MapReduce, jobs are constrained to only two processing
steps:
\begin{inparaenum}[(i)]
\item \emph{map} transforms each input element into a set of key-value pairs,
\item \emph{reduce} aggregates all values associated to a given key.
\end{inparaenum}
Both functions are data parallel: developers specify the map function for a
single input element and the reduce function for a single key, and the system
automatically applies them in parallel.
The data bus is implemented using a distributed filesystem.  The system
schedules map tasks as close as possible to the physical location of their
input in the filesystem.  It then automatically redistributes intermediate
results by key before scheduling the subsequent reduce tasks.
Fault detection is implemented using a leader-worker approach.  Tasks that did
not complete due to a failure are simply rescheduled.  The same approach is
used for tasks that take long to complete (stragglers): they are scheduled
multiple times if some workers are available, to increase the probability of
successful completion.
Dynamic scheduling at the granularity of tasks simplifies implementation of
dynamic reconfiguration mechanisms to promote elasticity.  For instance,
Hadoop supports scheduling policies based on user-defined quality of
service requirements, such as expected termination time: the scheduler tunes
the use of resources (possibly shared with other applications) to meet the
expectation while minimizing the use of resources.
Several works build on top of MapReduce to offer a declarative language
similar to SQL to express analytical queries that are automatically translated
into MapReduce
jobs~\cite{abouzeid:VLDB:2009:HadoopDB,camacho-rodriguez:SIGMOD:2019:ApacheHive}.

\fakeparagraph{Dryad} Several systems developed in parallel and after
MapReduce inherit and extend the core concepts in its programming and
execution models.  Dryad~\cite{isard:EuroSys:2007:Dryad} generalizes the
programming model, representing jobs as arbitrary acyclic dataflow graphs.
Like MapReduce, it uses a leader-worker approach, where a leader schedules
individual tasks (operators in the dataflow graph), but it enables different
types of channels (data bus in our model), including shared memory on the same
machine, TCP channels across machines, or distributed filesystems.  The leader
is also responsible for fault detection.  Jobs are assumed to be
deterministic, and in the case of failure the failing task is re-executed: in
the case of ephemeral channels, also upstream tasks in the dataflow graph are
re-executed to re-create the input for the failing task.

\fakeparagraph{HaLoop} Systems like MapReduce and Dryad are constrained to
acyclic job plans and cannot natively support iterative algorithms.
HaLoop~\cite{bu:VLDB:2010:HaLoop} addresses this limitation with a modified
version of MapReduce that:
\begin{inparaenum}[(i)]
\item integrates iterative MapReduce jobs as first class programming concepts;
\item optimizes task scheduling by co-locating tasks that reuse the same data
  across iterations;
\item caches and indexes loop-invariant data to optimize access across
  iterations;
\item caches and indexes results across iterations to optimize the evaluation
  of fixed point conditions for termination.
\end{inparaenum}

\fakeparagraph{CIEL} CIEL~\cite{murray:NSDI:2011:CIEL} extends the dataflow
programming model of Dryad by allowing tasks to dynamically create other
tasks.  This enables defining the job plan dynamically based on the results of
data computation, which can be used to implement iterative algorithms.  The
programming model ensures that tasks cannot have cyclic dependencies, thus
avoiding deadlocks in the execution.  The execution model and the fault
tolerance mechanism remain identical to Dryad, the only exception being that
the list of tasks that the leader schedules and tracks can dynamically change
during the execution, thus allowing the execution flow to be defined at
runtime based on the actual content of input data.

\fakeparagraph{Spark} Spark~\cite{zaharia:CACM:2016:spark} inherits the
dataflow programming model of Dryad and supports iterative execution and data
caching like HaLoop.
\begin{inparaenum}
\item Spark jobs are split in sequences of operations that do not alter data
  partitioning, called stages.  Multiple tasks are scheduled for each stage,
  to implement data parallelism.
\item The driver program can run either client-side or system-dice, and can
  dynamically spawn new jobs based on the results collected from previous
  jobs.  This enables data-dependent control flow under the assumption that
  control flow conditions are evaluated in the driver program, and is used to
  implement iterative algorithms.
\item As in HaLoop, intermediate results that are re-used by the same or
  different jobs (as in the case of iterative computations) can be cached in
  main memory to improve efficiency.
\end{inparaenum}
Spark provides domain specific libraries and languages for structured
(relational) data~\cite{armbrust:SIGMOD:2015:SparkSQL},
graphs~\cite{gonzalez:OSDI:2014:GraphX}, and machine learning
computations~\cite{meng:JMLR:2016:MLlib}, which often include ad-hoc job
optimizers.
Spark also inherits the fault tolerance model of MapReduce: as tasks are
stateless, failing tasks are simply re-executed starting from their input
data; if the input data is not available anymore (for instance, in the case of
intermediate results not persisted on any other node), all tasks necessary to
reconstruct the input are also re-executed.
Task-level deployment also enables runtime reconfiguration to provide
elasticity.

\fakeparagraph{Spark Streaming} Spark
Streaming~\cite{zaharia:SOSP:2013:Discretized_Streams} implements streaming
computations on top of Spark by splitting the input stream into small batches
and by running the same jobs for each batch.
Spark Streaming implements task state using native Spark features: the
state of a task after a given invocation is implicitly stored as a special
data item that the task receives as input in the subsequent invocation.  This
also enables reusing the fault tolerance mechanism of Spark to persist or
recompute state as any other data element.
Another benefit of using the Spark system is simple integration of static and
streaming input data.  The main drawback of the approach is latency: since
input data needs to be accumulated in batches before processing, Spark
Streaming can only provide latency in the range of seconds.

\subsection{Job-level deployment}

\fakeparagraph{MillWheel} MillWheel~\cite{akidau:VLDB:2013:MillWheel} is a
framework to build general-purpose and large-scale stream processing systems.
Jobs consist of data-parallel tasks expressed using an imperative language.
As part of their computation, tasks can use the MillWheel API to access (task
local) state, logical time information, and to produce data for downstream
tasks.
MillWheel implements the communication between tasks (the data bus) using
point-to-point remote procedure calls, and uses an external storage service to
persist task state and metadata about global progress (watermarks).  Upstream
tasks are acknowledged when downstream tasks complete, and the storage service
is kept consistent with atomic updates of state and watermarks.  In the case
of failure, individual tasks can rely on the external storage to restore a
consistent view of their state, and they can discard duplicate invocations
from upstream tasks in the case some acknowledgements are lost.  The same
approach is also used for dynamic load distribution and balancing.

\fakeparagraph{Flink} Flink~\cite{carbone:IEEEB:2015:flink} is a unified
execution engine for batch and stream processing.  In terms of programming
model, it strongly resembles Spark, with a core API to explicitly define job
plans and domain specific libraries for structural (relational) data, graph
processing, and machine learning.  One notable difference involves iterative
computations: Flink supports them with native operators (within jobs) rather
than controlling them from the driver program.
Flink adopts job-level deployment and implements the data bus using direct TCP
channels.
This brings several implications.
\begin{inparaenum}[(i)]
\item Tasks are always active and compete for workers resources.  For each
  processing step in the execution plan, Flink instantiates one (data
  parallel) task for each CPU core available on workers.  In practice, each
  CPU core receives one task for each step and scheduling is delegated to the
  operating system that hosts the worker.
\item In the case of stream processing, tasks can continuously exchange data,
  which traverses the graph of computation leading to a pipelined execution.
  As there is no need to accumulate input data in batches, this processing
  model may reduce latency.
\item As tasks cannot be deployed independently, fault tolerance requires
  restarting an entire job.  In the case of stream processing jobs, Flink
  takes periodic snapshots of task states: in the case of a failure, it
  restores the last snapshot and replays all input data that was not part of
  the snapshot (assuming it remains available).
\item Dynamic reconfiguration builds on the same mechanism: when the number of
  available slots changes, the system needs to restart and resume jobs from a
  recent snapshot.
\end{inparaenum}

\fakeparagraph{Storm} Storm~\cite{toshniwal:SIGMOD:2014:Storm} and its
successor Heron~\cite{kulkarni:SIGMOD:2015:Twitter_Heron} are stream
processing systems developed at Twitter.
They offer lower-level programming API than previously discussed dataflow
systems, where developers fully implement the logic of each processing step.
At the time of writing, both Storm and Heron have experimental higher-level
API that mimic the functional approach of Spark and Flink.
Storm and Heron adopt job-level deployment, and implement the data bus as
direct network channels between workers.
They implement fault tolerance by acknowledging every message.  If a message
is not acknowledged within a given timeout, the sender replays it, which leads
to at least once delivery (messages may be duplicated).  The same applies in
the case of state: state is checkpointed, but may be modified more than once
in the case of duplication.
Dynamic reconfiguration is possible, but required redeploying and restarting
the entire job.

\fakeparagraph{Kafka Streams} Kafka~\cite{kreps:NetDB:2011:kafka} is a
distributed communication platform designed to scale in terms of clients, data
volume and production rate.
Kafka offers logical communication channels named \emph{topics}: producers
append immutable data to topics and consumers read data from topics.  Topics
are persistent, which decouples data production and consumption times.  Topics
may be partitioned to improve scalability: multiple consumers may read in
parallel from different partitions, possibly hosted on different physical
nodes.  Topics may also be (semi-synchronously) replicated for fault
tolerance.
With respect to our model in \s{model}, Kafka represents the implementation of
a persistent data bus that external clients can use to exchange immutable
data.
Kafka Streams~\cite{bejeck:2018:KafkaStreams} implements batch and stream
processing functionalities on top of Kafka.  Its programming model is similar
to that of Spark and Flink, with core functional API and a higher-level domain
specific language for relational data processing (KSQL).  As in Flink, all
tasks for a job are instantiated and scheduled when the job starts and
continuously communicate in a pipelined fashion using Kafka as data bus.  Each
channel in the job logical plan is implemented as a Kafka topic, and data
parallelism exploits topic partitioning, allowing multiple tasks to
simultaneously read from different partitions.
Interestingly, Kafka Streams does not offer resource management
functionalities but runs the driver program and the tasks within clients: each
job definition is associated with a unique identifier, and clients can offer
resources for a job (that is, become workers) by referring to the job
identifier.  The policy for allocating tasks to slots is similar to that of
Flink: each slot represents a physical CPU core and receives one task for each
computation step in the logical plan.
Task state for streaming jobs is stored on Kafka, following the same idea of
persisting state as a special element in the data bus that we already found in
Spark Streaming.
Fault detection relies on Kafka ability to detect when consumers disconnect.
For fault recovery Kafka Streams adopts a two-phase commit protocol to ensure
that upon activation a task consumes its input, updates its state, and
produces results for downstream tasks atomically.  In the case of failure, a
task can resume from the input elements that were not successfully processed,
providing exactly once delivery of individual elements.  We still classify the
system as offering at least once delivery, because, in the case of timestamped
elements, it does not implement any mechanism to process them in timestamp
order, but retracts and updates its results upon receiving elements
out-of-order (resulting in visible changes at the sinks).
Storing both data and task state on Kafka enables for dynamic reconfiguration
that involve addition and removal of clients (workers) at runtime.
The same approach of Kafka Streams is used at LinkedIn in the Samza
system~\cite{noghabi:VLDB:2017:Samza}, which is the core for the platform to
integrate data from multiple sources and offer a unique view to the back-end
system, updated incrementally as new data becomes available.

\fakeparagraph{Timely dataflow} Timely dataflow~\cite{murray:SOSP:2013:Naiad}
is a unified programming model for batch and stream processing, which is
lower-level and more general than the dataflow model of systems such as Flink.
In timely dataflow, jobs are expressed as a graph of (data parallel) operators
and data elements carry a logical timestamp that tracks global progress.
Management of timestamps is explicit, and developers control how operators
handle and propagate them.  This enables implementing various execution
strategies: for instance, developers may choose to complete a given
computation step before letting the subsequent one start (mimicking a batch
processing strategy as implemented in MapReduce or Spark), or they may allow
overlapping of steps (as it happens in Storm or Flink). 
The flexibility of the model allows for complex workflows, including streaming
computations with nested iterations, which are hard or even impossible to
express in other systems.
Timely dataflow is currently implemented as a Rust
library\footnote{\url{https://github.com/TimelyDataflow}}: as in Kafka
Streams, developers write a program that defines the graph of computation
using the library API, and run multiple instances of the program, each of them
representing a worker in our model.  At runtime, the program instantiates the
concrete tasks, which communicate with each other either using shared memory
(within one worker) or TCP channels (across workers).
Timely dataflow provides API to checkpoint task state and to restore the last
checkpoint for a job.
Dynamic reconfiguration is currently not supported.

\subsection{Graph processing}

\fakeparagraph{Pregel} Pregel~\cite{malewicz:SIGMOD:2010:Pregel} is a
programming and execution model for computations on large-scale graph data
structures.  Pregel jobs are iterative: developers provide a single function
that encodes the behavior of each vertex $v$ at each iteration.  The function
takes in input the current (local) state of $v$ and the set of messages
produced for $v$ during the previous iteration; it outputs the new state of
$v$ and a set of messages to be delivered to connected vertices, which will be
evaluated during the next iteration.  The job terminates when vertices do
not produce any message at a given iteration.
Vertices are partitioned across workers and each task is responsible for a
given partition.  Jobs are continuous, as tasks are activated multiple times
(once for each iteration) and store the vertex state across activations (in
their task state).  Tasks only communicate by exchanging data (messages
between vertices) over the data bus, which is implemented as direct channels.
One worker acts as a leader and is responsible for coordinating the iterations
within the job and for detecting possible failures of other workers.  Workers
persist their state (task state and input messages) at each iteration: in the
case of a failure, the computation restarts from the last completed iteration.
Several systems inherit and improve the original Pregel model in various ways:
we discuss some key variants through the systems that introduced them.  The
interested reader can find more details and systems in the survey by McCune et
al.~\cite{mcCune:CSur:2015:TLaV}.

\fakeparagraph{GraphLab} GraphLab~\cite{low:VLDB:2012:GraphLab} abandons the
synchronous model of Pregel, where all vertices execute an iteration before
any can move to the subsequent one.
GraphLab schedules the execution of tasks that update vertices.  During
execution, tasks can read the value of neighboring vertices and edges (rather
than receiving update messages, as in Pregel) and can update the value of
outgoing edges.  We model this style of communication as a pull-based
persistent data bus\footnote{Another way to model this communication paradigm
  is by saying that tasks have access to a global shared state representing
  vertices and edges.  However, this would not capture the strong connection
  between tasks and the vertices they update.}.
Tasks are scheduled and execute asynchronously, without barriers between
iterations.  This paradigm is suitable for machine learning and data mining
computations that do not require synchronous execution for correctness and can
benefit from asynchronous executions for performance.
GraphLab still supports some form of synchronization between tasks: for
instance, users can grant exclusive or non-exclusive access to neighboring
edges and vertices.  GraphLab implements this synchronization constraints
either with a locking protocol or with scheduling policies that prevent
execution of potentially conflicting tasks.

\fakeparagraph{PowerGraph} PowerGraph~\cite{gonzalez:OSDI:2012:PowerGraph}
observes that vertex-centric execution may lead to unbalanced work in the
(frequent) scenario of skewed graphs.  It proposes a solution that splits each
iteration into four steps:
\begin{inparaenum}[(i)]
\item \emph{gather} collects data from adjacent vertices and edges;
\item \emph{sum} combines the collected data;
\item \emph{apply} updates the state of the local vertex;
\item \emph{scatter} distributes data to adjacent edges for the next
  iteration.
\end{inparaenum}
These steps can be distributed across all workers, and executed in a MapReduce
fashion.  PowerGraph tasks can be executed synchronously, as in Pregel, or
asynchronously, as in GraphLab, depending on the specific problem at hand.

\fakeparagraph{Sub-graph centric systems} Graph mining problems typically
require retrieving sub-graphs with given characteristics.  A class of systems
designed to tackle these problems uses a sub-graph centric approach.
We model these systems by considering the input graph as a static data source
and by storing the state of each sub-graph in task state.
Arabesque~\cite{teixeira:SOSP:2015:Arabesque} explores the graph in
synchronous rounds, it starts with candidate sub-graphs consisting of a single
vertex and at each round it expands the exploration by adding one neighboring
vertex or edge to a candidate.
G-Miner~\cite{chen:EuroSys:2018:G-Miner} spawns a new task for each candidate
sub-graph, allowing tasks to proceeed asynchronously.  When scheduled for
execution, a task can update its (task) state.  G-Miner supports dynamic
load-balancing with task stealing.


  \section{Other Systems}
\label{sec:systems_other_list}

\begin{table}[htpb]
  \centering
  \tiny
  \rowcolors{2}{gray!25}{white}
\begin{tabular}[t]{|l|c|c|c|c|c|c|c|}
    \hline
    \rowcolor{gray!50}
           & Driver exec  & Driver exec time & Invoc of jobs & Sources & Sinks & State & Deployment         \\ \hline

\rowcolor{gray!50}
\multicolumn{8}{|c|}{Computations on data management systems} \\ \hline

Percolator & sys          & start            & sync          & active  & yes   & yes   & cluster            \\
F1         & client       & reg              & sync+async    & both    & yes   & yes   & cluster + wide     \\
Trinity    & client       & reg              & unknown       & no      & no    & yes   & cluster            \\

\rowcolor{gray!50}
\hline \multicolumn{8}{|c|}{New programming models} \\ \hline

SDG        & sys          & reg              & sync          & active  & yes   & yes   & cluster            \\
TensorFlow & client       & reg              & sync          & passive & yes   & yes   & cluster            \\
Tangram    & sys          & reg              & sync          & passive & yes   & no    & cluster            \\
ReactDB    & sys          & start            & sync+async    & no      & no    & yes   & cluster (not impl) \\

\rowcolor{gray!50}
\hline \multicolumn{8}{|c|}{Hybrid systems} \\ \hline

S-Store    & sys          & reg              & sync+async    & active  & yes   & yes   & cluster (not impl) \\
SnappyData & sys          & reg              & sync+async    & both    & yes   & yes   & cluster            \\
StreamDB   & sys          & start            & async         & active  & yes   & yes   & cluster            \\
Tspoon     & configurable & reg              & sync+async    & active  & yes   & yes   & cluster            \\
Hologres   & client       & reg              & unknown       & both    & yes   & yes   & cluster            \\ \hline
  \end{tabular}
  \caption{Other systems: functional model.}
  \label{tab:other:functional}
\end{table}

\begin{table}[tpb]
  \centering
  \tiny
  \rowcolors{2}{gray!25}{white}
  \begin{tabular}[t]{|l|c|c|c|c|c|c|c|c|c|c|}
    \hline
    \rowcolor{gray!50}
           & Jobs def API                                 & \Gape{\makecell{Exec plan\\def}} & \Gape{\makecell{Task\\comm}} & \Gape{\makecell{Exec plan\\struct}}      & Iter    & \Gape{\makecell{Dyn\\creat.}} & \Gape{\makecell{Nature\\of jobs}}             & State man. & \Gape{\makecell{Data par\\API}} & \Gape{\makecell{Placem.\\-aware API}} \\ \hline

\rowcolor{gray!50}
\multicolumn{11}{|c|}{Computations on data management systems} \\ \hline

Percolator & \Gape{\makecell{imperative\\+BigTable jobs}} & impl                             & impl                         & task (put/get)                           & no      & no                            & one-shot                                      & expl       & no                              & no  \\
F1         & SQL                                          & impl                             & impl                         & datafl+coord                             & no      & no                            & one-shot                                      & expl       & no                              & no  \\
Trinity    & crud+lib                                     & impl+expl                        & impl + expl                  & \Gape{\makecell{task (put/get)\\+graph}} & yes     & no                            & one-shot                                      & expl       & yes                             & no  \\

\rowcolor{gray!50}
\hline \multicolumn{11}{|c|}{New programming models} \\ \hline

SDG        & imperative                                   & impl                             & impl                         & stateful datafl                          & yes     & no                            & cont                                          & expl       & yes                             & no  \\
\hline     
  \end{tabular}
  \caption{Other systems: jobs definition. Legend: \batch = in batch processing; \stream = in stream processing.}
  \label{tab:other:job_definition}
\end{table}

\begin{table}[tpb]
  \centering
  \tiny
  \rowcolors{2}{gray!25}{white}
  \begin{tabular}[t]{|l|c|c|c|c|c|c|c|}
    \hline
    \rowcolor{gray!50}
           & Jobs comp time & Use resources info (comp) & Granul of depl & Depl time  & Use resources info (depl) & Manag of res \\ \hline

\rowcolor{gray!50}
\multicolumn{7}{|c|}{Computations on data management systems} \\ \hline

Percolator & reg             & static               & job            & job compil & static               & shared       \\
F1         & exec            & static               & job            & job compil & static               & sys-only     \\
Trinity    & exec            & static               & job            & job compil & static               & sys-only     \\

\rowcolor{gray!50}
\hline \multicolumn{7}{|c|}{New programming models} \\ \hline

SDG        & exec            & static               & job            & job compil & static               & sys-only     \\
TensorFlow & exec            & static               & task           & task activ & static               & sys-only     \\
Tangram    & exec            & dynamic              & task           & task activ & dynamic              & shared       \\
ReactDB    & reg             & static               & job            & job compil & static               & sys-only     \\

\rowcolor{gray!50}
\hline \multicolumn{7}{|c|}{Hybrid systems} \\ \hline

S-Store    & exec            & static               & job            & job compil & static               & sys-only     \\
SnappyData & exec            & static               & task           & task activ & dynamic              & sys-only     \\
StreamDB   & reg             & static               & job            & job compil & static               & sys-only     \\
Tspoon     & exec            & static               & job            & job compil & static               & sys-only     \\
Hologres   & exec            & static               & task           & task activ & dynamic              & sys-only     \\ \hline
  \end{tabular}
  \caption{Other systems: jobs compilation and execution.}
  \label{tab:other:jobs_compilation_execution}
\end{table}

\begin{table}[tpb]
  \centering
  \tiny
  \rowcolors{2}{gray!25}{white}
  \begin{tabular}[t]{|l|c|c|c|c|c|c|c|c|}
    \hline
    \rowcolor{gray!50}
           & Elem struc  & Temp elem & Bus conn                                   & Bus impl                               & Bus persist                              & Bus partition & Bus repl & Bus inter                         \\ \hline

\rowcolor{gray!50}
\multicolumn{9}{|c|}{Computations on data management systems} \\ \hline
    
Percolator & wide-column & yes       & direct                                     & net chan + RPC                         & ephem                                    & yes           & no       & pull                              \\
F1         & relational  & no        & direct                                     & net chan                               & ephem                                    & yes           & no       & push                              \\
Trinity    & typed graph & no        & direct                                     & net chan                               & ephem                                    & yes           & no       & push                              \\

\rowcolor{gray!50}
\hline \multicolumn{9}{|c|}{New programming models} \\ \hline

SDG        & general     & no        & direct                                     & net chan                               & ephem                                    & yes           & no       & push                              \\
TensorFlow & tensors     & no        & direct                                     & net chan                               & ephem                                    & yes           & no       & push                              \\
\hline
  \end{tabular}
  \caption{Other systems: data management. Legend: \stream = in stream processing.}
  \label{tab:other:data}
\end{table}

\begin{table}[tpb]
  \centering
  \tiny
  \rowcolors{2}{gray!25}{white}
  \begin{tabular}[t]{|l|c|c|c|c|c|c|c|c|c|c|}
    \hline
    \rowcolor{gray!50}
    & Elem struct & \Gape{\makecell{Stor\\medium}} & \Gape{\makecell{Stor\\struct}}  & Task st & Shared st & St part & Repl & Repl consist  & Repl prot & \Gape{\makecell{Update\\propag}} \\ \hline
    
\rowcolor{gray!50}
\multicolumn{11}{|c|}{Computations on data management systems} \\ \hline

Percolator & wide-col    & hybrid  & LSM trees  & no         & yes       & yes     & backup & n.a.         & n.a.           & n.a.    \\
F1         & relational  & service & Spanner    & no         & yes       & yes     & yes    & strong       & leader         & op      \\
Trinity    & typed graph & hybrid  & map        & no         & yes       & yes     & yes    & weak         & leader         & unknown \\

\rowcolor{gray!50}
\hline \multicolumn{11}{|c|}{New programming models} \\ \hline

SDG        & general     & mem     & user-def   & no         & yes       & yes     & no     & n.a.         & n.a.           & n.a.    \\
TensorFlow & general     & mem     & user-def   & no         & yes       & yes     & no     & n.a.         & n.a.           & n.a.    \\
\hline
  \end{tabular}
  \caption{Other systems: state management. Legend: \stream = in stream processing.}
  \label{tab:other:state}
\end{table}

\begin{table}[tpb]
  \centering
  \tiny
  \rowcolors{2}{gray!25}{white}
  \begin{tabular}[t]{|l|c|c|c|c|c|c|c|c|c|}
    \hline
    \rowcolor{gray!50}
           & Aborts        & Protocol   & Assumptions & Level      & Impl    & Assumptions & Delivery & Nature of ts & Order \\ \hline

\rowcolor{gray!50}
\multicolumn{10}{|c|}{Computations on data management systems} \\ \hline

Percolator & sys+job       & blocking   & n.a.        & blocking   & ts      & none        & exact    & event        & n.a.   \\
F1         & sys+job       & blocking   & none        & blocking   & lock+ts & none        & exact    & no           & n.a.   \\
Trinity    & n.a.          & n.a.       & n.a.        & blocking   & \seq    & \oneportion & most     & no           & n.a.   \\

\rowcolor{gray!50}
\hline \multicolumn{10}{|c|}{New programming models} \\ \hline

SDG        & n.a.          & n.a.       & n.a.        & n.a.       & n.a.    & n.a.        & exact    & no           & n.a.   \\
TensorFlow & n.a.          & n.a.       & n.a.        & config     & barrier & n.a.        & exact    & no           & n.a.   \\
\hline
  \end{tabular}
  \caption{Other systems: group atomicity, group isolation, delivery, order.
    Legend: \detjob = jobs are deterministic; \seq = jobs are executed
    sequentially, with no interleaving; \oneportion = jobs access a single
    state portion.}
  \label{tab:other:guarantees}
\end{table}

\begin{table}[tpb]
  \centering
  \tiny
  \rowcolors{2}{gray!25}{white}
  \begin{tabular}[t]{|l|c|c|c|c|c|c|}
    \hline
    \rowcolor{gray!50}
           & Detection & Scope                    & Comput recov & State recov     & Guarantees for state & Assumptions   \\ \hline

\rowcolor{gray!50}
\multicolumn{7}{|c|}{Computations on data management systesms} \\ \hline

Percolator & lead-work & shared st                & n.a.         & log+repl        & same                 & \durstore     \\
F1         & lead-work & shared st                & n.a.         & log+checkp+repl & same                 & \durstore     \\
Trinity    & p2p       & comput+shared st         & job          & checkp+repl     & none                 & none          \\

\rowcolor{gray!50}
\hline \multicolumn{7}{|c|}{New programming models} \\ \hline
    
SDG        & lead-work & comput+task st           & job          & log+checkp      & valid                & \replaysource \\
TensorFlow & n.a.      & shared st                & n.a.         & checkp          & valid                & \durstore     \\
Tangram    & lead-work & comput+shared st         & task         & checkp          & valid                & \replaysource \\
ReactDB    & n.a.      & n.a.                     & n.a.         & n.a.            & n.a.                 & n.a.          \\

\rowcolor{gray!50}
\hline \multicolumn{7}{|c|}{Hybrid systems} \\ \hline

S-Store    & lead-work & comput+shared st         & job          & log+checkp      & valid or same        & \replaysource \\
SnappyData & p2p       & comput+task st+shared st & task         & checkp+repl     & valid or same        & \replaysource \\
StreamDB   & n.a.      & n.a.                     & n.a.         & n.a.            & n.a.                 & n.a.          \\
Tspoon     & lead-work & comput+task st           & job          & log+checkp      & valid or same        & \replaysource \\
Hologres   & lead-work & shared st                & unknown      & log             & same                 & \durstore     \\ \hline
  \end{tabular}
  \caption{Other systems: fault tolerance.  Legend: \durstore = storage layer
    is durable; \replaysource = sources are replayable.}
  \label{tab:other:fault}
\end{table}

\begin{table}[tpb]
  \centering
  \tiny
  \rowcolors{2}{gray!25}{white}
  \begin{tabular}[t]{|l|c|c|c|c|c|c|}
    \hline
    \rowcolor{gray!50}
           & Goal                     & Automated & State migration & Task migration & Add/del slots & Restart \\ \hline

\rowcolor{gray!50}
\multicolumn{7}{|c|}{Computations on data management systems} \\ \hline

Percolator & load balan               & yes       & yes             & n.a.           & yes           & no   \\
F1         & change schema+load balan & yes       & yes             & no             & yes           & no   \\
Trinity    & avail                    & yes       & yes             & yes            & unknown       & no   \\

\rowcolor{gray!50}
\hline \multicolumn{7}{|c|}{New programming models} \\ \hline
    
SDG        & elast                    & yes       & yes             & yes            & yes           & no   \\
TensorFlow & n.a.                     & n.a.      & n.a.            & n.a.           & n.a.          & n.a. \\
Tangram    & n.a.                     & n.a.      & n.a.            & n.a.           & n.a.          & n.a. \\
ReactDB    & n.a.                     & n.a.      & n.a.            & n.a.           & n.a.          & n.a. \\

\rowcolor{gray!50}
\hline \multicolumn{7}{|c|}{Hybrid systems} \\ \hline
    
S-Store    & n.a.                     & n.a.      & n.a.            & n.a.           & n.a.          & n.a. \\
SnappyData & n.a.                     & n.a.      & n.a.            & n.a.           & n.a.          & n.a. \\
StreamDB   & n.a.                     & n.a.      & n.a.            & n.a.           & n.a.          & n.a. \\
Tspoon     & elast                    & no        & yes             & yes            & yes           & yes  \\ 
Hologres   & load balan+elast         & yes       & yes             & yes            & yes           & no   \\ \hline
  \end{tabular}
  \caption{Other systems: dynamic reconfiguration.}
  \label{tab:other:reconf}
\end{table}


\subsection{Computations on data management systems}

\fakeparagraph{Percolator} Percolator~\cite{peng:OSDI:2010:LargeScale} builds
on top of BigTable and is used to automatically and incrementally maintain
views when BigTable gets updated.  For instance, it is used within Google to
incrementally maintain the indexes of its search engine as Web pages and links
change.
Percolator enables developers to register driver programs within the system,
which are invoked when a given BigTable column changes.  Each driver program
is executed on a single process server-side, and can start multiple jobs that
read and modify BigTable columns.  All these jobs are executed as a group
ensuring group atomicity through two-phase commit and group isolation
(snapshot isolation) through timestamps.
Percolator relies on an external service to obtain valid timestamps to
interact with BigTable, and saves metadata about running transactions on
additional BigTable columns.
Changes performed during the execution of a driver program may trigger the
execution of other driver programs.  However, these executions are
independent, and atomicity and isolation are not guaranteed across them.

\fakeparagraph{F1} F1~\cite{shute:VLDB:2013:F1} builds a relational database
on top of the storage, replication, and transactional features of Spanner.
F1 inherits all features of Spanner and adds distributed SQL query evaluation,
support for external data sources and sinks, and optimistic transactions.
F1 converts SQL queries into a plan that can be either fully executed on a
single coordinator worker, or include dataflow sub-plans that are executed on
multiple workers and managed by the coordinator.  This execution mode mimics
dataflow DPSs.
To optimize read-intensive, analytical jobs, F1 introduces optimistic
transactions.  They are split into two phases, the first one reads all data
needed for processing, the second one attempts to write results.  The read
phase does not block any other transaction, and so it can be arbitrary long
(as in the case of complex data analytics).  The subsequent write phase will
complete only if no conflicting updates from other transactions occurred
during the read phase.

\fakeparagraph{Trinity} Trinity~\cite{shao:SIGMOD:2013:Trinity} is a graph
data store developed at Microsoft with similar characteristics as TAO in terms
of data model (typed graphs), storage model (in-memory key-value store
backed up in a shared distributed file system), and guarantees (weak
consistency without transactions).
The main distinguishing feature of Trinity is the ability to perform more
complex computations on graphs, including those that require traversing
multiple hops of the graph (such as graph pattern matching) and iterative
analytical jobs (such as vertex-centric computations, as introduced by
Pregel~\cite{malewicz:SIGMOD:2010:Pregel}).
To support these computations, Trinity lets users define different
communication protocols that govern data exchange over the data bus during job
execution.  For instance, data may be buffered and aggregated at sender side
or at receiver side.
For fault tolerance, Trinity stores the association between shared state
portions and workers on the distributed file system, and updates it in the
case of failure.  It also uses checkpoints within long lasting iterative
computations to resume them in the case of failure.

\subsection{New programming models}

\subsubsection{Stateful dataflow}

\fakeparagraph{SDG} Stateful dataflow graphs
(SDG)~\cite{fernandez:ATC:2014:MakingStateExplicit} is a programming model
that extracts a dataflow graph of computation from imperative code (Java
programs).  In SDG, developers write driver programs that include mutable
state and methods to access and modify it.  Code annotations are used to
specify state access patterns within methods.
The program executor (a client-side compiler) analyzes the program to extract
state elements and task elements, representing shared state and
data-parallel tasks in our model.  If possible, state elements are
partitioned across workers.  Similarly, task elements are converted into
multiple concrete tasks, each accessing one single state element from the
shared state portion of the worker it is deployed on.
For instance, consider a program including a matrix of numbers and two methods
to update a value in the matrix and to return the sum of a row.  The matrix
would be converted into a state element partitioned by row (since both methods
can work on individual rows).  Each method would be converted into a
data-parallel task, with one instance of the task per matrix partition.
State elements that cannot be partitioned are replicated in each worker, and
the programming model supports user-defined functions to merge changes applied
to different replicas.
In terms of execution, SDGs are similar to stream processing systems such as
Storm or Flink: jobs are continuous and tasks communicate through direct TCP
channels.  SDGs rely on periodic snapshots and re-execution for fault
tolerance: the same mechanism is adopted to dynamically scale in and out
individual task elements depending on the input load they
receive~\cite{castro:SIGMOD:2013:SEEP}.

\fakeparagraph{TensorFlow} TensorFlow~\cite{abadi:ATC:2016:TensorFlow} is a
system for large-scale machine learning that extends the dataflow model with
explicit shared mutable state.
Jobs represent machine learning models and include operations (data
transformations) and variables (shared mutable state elements representing the
parameters of the machine learning model).  Frequently, jobs are iterative and
update variables at each iteration.
The specific application scenario does not require strong consistency
guarantees for accessing shared state, so tasks are allowed to execute and
read/write variables asynchronously.  If needed, TensorFlow permits some form
of barrier synchronization, for instance to guarantee that all tasks perform
an iteration step using a given value of variables before they get updated
with the results of that step.
TensorFlow tasks can be executed on heterogeneous devices (e.g., hardware
accelerators) and users can express explicit placement constraints.
Since version 2, part of the dataflow plan may be defined at runtime, meaning
that tasks can dynamically define and spawn downstream tasks based on the
input data.
Variables can be periodically checkpointed to durable storage for
faul-tolerance.  In the case of failure, workers can be restarted and they
restore the latest checkpoing available, with no further consistency
guarantees.
TensorFlow has been conceived from the very beginning as a distributed
platform, but many other libraries for machine learning, initially designed
for a single machine, inherited its stateful dataflow execution model: the
most prominent and most widely adopted example is
PyTorch\footnote{\url{https://pytorch.org}}.

\fakeparagraph{Tangram} Tangram~\cite{huang:ATC:2019:Tangram} is a data
processing framework that extends the dataflow model with explicit shared
mutable state.
It implements task-based deployment but allows tasks to access and update an
in-memory key-value store as part of their execution, which enables optimizing
algorithms that benefit from fine-grain updates of intermediate states of
computations (e.g., iterative algorithms or graph processing algorithms).
By analyzing the execution plan, Tangram can understand which parts of the
computation depend on mutable state and which parts do not, and optimizes
fault tolerance for the job at hand.  Immutable data is recomputed using the
same (lineage) approach of MapReduce and Spark, thus re-executing only
tasks that are necessary to rebuild the data.  Mutable state is periodically
checkpointed.
In general, Tangram does not provide group atomicity or isolation for state:
simultaneous accesses to the same state portions from multiple tasks may be
executed in any order.

\subsubsection{Relational actors}

\fakeparagraph{ReactDB} ReactDB~\cite{shah:SIGMOD:2018:reactors} extends the
actor-based programming model 
with database concepts such as relational tables, declarative queries, and
transactional execution semantics.
ReactDB builds on the abstraction of reactors, which are logical actors that
embed state in the form of relational tables.  Each reactor can query its
internal state using a declarative language (SQL) or can explicitly invoke
other reactors.  Invocations across reactors are asynchronous and retain
transactional semantics: clients invoke a root reactor and all invocations it
makes belong to the same root-level transaction, and are atomic and isolated.
The core idea of ReactDB is that developers control data partitioning across
reactors and distributed execution: on one extreme, they can place all state
in a single reactor and mimic a centralized database; on the other extreme,
they can assign a single table (or a single partition of a table) to each
reactor, which maximizes distributed execution.  With this model, the
execution plan is partly implicit (within a single reactor) and partly
explicit (calls between reactors).
ReactDB is currently a research prototype.  As such, it lacks a distributed
implementation, replication, fault tolerance, and dynamic reconfiguration
mechanisms.

\subsection{Hybrid systems}

\fakeparagraph{S-Store} S-Store~\cite{cetintemel:VLDB:2014:sstore} integrates
stream processing capabilities withing a transactional database system.  It
builds on H-Store~\cite{stonebraker:VLDB:2007:hstore}, the research prototype
that later evolved into VoltDB~\cite{stonebraker:IEEEB:2013:voltdb}, and
adopts the same approach to implement transactional guarantees with limited
overhead.
It extends H-Store by enabling stream processing jobs, represented as a
dataflow graph of tasks that may access local task state as part of their
processing.  S-Store exploits the database state to implement the shared state
(visible to all tasks), the task state (visible only to individual tasks of
stream processing jobs), and the data bus (that stream processing tasks use to
exchange data streams).  Input data (for streaming jobs) and transaction
invocations (for data management jobs) are handled by the same engine, which
schedules task execution ensuring that dataflow order is preserved for stream
processing jobs.
S-Store supports two fault tolerance mechanisms, one ensuring same state
recovery through a command log and a periodic snapshot, and one ensuring valid
state by replaying streaming data.
While the S-Store prototype is not distributed, we included it in the survey
for its original integration of data management and data stream processing and
because its core concepts can easily lead to a distributed implementation.

\fakeparagraph{SnappyData} SnappyData~\cite{barzan:CIDR:17:SnappyData} aims to
unify data processing abstractions (both for static and for streaming data)
with mutable shared state.  To do so, it builds on Spark and Spark
Streaming as job execution engines, but extends them to enable
writing to a distributed key-value store.
Users write jobs as declarative SQL queries.  SnappyData analyzes jobs and
classifies them as lightweight (transactional) or heavy (analytical): in the
first case, it directly interacts with the underlying key-value store, while
in the latter case it compiles them into an Spark dataflow execution
plan.
Based on the application at hand, users can decide how to store shared state
(for example, in row or in column format) and how to partition and replicate
it, and how to associate shared state portions to workers, to maximize
co-location of state elements that are frequently accessed together.
Interestingly, SnappyData also supports probabilistic data and query models
that may sacrifice precision to reduce latency.
SnappyData supports group atomicity and group isolation (up to the
repeatable-read isolation model) using two-phase commit and multiversion
concurrency control.
Fault detection is performed in a distributed manner, to avoid single points
of failures.  Fault recovery is based on replication of the key-value store,
which is also used to persist the checkpoints of the data used by Spark
during long-running one-shot and continuous jobs.

\fakeparagraph{StreamDB} StreamDB~\cite{chen:SCC:18:streamDB} integrates
shared state and transactional semantics within a distributed stream
processing system.  From stream processing systems, StreamDB inherits a
dataflow execution plan with job-level deployment.  As in data management
systems, tasks have access to a shared state, which represents relational
tables that can be replicated and partitioned horizontally and/or vertically
across workers.  Each worker is responsible for reading and updating its
portion of the shared state.
Input requests represent invocations of jobs: they are timestamped when
received by the system and each worker executes tasks from multiple jobs in
timestamp order, thus ensuring that jobs are executed in a sequential order
without using explicit locks (group isolation).  The results of a job are
provided to sinks that subscribed to them.
StreamDB is a research prototype and currently requires developers to
explicitly define how to partition the shared state and to write the dataflow
execution plan for all the jobs.
It lacks fault tolerance and reconfiguration mechanisms.

\fakeparagraph{TSpoon} Like StreamDB, TSpoon~\cite{affetti:JPDC:2020:tspoon}
also integrates data management capabilities within a distributed stream
processing system.  Unlike StreamDB, it does not provide a shared state.
Instead, it builds on the programming and execution models of Flink and
enriches them with
\begin{inparaenum}[(i)]
\item the possibility to read (query) task state on demand;
\item transactional guarantees in the access to task state.
\end{inparaenum}
TSpoon considers each input data element as a notification of some change
occurred in the environment in which the system operates.  Developers can
identify portions of the dataflow graph (denoted as transactional subgraphs)
that need to be read and modified in a consistent way, meaning that each
change should be reflected in all task states or none (group atomicity) and
the effects of changes should not overlap in unexpected ways (group
isolation).
TSpoon implements atomicity and isolation by decorating the dataflow graph
with additional operators that act as transaction managers.  It supports
atomicity under the assumption that jobs abort either due to a system failure
or due to some inconsistency during the update to the state of individual
tasks.  It supports different levels of isolation (from read committed to
serializable) with different tradeoffs between guarantees and runtime
overhead.  It supports different isolation protocols, both based on locks and
on timestamps.

\fakeparagraph{Hologres} Hologres~\cite{jiang:VLDB:2020:Hologres} is a system
developed at Alibaba to integrate analytical (long-running) and interactive
(lightweight) jobs.  The system is designed to support high volume data
ingestion from external sources, continuously compute derived information,
store it into a shared state, and make it available to external sinks.
Hologres uses a modular approach, where the storage layer is decoupled from
the processing layer and delegated to external services (e.g., a distributed
file system).
It adopts a leader-worker approach: the shared state is partitioned across
workers and each worker stores a log of updates for the partition it is
responsible for and an in-memory store that is periodically flushed on the
durable storage service.
Hologres supports a structured data model, where data is organized into tables
that can be stored row-wise or column-wise or both depending on the access
pattern, element by element rather than for range scan or aggregation.
A distinctive feature of the system is its scheduling mechanism.  Each job is
decomposed into tasks, and jobs execution is orchestrated by a coordinator,
which assigns tasks to workers based on their current load and their priority:
for instance analytical tasks may be assigned a lower priority to guarantee
low response time for interactive queries.
Hologres supports group atomicity using two-phase commit, but it does not
support group isolation.  Replication of the shared state is not currently
implemented.  Fault tolerance relies on logging and checkpointing and assumes
that the storage layer is durable.  Dynamic reconfiguration is a design
concern, and includes migration of shared state portions but also
reconfiguration of execution slots to enforce load balancing or user-defined
specification of priorities.


  \section{Related surveys and studies}
\label{sec:related}

This section presents related surveys and studies that complement our work,
putting it in a broader context.

The book by Kleppmann~\cite{kleppmann:2016:data-intensive} presents the main
design and implementation strategies to build data-intensive applications.  It
is complementary to our work, as it discusses key design concepts in greater
details, but it does not provide a unifying model nor a taxonomy of systems.

The work on highly available transactions by Bailis et
al.~\cite{bailis:2013:VLDB:HAT} provides a conceptual framework that unifies
various guarantees associated to data management in distributed systems.  It
influenced our discussion of task grouping and replication management.

Various works by Stonebraker and
colleagues~\cite{stonebraker:ICDE:2005:one_size_fits_all,
  stonebraker:VLDB:2007:hstore, stonebraker:CACM:2010:SQLvsNoSQL,
  stonebraker:CACM:2012:NewSQL} guided our classification of DMSs and the
terminology we adopted for the NoSQL and NewSQL classes.
In this area, Davoudian et al.~\cite{davoudian:CSur:2018:NoSQL} present a
survey of NoSQL stores, which expands some of the concepts presented in this
paper, in particular related to data models and storage structures.
Other works explore approaches dedicated to specific data models, such as time
series management~\cite{jensen:TKDE:2017:time_series} and graph
processing~\cite{mcCune:CSur:2015:TLaV}, or strategies to adapt to multiple
data models~\cite{lu:CSur:2019:multi-model}.

In the domain of DPSs, the dataflow model has received large attention, with
work focusing on state management~\cite{to:VLDBJ:2018:survey_state}, handling
of iterations~\cite{gevay:CSur:2021:iterations}, parallelization and
elasticity strategies~\cite{roger:CSur:2019:ParallelizationElasticity},
optimizations for stream processing~\cite{hirzel:CSur:2014:Catalog}.


\end{appendix}


\begin{thebibliography}{109}


\ifx \showCODEN    \undefined \def \showCODEN     #1{\unskip}     \fi
\ifx \showDOI      \undefined \def \showDOI       #1{#1}\fi
\ifx \showISBNx    \undefined \def \showISBNx     #1{\unskip}     \fi
\ifx \showISBNxiii \undefined \def \showISBNxiii  #1{\unskip}     \fi
\ifx \showISSN     \undefined \def \showISSN      #1{\unskip}     \fi
\ifx \showLCCN     \undefined \def \showLCCN      #1{\unskip}     \fi
\ifx \shownote     \undefined \def \shownote      #1{#1}          \fi
\ifx \showarticletitle \undefined \def \showarticletitle #1{#1}   \fi
\ifx \showURL      \undefined \def \showURL       {\relax}        \fi
\providecommand\bibfield[2]{#2}
\providecommand\bibinfo[2]{#2}
\providecommand\natexlab[1]{#1}
\providecommand\showeprint[2][]{arXiv:#2}

\bibitem[Abadi et~al\mbox{.}(2016)]%
        {abadi:ATC:2016:TensorFlow}
\bibfield{author}{\bibinfo{person}{Mart\'{\i}n Abadi}, \bibinfo{person}{Paul
  Barham}, \bibinfo{person}{Jianmin Chen}, \bibinfo{person}{Zhifeng Chen},
  \bibinfo{person}{Andy Davis}, \bibinfo{person}{Jeffrey Dean},
  \bibinfo{person}{Matthieu Devin}, \bibinfo{person}{Sanjay Ghemawat},
  \bibinfo{person}{Geoffrey Irving}, \bibinfo{person}{Michael Isard},
  \bibinfo{person}{Manjunath Kudlur}, \bibinfo{person}{Josh Levenberg},
  \bibinfo{person}{Rajat Monga}, \bibinfo{person}{Sherry Moore},
  \bibinfo{person}{Derek~G. Murray}, \bibinfo{person}{Benoit Steiner},
  \bibinfo{person}{Paul Tucker}, \bibinfo{person}{Vijay Vasudevan},
  \bibinfo{person}{Pete Warden}, \bibinfo{person}{Martin Wicke},
  \bibinfo{person}{Yuan Yu}, {and} \bibinfo{person}{Xiaoqiang Zheng}.}
  \bibinfo{year}{2016}\natexlab{}.
\newblock \showarticletitle{TensorFlow: A System for Large-Scale Machine
  Learning}. In \bibinfo{booktitle}{\emph{Proc of the Conf on Operating Systems
  Design and Impl}} \emph{(\bibinfo{series}{OSDI'16})}.
  \bibinfo{publisher}{USENIX}, \bibinfo{pages}{265–283}.
\newblock


\bibitem[Abouzeid et~al\mbox{.}(2009)]%
        {abouzeid:VLDB:2009:HadoopDB}
\bibfield{author}{\bibinfo{person}{Azza Abouzeid}, \bibinfo{person}{Kamil
  Bajda-Pawlikowski}, \bibinfo{person}{Daniel Abadi}, \bibinfo{person}{Avi
  Silberschatz}, {and} \bibinfo{person}{Alexander Rasin}.}
  \bibinfo{year}{2009}\natexlab{}.
\newblock \showarticletitle{HadoopDB: An Architectural Hybrid of MapReduce and
  DBMS Technologies for Analytical Workloads}.
\newblock \bibinfo{journal}{\emph{Proc VLDB}} \bibinfo{volume}{2},
  \bibinfo{number}{1} (\bibinfo{year}{2009}), \bibinfo{pages}{922–933}.
\newblock


\bibitem[Adams et~al\mbox{.}(2020)]%
        {adams:VLDB:2020:Monarch}
\bibfield{author}{\bibinfo{person}{Colin Adams}, \bibinfo{person}{Luis Alonso},
  \bibinfo{person}{Benjamin Atkin}, \bibinfo{person}{John Banning},
  \bibinfo{person}{Sumeer Bhola}, \bibinfo{person}{Rick Buskens},
  \bibinfo{person}{Ming Chen}, \bibinfo{person}{Xi Chen}, \bibinfo{person}{Yoo
  Chung}, \bibinfo{person}{Qin Jia}, \bibinfo{person}{Nick Sakharov},
  \bibinfo{person}{George Talbot}, \bibinfo{person}{Adam Tart}, {and}
  \bibinfo{person}{Nick Taylor}.} \bibinfo{year}{2020}\natexlab{}.
\newblock \showarticletitle{Monarch: Google's Planet-Scale in-Memory Time
  Series Database}.
\newblock \bibinfo{journal}{\emph{Proc of VLDB}} \bibinfo{volume}{13},
  \bibinfo{number}{12} (\bibinfo{year}{2020}), \bibinfo{pages}{3181–3194}.
\newblock


\bibitem[Adya et~al\mbox{.}(2000)]%
        {Adya:ICDE:2000:Generalized_isolation}
\bibfield{author}{\bibinfo{person}{A. Adya}, \bibinfo{person}{B. Liskov}, {and}
  \bibinfo{person}{P. O'Neil}.} \bibinfo{year}{2000}\natexlab{}.
\newblock \showarticletitle{Generalized isolation level definitions}. In
  \bibinfo{booktitle}{\emph{Proc of the Intl Conf on Data Engineering}}
  \emph{(\bibinfo{series}{ICDE '00})}. \bibinfo{publisher}{IEEE},
  \bibinfo{pages}{67--78}.
\newblock


\bibitem[Affetti et~al\mbox{.}(2020)]%
        {affetti:JPDC:2020:tspoon}
\bibfield{author}{\bibinfo{person}{Lorenzo Affetti},
  \bibinfo{person}{Alessandro Margara}, {and} \bibinfo{person}{Gianpaolo
  Cugola}.} \bibinfo{year}{2020}\natexlab{}.
\newblock \showarticletitle{TSpoon: Transactions on a stream processor}.
\newblock \bibinfo{journal}{\emph{J. Parallel and Distrib. Comput.}}
  \bibinfo{volume}{140} (\bibinfo{year}{2020}), \bibinfo{pages}{65--79}.
\newblock


\bibitem[Agha(1990)]%
        {agha:90:actors}
\bibfield{author}{\bibinfo{person}{Gul~A. Agha}.}
  \bibinfo{year}{1990}\natexlab{}.
\newblock \bibinfo{booktitle}{\emph{{ACTORS} - a model of concurrent
  computation in distributed systems}}.
\newblock \bibinfo{publisher}{{MIT} Press}.
\newblock


\bibitem[Ajoux et~al\mbox{.}(2015)]%
        {ajoux:HOTOS:2015:Challenges}
\bibfield{author}{\bibinfo{person}{Phillipe Ajoux}, \bibinfo{person}{Nathan
  Bronson}, \bibinfo{person}{Sanjeev Kumar}, \bibinfo{person}{Wyatt Lloyd},
  {and} \bibinfo{person}{Kaushik Veeraraghavan}.}
  \bibinfo{year}{2015}\natexlab{}.
\newblock \showarticletitle{Challenges to Adopting Stronger Consistency at
  Scale}. In \bibinfo{booktitle}{\emph{Proc of the Conf on Hot Topics in
  Operating Systems}} \emph{(\bibinfo{series}{HOTOS'15})}.
  \bibinfo{publisher}{USENIX}, \bibinfo{pages}{13:1--13:7}.
\newblock


\bibitem[Akidau et~al\mbox{.}(2013)]%
        {akidau:VLDB:2013:MillWheel}
\bibfield{author}{\bibinfo{person}{Tyler Akidau}, \bibinfo{person}{Alex
  Balikov}, \bibinfo{person}{Kaya Bekiro\u{g}lu}, \bibinfo{person}{Slava
  Chernyak}, \bibinfo{person}{Josh Haberman}, \bibinfo{person}{Reuven Lax},
  \bibinfo{person}{Sam McVeety}, \bibinfo{person}{Daniel Mills},
  \bibinfo{person}{Paul Nordstrom}, {and} \bibinfo{person}{Sam Whittle}.}
  \bibinfo{year}{2013}\natexlab{}.
\newblock \showarticletitle{MillWheel: Fault-Tolerant Stream Processing at
  Internet Scale}.
\newblock \bibinfo{journal}{\emph{Proc of VLDB}} \bibinfo{volume}{6},
  \bibinfo{number}{11} (\bibinfo{year}{2013}), \bibinfo{pages}{1033–1044}.
\newblock


\bibitem[Akidau et~al\mbox{.}(2015)]%
        {akidau:VLDB:2015:dataflow}
\bibfield{author}{\bibinfo{person}{Tyler Akidau}, \bibinfo{person}{Robert
  Bradshaw}, \bibinfo{person}{Craig Chambers}, \bibinfo{person}{Slava
  Chernyak}, \bibinfo{person}{Rafael~J. Fern\'{a}ndez-Moctezuma},
  \bibinfo{person}{Reuven Lax}, \bibinfo{person}{Sam McVeety},
  \bibinfo{person}{Daniel Mills}, \bibinfo{person}{Frances Perry},
  \bibinfo{person}{Eric Schmidt}, {and} \bibinfo{person}{Sam Whittle}.}
  \bibinfo{year}{2015}\natexlab{}.
\newblock \showarticletitle{The Dataflow Model: A Practical Approach to
  Balancing Correctness, Latency, and Cost in Massive-scale, Unbounded,
  Out-of-order Data Processing}.
\newblock \bibinfo{journal}{\emph{Proc of VLDB}} \bibinfo{volume}{8},
  \bibinfo{number}{12} (\bibinfo{year}{2015}), \bibinfo{pages}{1792--1803}.
\newblock


\bibitem[Alsubaiee et~al\mbox{.}(2014)]%
        {alsubaiee:VLDB:2014:AsterixDB}
\bibfield{author}{\bibinfo{person}{Sattam Alsubaiee}, \bibinfo{person}{Yasser
  Altowim}, \bibinfo{person}{Hotham Altwaijry}, \bibinfo{person}{Alexander
  Behm}, \bibinfo{person}{Vinayak Borkar}, \bibinfo{person}{Yingyi Bu},
  \bibinfo{person}{Michael Carey}, \bibinfo{person}{Inci Cetindil},
  \bibinfo{person}{Madhusudan Cheelangi}, \bibinfo{person}{Khurram Faraaz},
  \bibinfo{person}{Eugenia Gabrielova}, \bibinfo{person}{Raman Grover},
  \bibinfo{person}{Zachary Heilbron}, \bibinfo{person}{Young-Seok Kim},
  \bibinfo{person}{Chen Li}, \bibinfo{person}{Guangqiang Li},
  \bibinfo{person}{Ji~Mahn Ok}, \bibinfo{person}{Nicola Onose},
  \bibinfo{person}{Pouria Pirzadeh}, \bibinfo{person}{Vassilis Tsotras},
  \bibinfo{person}{Rares Vernica}, \bibinfo{person}{Jian Wen}, {and}
  \bibinfo{person}{Till Westmann}.} \bibinfo{year}{2014}\natexlab{}.
\newblock \showarticletitle{AsterixDB: A Scalable, Open Source BDMS}.
\newblock \bibinfo{journal}{\emph{Proc of VLDB}} \bibinfo{volume}{7},
  \bibinfo{number}{14} (\bibinfo{year}{2014}), \bibinfo{pages}{1905–1916}.
\newblock


\bibitem[Anderson et~al\mbox{.}(2010)]%
        {anderson:2010:CouchDB}
\bibfield{author}{\bibinfo{person}{J~Chris Anderson}, \bibinfo{person}{Jan
  Lehnardt}, {and} \bibinfo{person}{Noah Slater}.}
  \bibinfo{year}{2010}\natexlab{}.
\newblock \bibinfo{booktitle}{\emph{CouchDB: the definitive guide: time to
  relax}}.
\newblock \bibinfo{publisher}{O'Reilly}.
\newblock


\bibitem[Antonopoulos et~al\mbox{.}(2019)]%
        {antonopoulos:SIGMOD:2019:Socrates}
\bibfield{author}{\bibinfo{person}{Panagiotis Antonopoulos},
  \bibinfo{person}{Alex Budovski}, \bibinfo{person}{Cristian Diaconu},
  \bibinfo{person}{Alejandro Hernandez~Saenz}, \bibinfo{person}{Jack Hu},
  \bibinfo{person}{Hanuma Kodavalla}, \bibinfo{person}{Donald Kossmann},
  \bibinfo{person}{Sandeep Lingam}, \bibinfo{person}{Umar~Farooq Minhas},
  \bibinfo{person}{Naveen Prakash}, \bibinfo{person}{Vijendra Purohit},
  \bibinfo{person}{Hugh Qu}, \bibinfo{person}{Chaitanya~Sreenivas Ravella},
  \bibinfo{person}{Krystyna Reisteter}, \bibinfo{person}{Sheetal Shrotri},
  \bibinfo{person}{Dixin Tang}, {and} \bibinfo{person}{Vikram Wakade}.}
  \bibinfo{year}{2019}\natexlab{}.
\newblock \showarticletitle{Socrates: The New SQL Server in the Cloud}. In
  \bibinfo{booktitle}{\emph{Proc of the Intl Conf on Management of Data}}
  \emph{(\bibinfo{series}{SIGMOD '19})}. \bibinfo{publisher}{ACM},
  \bibinfo{pages}{1743–1756}.
\newblock


\bibitem[Armbrust et~al\mbox{.}(2015)]%
        {armbrust:SIGMOD:2015:SparkSQL}
\bibfield{author}{\bibinfo{person}{Michael Armbrust},
  \bibinfo{person}{Reynold~S. Xin}, \bibinfo{person}{Cheng Lian},
  \bibinfo{person}{Yin Huai}, \bibinfo{person}{Davies Liu},
  \bibinfo{person}{Joseph~K. Bradley}, \bibinfo{person}{Xiangrui Meng},
  \bibinfo{person}{Tomer Kaftan}, \bibinfo{person}{Michael~J. Franklin},
  \bibinfo{person}{Ali Ghodsi}, {and} \bibinfo{person}{Matei Zaharia}.}
  \bibinfo{year}{2015}\natexlab{}.
\newblock \showarticletitle{Spark SQL: Relational Data Processing in Spark}. In
  \bibinfo{booktitle}{\emph{Proc of the Intl Conf on Management of Data}}
  \emph{(\bibinfo{series}{SIGMOD '15})}. \bibinfo{publisher}{ACM},
  \bibinfo{pages}{1383–1394}.
\newblock


\bibitem[Arulraj and Pavlo(2017)]%
        {arulraj:SIGMOD:2017:HowTo}
\bibfield{author}{\bibinfo{person}{Joy Arulraj} {and} \bibinfo{person}{Andrew
  Pavlo}.} \bibinfo{year}{2017}\natexlab{}.
\newblock \showarticletitle{How to Build a Non-Volatile Memory Database
  Management System}. In \bibinfo{booktitle}{\emph{Proc of the Intl Conf on
  Management of Data}} \emph{(\bibinfo{series}{SIGMOD '17})}.
  \bibinfo{publisher}{ACM}, \bibinfo{pages}{1753–1758}.
\newblock


\bibitem[Bacon et~al\mbox{.}(2017)]%
        {bacon:SIGMOD:2017:Spanner}
\bibfield{author}{\bibinfo{person}{David~F. Bacon}, \bibinfo{person}{Nathan
  Bales}, \bibinfo{person}{Nico Bruno}, \bibinfo{person}{Brian~F. Cooper},
  \bibinfo{person}{Adam Dickinson}, \bibinfo{person}{Andrew Fikes},
  \bibinfo{person}{Campbell Fraser}, \bibinfo{person}{Andrey Gubarev},
  \bibinfo{person}{Milind Joshi}, \bibinfo{person}{Eugene Kogan},
  \bibinfo{person}{Alexander Lloyd}, \bibinfo{person}{Sergey Melnik},
  \bibinfo{person}{Rajesh Rao}, \bibinfo{person}{David Shue},
  \bibinfo{person}{Christopher Taylor}, \bibinfo{person}{Marcel van~der Holst},
  {and} \bibinfo{person}{Dale Woodford}.} \bibinfo{year}{2017}\natexlab{}.
\newblock \showarticletitle{Spanner: Becoming a SQL System}. In
  \bibinfo{booktitle}{\emph{Proc of the Intl Conf on Management of Data}}
  \emph{(\bibinfo{series}{SIGMOD '17})}. \bibinfo{publisher}{ACM},
  \bibinfo{pages}{331–343}.
\newblock


\bibitem[Bailis et~al\mbox{.}(2013)]%
        {bailis:2013:VLDB:HAT}
\bibfield{author}{\bibinfo{person}{Peter Bailis}, \bibinfo{person}{Aaron
  Davidson}, \bibinfo{person}{Alan Fekete}, \bibinfo{person}{Ali Ghodsi},
  \bibinfo{person}{Joseph~M. Hellerstein}, {and} \bibinfo{person}{Ion Stoica}.}
  \bibinfo{year}{2013}\natexlab{}.
\newblock \showarticletitle{Highly Available Transactions: Virtues and
  Limitations}.
\newblock \bibinfo{journal}{\emph{Proc of VLDB}} \bibinfo{volume}{7},
  \bibinfo{number}{3} (\bibinfo{year}{2013}), \bibinfo{pages}{181--192}.
\newblock


\bibitem[Balakrishnan et~al\mbox{.}(2013)]%
        {balakrishnan:SOSP:2013:Tango}
\bibfield{author}{\bibinfo{person}{Mahesh Balakrishnan},
  \bibinfo{person}{Dahlia Malkhi}, \bibinfo{person}{Ted Wobber},
  \bibinfo{person}{Ming Wu}, \bibinfo{person}{Vijayan Prabhakaran},
  \bibinfo{person}{Michael Wei}, \bibinfo{person}{John~D. Davis},
  \bibinfo{person}{Sriram Rao}, \bibinfo{person}{Tao Zou}, {and}
  \bibinfo{person}{Aviad Zuck}.} \bibinfo{year}{2013}\natexlab{}.
\newblock \showarticletitle{Tango: Distributed Data Structures over a Shared
  Log}. In \bibinfo{booktitle}{\emph{Proc of the Symposium on Operating Systems
  Principles}} \emph{(\bibinfo{series}{SOSP '13})}. \bibinfo{publisher}{ACM},
  \bibinfo{pages}{325–340}.
\newblock


\bibitem[Banciihon(1988)]%
        {banciihon:PODS:1988:OO}
\bibfield{author}{\bibinfo{person}{Fran\c{c}ois Banciihon}.}
  \bibinfo{year}{1988}\natexlab{}.
\newblock \showarticletitle{Object-Oriented Database Systems}. In
  \bibinfo{booktitle}{\emph{Proc of the Symposium on Principles of Database
  Systems}} \emph{(\bibinfo{series}{PODS '88})}. \bibinfo{publisher}{ACM},
  \bibinfo{pages}{152–162}.
\newblock
\showISBNx{0897912632}


\bibitem[Baresi et~al\mbox{.}(2021)]%
        {baresi:TSE:2021:allocation}
\bibfield{author}{\bibinfo{person}{Luciano Baresi}, \bibinfo{person}{Alberto
  Leva}, {and} \bibinfo{person}{Giovanni Quattrocchi}.}
  \bibinfo{year}{2021}\natexlab{}.
\newblock \showarticletitle{Fine-Grained Dynamic Resource Allocation for
  Big-Data Applications}.
\newblock \bibinfo{journal}{\emph{IEEE Transactions on Software Engineering}}
  \bibinfo{volume}{47}, \bibinfo{number}{8} (\bibinfo{year}{2021}),
  \bibinfo{pages}{1668--1682}.
\newblock


\bibitem[Bejeck(2018)]%
        {bejeck:2018:KafkaStreams}
\bibfield{author}{\bibinfo{person}{Bill Bejeck}.}
  \bibinfo{year}{2018}\natexlab{}.
\newblock \bibinfo{booktitle}{\emph{Kafka Streams in Action: Real-time apps and
  microservices with the Kafka Streams API}}.
\newblock \bibinfo{publisher}{Manning}.
\newblock


\bibitem[Bernstein and Goodman(1981)]%
        {bernstein:CSUR:81:concurrency}
\bibfield{author}{\bibinfo{person}{Philip~A. Bernstein} {and}
  \bibinfo{person}{Nathan Goodman}.} \bibinfo{year}{1981}\natexlab{}.
\newblock \showarticletitle{Concurrency Control in Distributed Database
  Systems}.
\newblock \bibinfo{journal}{\emph{{ACM} Comput. Surv.}} \bibinfo{volume}{13},
  \bibinfo{number}{2} (\bibinfo{year}{1981}), \bibinfo{pages}{185--221}.
\newblock


\bibitem[Borkar et~al\mbox{.}(2011)]%
        {borkar:ICDE:2011:Hyracks}
\bibfield{author}{\bibinfo{person}{Vinayak Borkar}, \bibinfo{person}{Michael
  Carey}, \bibinfo{person}{Raman Grover}, \bibinfo{person}{Nicola Onose}, {and}
  \bibinfo{person}{Rares Vernica}.} \bibinfo{year}{2011}\natexlab{}.
\newblock \showarticletitle{Hyracks: A Flexible and Extensible Foundation for
  Data-Intensive Computing}. In \bibinfo{booktitle}{\emph{Proc of the Intl Conf
  on Data Engineering}} \emph{(\bibinfo{series}{ICDE '11})}.
  \bibinfo{publisher}{IEEE}, \bibinfo{pages}{1151–1162}.
\newblock


\bibitem[Bronson et~al\mbox{.}(2013)]%
        {bronson:ATC:2013:TAO}
\bibfield{author}{\bibinfo{person}{Nathan Bronson}, \bibinfo{person}{Zach
  Amsden}, \bibinfo{person}{George Cabrera}, \bibinfo{person}{Prasad Chakka},
  \bibinfo{person}{Peter Dimov}, \bibinfo{person}{Hui Ding},
  \bibinfo{person}{Jack Ferris}, \bibinfo{person}{Anthony Giardullo},
  \bibinfo{person}{Sachin Kulkarni}, \bibinfo{person}{Harry Li},
  \bibinfo{person}{Mark Marchukov}, \bibinfo{person}{Dmitri Petrov},
  \bibinfo{person}{Lovro Puzar}, \bibinfo{person}{Yee~Jiun Song}, {and}
  \bibinfo{person}{Venkat Venkataramani}.} \bibinfo{year}{2013}\natexlab{}.
\newblock \showarticletitle{{TAO}: Facebook{\textquoteright}s Distributed Data
  Store for the Social Graph}. In \bibinfo{booktitle}{\emph{Proc of the USENIX
  Annual Technical Conf}} \emph{(\bibinfo{series}{ATC '13})}.
  \bibinfo{publisher}{{USENIX} Assoc.}, \bibinfo{pages}{49--60}.
\newblock


\bibitem[Bu et~al\mbox{.}(2010)]%
        {bu:VLDB:2010:HaLoop}
\bibfield{author}{\bibinfo{person}{Yingyi Bu}, \bibinfo{person}{Bill Howe},
  \bibinfo{person}{Magdalena Balazinska}, {and} \bibinfo{person}{Michael~D.
  Ernst}.} \bibinfo{year}{2010}\natexlab{}.
\newblock \showarticletitle{HaLoop: Efficient Iterative Data Processing on
  Large Clusters}.
\newblock \bibinfo{journal}{\emph{Proceeding of VLDB}} \bibinfo{volume}{3},
  \bibinfo{number}{1–2} (\bibinfo{year}{2010}), \bibinfo{pages}{285–296}.
\newblock


\bibitem[Buragohain et~al\mbox{.}(2020)]%
        {buragohain:SIGMOD:2020:A1}
\bibfield{author}{\bibinfo{person}{Chiranjeeb Buragohain},
  \bibinfo{person}{Knut~Magne Risvik}, \bibinfo{person}{Paul Brett},
  \bibinfo{person}{Miguel Castro}, \bibinfo{person}{Wonhee Cho},
  \bibinfo{person}{Joshua Cowhig}, \bibinfo{person}{Nikolas Gloy},
  \bibinfo{person}{Karthik Kalyanaraman}, \bibinfo{person}{Richendra Khanna},
  \bibinfo{person}{John Pao}, \bibinfo{person}{Matthew Renzelmann},
  \bibinfo{person}{Alex Shamis}, \bibinfo{person}{Timothy Tan}, {and}
  \bibinfo{person}{Shuheng Zheng}.} \bibinfo{year}{2020}\natexlab{}.
\newblock \showarticletitle{A1: A Distributed In-Memory Graph Database}. In
  \bibinfo{booktitle}{\emph{Proc of the Intl Conf on Management of Data}}
  \emph{(\bibinfo{series}{SIGMOD '20})}. \bibinfo{publisher}{ACM},
  \bibinfo{pages}{329–344}.
\newblock


\bibitem[Camacho-Rodr\'{\i}guez et~al\mbox{.}(2019)]%
        {camacho-rodriguez:SIGMOD:2019:ApacheHive}
\bibfield{author}{\bibinfo{person}{Jes\'{u}s Camacho-Rodr\'{\i}guez},
  \bibinfo{person}{Ashutosh Chauhan}, \bibinfo{person}{Alan Gates},
  \bibinfo{person}{Eugene Koifman}, \bibinfo{person}{Owen O'Malley},
  \bibinfo{person}{Vineet Garg}, \bibinfo{person}{Zoltan Haindrich},
  \bibinfo{person}{Sergey Shelukhin}, \bibinfo{person}{Prasanth Jayachandran},
  \bibinfo{person}{Siddharth Seth}, \bibinfo{person}{Deepak Jaiswal},
  \bibinfo{person}{Slim Bouguerra}, \bibinfo{person}{Nishant Bangarwa},
  \bibinfo{person}{Sankar Hariappan}, \bibinfo{person}{Anishek Agarwal},
  \bibinfo{person}{Jason Dere}, \bibinfo{person}{Daniel Dai},
  \bibinfo{person}{Thejas Nair}, \bibinfo{person}{Nita Dembla},
  \bibinfo{person}{Gopal Vijayaraghavan}, {and} \bibinfo{person}{G\"{u}nther
  Hagleitner}.} \bibinfo{year}{2019}\natexlab{}.
\newblock \showarticletitle{Apache Hive: From MapReduce to Enterprise-Grade Big
  Data Warehousing}. In \bibinfo{booktitle}{\emph{Proc of the Intl Conf on
  Management of Data}} \emph{(\bibinfo{series}{SIGMOD '19})}.
  \bibinfo{publisher}{ACM}, \bibinfo{pages}{1773–1786}.
\newblock


\bibitem[Carbone et~al\mbox{.}(2015)]%
        {carbone:IEEEB:2015:flink}
\bibfield{author}{\bibinfo{person}{Paris Carbone}, \bibinfo{person}{Asterios
  Katsifodimos}, \bibinfo{person}{Stephan Ewen}, \bibinfo{person}{Volker
  Markl}, \bibinfo{person}{Seif Haridi}, {and} \bibinfo{person}{Kostas
  Tzoumas}.} \bibinfo{year}{2015}\natexlab{}.
\newblock \showarticletitle{Apache Flink{\texttrademark}: Stream and Batch
  Processing in a Single Engine}.
\newblock \bibinfo{journal}{\emph{IEEE Data Engineering Bulletin}}
  \bibinfo{volume}{38}, \bibinfo{number}{4} (\bibinfo{year}{2015}),
  \bibinfo{pages}{28--38}.
\newblock


\bibitem[Cardellini et~al\mbox{.}(2022)]%
        {cardellini:CSur:2022:adaptation}
\bibfield{author}{\bibinfo{person}{Valeria Cardellini},
  \bibinfo{person}{Francesco Lo~Presti}, \bibinfo{person}{Matteo Nardelli},
  {and} \bibinfo{person}{Gabriele Russo~Russo}.}
  \bibinfo{year}{2022}\natexlab{}.
\newblock \showarticletitle{Run-Time Adaptation of Data Stream Processing
  Systems: The State of the Art}.
\newblock \bibinfo{journal}{\emph{Comput. Surveys}} (\bibinfo{year}{2022}).
\newblock


\bibitem[Castro~Fernandez et~al\mbox{.}(2013)]%
        {castro:SIGMOD:2013:SEEP}
\bibfield{author}{\bibinfo{person}{Raul Castro~Fernandez},
  \bibinfo{person}{Matteo Migliavacca}, \bibinfo{person}{Evangelia
  Kalyvianaki}, {and} \bibinfo{person}{Peter Pietzuch}.}
  \bibinfo{year}{2013}\natexlab{}.
\newblock \showarticletitle{Integrating Scale out and Fault Tolerance in Stream
  Processing Using Operator State Management}. In
  \bibinfo{booktitle}{\emph{Proc of the Intl Conf on Management of Data}}
  \emph{(\bibinfo{series}{SIGMOD '13})}. \bibinfo{publisher}{ACM},
  \bibinfo{pages}{725–736}.
\newblock


\bibitem[Cetintemel et~al\mbox{.}(2014)]%
        {cetintemel:VLDB:2014:sstore}
\bibfield{author}{\bibinfo{person}{Ugur Cetintemel}, \bibinfo{person}{Jiang
  Du}, \bibinfo{person}{Tim Kraska}, \bibinfo{person}{Samuel Madden},
  \bibinfo{person}{David Maier}, \bibinfo{person}{John Meehan},
  \bibinfo{person}{Andrew Pavlo}, \bibinfo{person}{Michael Stonebraker},
  \bibinfo{person}{Erik Sutherland}, \bibinfo{person}{Nesime Tatbul},
  {et~al\mbox{.}}} \bibinfo{year}{2014}\natexlab{}.
\newblock \showarticletitle{S-Store: a streaming NewSQL system for big velocity
  applications}.
\newblock \bibinfo{journal}{\emph{Proc of VLDB}} \bibinfo{volume}{7},
  \bibinfo{number}{13} (\bibinfo{year}{2014}), \bibinfo{pages}{1633--1636}.
\newblock


\bibitem[Chandy and Lamport(1985)]%
        {chandi:TOCS:1985:distributed}
\bibfield{author}{\bibinfo{person}{K.~Mani Chandy} {and}
  \bibinfo{person}{Leslie Lamport}.} \bibinfo{year}{1985}\natexlab{}.
\newblock \showarticletitle{Distributed Snapshots: Determining Global States of
  Distributed Systems}.
\newblock \bibinfo{journal}{\emph{Trans on Computer Systems}}
  \bibinfo{volume}{3}, \bibinfo{number}{1} (\bibinfo{year}{1985}),
  \bibinfo{pages}{63–75}.
\newblock


\bibitem[Chang et~al\mbox{.}(2006)]%
        {chang:OSDI:06:bigtable}
\bibfield{author}{\bibinfo{person}{Fay Chang}, \bibinfo{person}{Jeffrey Dean},
  \bibinfo{person}{Sanjay Ghemawat}, \bibinfo{person}{Wilson~C. Hsieh},
  \bibinfo{person}{Deborah~A. Wallach}, \bibinfo{person}{Michael Burrows},
  \bibinfo{person}{Tushar Chandra}, \bibinfo{person}{Andrew Fikes}, {and}
  \bibinfo{person}{Robert Gruber}.} \bibinfo{year}{2006}\natexlab{}.
\newblock \showarticletitle{Bigtable: {A} Distributed Storage System for
  Structured Data}. In \bibinfo{booktitle}{\emph{Proc of the Symposium on
  Operating Systems Design and Implementation}} \emph{(\bibinfo{series}{OSDI
  '06})}, \bibfield{editor}{\bibinfo{person}{Brian~N. Bershad} {and}
  \bibinfo{person}{Jeffrey~C. Mogul}} (Eds.). \bibinfo{publisher}{{USENIX}
  Assoc.}, \bibinfo{pages}{205--218}.
\newblock


\bibitem[Chen et~al\mbox{.}(2018)]%
        {chen:EuroSys:2018:G-Miner}
\bibfield{author}{\bibinfo{person}{Hongzhi Chen}, \bibinfo{person}{Miao Liu},
  \bibinfo{person}{Yunjian Zhao}, \bibinfo{person}{Xiao Yan},
  \bibinfo{person}{Da Yan}, {and} \bibinfo{person}{James Cheng}.}
  \bibinfo{year}{2018}\natexlab{}.
\newblock \showarticletitle{G-Miner: An Efficient Task-Oriented Graph Mining
  System}. In \bibinfo{booktitle}{\emph{Proc of the EuroSys Conf}}
  \emph{(\bibinfo{series}{EuroSys '18})}. \bibinfo{publisher}{ACM}, Article
  \bibinfo{articleno}{32}.
\newblock


\bibitem[Chen and Migliavacca(2018)]%
        {chen:SCC:18:streamDB}
\bibfield{author}{\bibinfo{person}{Huankai Chen} {and} \bibinfo{person}{Matteo
  Migliavacca}.} \bibinfo{year}{2018}\natexlab{}.
\newblock \showarticletitle{StreamDB: {A} Unified Data Management System for
  Service-Based Cloud Application}. In \bibinfo{booktitle}{\emph{Proc of the
  Intl Conf on Services Computing}} \emph{(\bibinfo{series}{SCC '18})}.
  \bibinfo{publisher}{{IEEE}}, \bibinfo{pages}{169--176}.
\newblock


\bibitem[Chodorow(2013)]%
        {chodorow:2013:MongoDB}
\bibfield{author}{\bibinfo{person}{Kristina Chodorow}.}
  \bibinfo{year}{2013}\natexlab{}.
\newblock \bibinfo{booktitle}{\emph{MongoDB: the definitive guide: powerful and
  scalable data storage}}.
\newblock \bibinfo{publisher}{O'Reilly}.
\newblock


\bibitem[Cooper et~al\mbox{.}(2008)]%
        {cooper:VLDB:2008:PNUTS}
\bibfield{author}{\bibinfo{person}{Brian~F. Cooper}, \bibinfo{person}{Raghu
  Ramakrishnan}, \bibinfo{person}{Utkarsh Srivastava}, \bibinfo{person}{Adam
  Silberstein}, \bibinfo{person}{Philip Bohannon}, \bibinfo{person}{Hans-Arno
  Jacobsen}, \bibinfo{person}{Nick Puz}, \bibinfo{person}{Daniel Weaver}, {and}
  \bibinfo{person}{Ramana Yerneni}.} \bibinfo{year}{2008}\natexlab{}.
\newblock \showarticletitle{PNUTS: Yahoo!'s Hosted Data Serving Platform}.
\newblock \bibinfo{journal}{\emph{Proc of VLDB}} \bibinfo{volume}{1},
  \bibinfo{number}{2} (\bibinfo{year}{2008}), \bibinfo{pages}{1277–1288}.
\newblock


\bibitem[Corbett et~al\mbox{.}(2013)]%
        {corbett:TOCS:2013:Spanner}
\bibfield{author}{\bibinfo{person}{James~C. Corbett}, \bibinfo{person}{Jeffrey
  Dean}, \bibinfo{person}{Michael Epstein}, \bibinfo{person}{Andrew Fikes},
  \bibinfo{person}{Christopher Frost}, \bibinfo{person}{J.~J. Furman},
  \bibinfo{person}{Sanjay Ghemawat}, \bibinfo{person}{Andrey Gubarev},
  \bibinfo{person}{Christopher Heiser}, \bibinfo{person}{Peter Hochschild},
  \bibinfo{person}{Wilson Hsieh}, \bibinfo{person}{Sebastian Kanthak},
  \bibinfo{person}{Eugene Kogan}, \bibinfo{person}{Hongyi Li},
  \bibinfo{person}{Alexander Lloyd}, \bibinfo{person}{Sergey Melnik},
  \bibinfo{person}{David Mwaura}, \bibinfo{person}{David Nagle},
  \bibinfo{person}{Sean Quinlan}, \bibinfo{person}{Rajesh Rao},
  \bibinfo{person}{Lindsay Rolig}, \bibinfo{person}{Yasushi Saito},
  \bibinfo{person}{Michal Szymaniak}, \bibinfo{person}{Christopher Taylor},
  \bibinfo{person}{Ruth Wang}, {and} \bibinfo{person}{Dale Woodford}.}
  \bibinfo{year}{2013}\natexlab{}.
\newblock \showarticletitle{Spanner: Google's Globally Distributed Database}.
\newblock \bibinfo{journal}{\emph{Trans on Computer Systems}}
  \bibinfo{volume}{31}, \bibinfo{number}{3} (\bibinfo{year}{2013}),
  \bibinfo{pages}{8:1--8:22}.
\newblock


\bibitem[Curtiss et~al\mbox{.}(2013)]%
        {curtiss:VLDB:2013:Unicorn}
\bibfield{author}{\bibinfo{person}{Michael Curtiss}, \bibinfo{person}{Iain
  Becker}, \bibinfo{person}{Tudor Bosman}, \bibinfo{person}{Sergey Doroshenko},
  \bibinfo{person}{Lucian Grijincu}, \bibinfo{person}{Tom Jackson},
  \bibinfo{person}{Sandhya Kunnatur}, \bibinfo{person}{Soren Lassen},
  \bibinfo{person}{Philip Pronin}, \bibinfo{person}{Sriram Sankar},
  \bibinfo{person}{Guanghao Shen}, \bibinfo{person}{Gintaras Woss},
  \bibinfo{person}{Chao Yang}, {and} \bibinfo{person}{Ning Zhang}.}
  \bibinfo{year}{2013}\natexlab{}.
\newblock \showarticletitle{Unicorn: A System for Searching the Social Graph}.
\newblock \bibinfo{journal}{\emph{Proc of VLDB}} \bibinfo{volume}{6},
  \bibinfo{number}{11} (\bibinfo{year}{2013}), \bibinfo{pages}{1150–1161}.
\newblock
\showISSN{2150-8097}


\bibitem[Davoudian et~al\mbox{.}(2018)]%
        {davoudian:CSur:2018:NoSQL}
\bibfield{author}{\bibinfo{person}{Ali Davoudian}, \bibinfo{person}{Liu Chen},
  {and} \bibinfo{person}{Mengchi Liu}.} \bibinfo{year}{2018}\natexlab{}.
\newblock \showarticletitle{A Survey on NoSQL Stores}.
\newblock \bibinfo{journal}{\emph{ACM Comput. Surv.}} \bibinfo{volume}{51},
  \bibinfo{number}{2}, Article \bibinfo{articleno}{40} (\bibinfo{year}{2018}).
\newblock


\bibitem[Davoudian and Liu(2020)]%
        {davoudian:CSur:2020:BigData_SE}
\bibfield{author}{\bibinfo{person}{Ali Davoudian} {and}
  \bibinfo{person}{Mengchi Liu}.} \bibinfo{year}{2020}\natexlab{}.
\newblock \showarticletitle{Big Data Systems: A Software Engineering
  Perspective}.
\newblock \bibinfo{journal}{\emph{ACM Comput. Surv.}} \bibinfo{volume}{53},
  \bibinfo{number}{5}, Article \bibinfo{articleno}{110} (\bibinfo{year}{2020}),
  \bibinfo{numpages}{39}~pages.
\newblock


\bibitem[Dean and Ghemawat(2008)]%
        {dean:CACM:2008:mapreduce}
\bibfield{author}{\bibinfo{person}{Jeffrey Dean} {and} \bibinfo{person}{Sanjay
  Ghemawat}.} \bibinfo{year}{2008}\natexlab{}.
\newblock \showarticletitle{MapReduce: Simplified Data Processing on Large
  Clusters}.
\newblock \bibinfo{journal}{\emph{Commun. ACM}} \bibinfo{volume}{51},
  \bibinfo{number}{1} (\bibinfo{year}{2008}), \bibinfo{pages}{107--113}.
\newblock


\bibitem[DeCandia et~al\mbox{.}(2007)]%
        {deCandia:SOSP:2007:Dynamo}
\bibfield{author}{\bibinfo{person}{Giuseppe DeCandia}, \bibinfo{person}{Deniz
  Hastorun}, \bibinfo{person}{Madan Jampani}, \bibinfo{person}{Gunavardhan
  Kakulapati}, \bibinfo{person}{Avinash Lakshman}, \bibinfo{person}{Alex
  Pilchin}, \bibinfo{person}{Swaminathan Sivasubramanian},
  \bibinfo{person}{Peter Vosshall}, {and} \bibinfo{person}{Werner Vogels}.}
  \bibinfo{year}{2007}\natexlab{}.
\newblock \showarticletitle{Dynamo: Amazon's Highly Available Key-value Store}.
  In \bibinfo{booktitle}{\emph{Proc of the Symposium on Operating Systems
  Principles}} \emph{(\bibinfo{series}{SOSP '07})}. \bibinfo{publisher}{ACM},
  \bibinfo{pages}{205--220}.
\newblock


\bibitem[Del~Monte et~al\mbox{.}(2022)]%
        {delMonte:SIGMOD:2022:rethinking}
\bibfield{author}{\bibinfo{person}{Bonaventura Del~Monte},
  \bibinfo{person}{Steffen Zeuch}, \bibinfo{person}{Tilmann Rabl}, {and}
  \bibinfo{person}{Volker Markl}.} \bibinfo{year}{2022}\natexlab{}.
\newblock \showarticletitle{Rethinking Stateful Stream Processing with RDMA}.
  In \bibinfo{booktitle}{\emph{Proc of the Intl Conf on Management of Data}}
  \emph{(\bibinfo{series}{SIGMOD '22})}. \bibinfo{publisher}{ACM},
  \bibinfo{pages}{1078–1092}.
\newblock


\bibitem[Dragojevi{\'c} et~al\mbox{.}(2014)]%
        {dragojevic:NSDI:2014:FaRM}
\bibfield{author}{\bibinfo{person}{Aleksandar Dragojevi{\'c}},
  \bibinfo{person}{Dushyanth Narayanan}, \bibinfo{person}{Miguel Castro}, {and}
  \bibinfo{person}{Orion Hodson}.} \bibinfo{year}{2014}\natexlab{}.
\newblock \showarticletitle{FaRM: Fast Remote Memory}. In
  \bibinfo{booktitle}{\emph{{USENIX} Symposium on Networked Systems Design and
  Implementation}} \emph{(\bibinfo{series}{NSDI '14})}.
  \bibinfo{publisher}{{USENIX} Assoc.}, \bibinfo{pages}{401--414}.
\newblock


\bibitem[Fang et~al\mbox{.}(2020)]%
        {fang:VLDB:2020:in-memory}
\bibfield{author}{\bibinfo{person}{Jian Fang}, \bibinfo{person}{Yvo~TB Mulder},
  \bibinfo{person}{Jan Hidders}, \bibinfo{person}{Jinho Lee}, {and}
  \bibinfo{person}{H~Peter Hofstee}.} \bibinfo{year}{2020}\natexlab{}.
\newblock \showarticletitle{In-memory database acceleration on FPGAs: a
  survey}.
\newblock \bibinfo{journal}{\emph{The VLDB Journal}} \bibinfo{volume}{29},
  \bibinfo{number}{1} (\bibinfo{year}{2020}), \bibinfo{pages}{33--59}.
\newblock


\bibitem[Fernandez et~al\mbox{.}(2014)]%
        {fernandez:ATC:2014:MakingStateExplicit}
\bibfield{author}{\bibinfo{person}{Raul~Castro Fernandez},
  \bibinfo{person}{Matteo Migliavacca}, \bibinfo{person}{Evangelia
  Kalyvianaki}, {and} \bibinfo{person}{Peter Pietzuch}.}
  \bibinfo{year}{2014}\natexlab{}.
\newblock \showarticletitle{Making State Explicit for Imperative Big Data
  Processing}. In \bibinfo{booktitle}{\emph{Proc of the USENIX Annual Technical
  Conf}} \emph{(\bibinfo{series}{ATC'14})}. \bibinfo{publisher}{USENIX Assoc.},
  \bibinfo{pages}{49–60}.
\newblock


\bibitem[G\'{e}vay et~al\mbox{.}(2021)]%
        {gevay:CSur:2021:iterations}
\bibfield{author}{\bibinfo{person}{G\'{a}bor~E. G\'{e}vay},
  \bibinfo{person}{Juan Soto}, {and} \bibinfo{person}{Volker Markl}.}
  \bibinfo{year}{2021}\natexlab{}.
\newblock \showarticletitle{Handling Iterations in Distributed Dataflow
  Systems}.
\newblock \bibinfo{journal}{\emph{ACM Comput. Surv.}} \bibinfo{volume}{54},
  \bibinfo{number}{9}, Article \bibinfo{articleno}{26} (\bibinfo{year}{2021}).
\newblock


\bibitem[Gonzalez et~al\mbox{.}(2012)]%
        {gonzalez:OSDI:2012:PowerGraph}
\bibfield{author}{\bibinfo{person}{Joseph~E. Gonzalez},
  \bibinfo{person}{Yucheng Low}, \bibinfo{person}{Haijie Gu},
  \bibinfo{person}{Danny Bickson}, {and} \bibinfo{person}{Carlos Guestrin}.}
  \bibinfo{year}{2012}\natexlab{}.
\newblock \showarticletitle{PowerGraph: Distributed Graph-Parallel Computation
  on Natural Graphs}. In \bibinfo{booktitle}{\emph{Proc of the Conf on
  Operating Systems Design and Implementation}}
  \emph{(\bibinfo{series}{OSDI'12})}. \bibinfo{publisher}{USENIX},
  \bibinfo{pages}{17–30}.
\newblock


\bibitem[Gonzalez et~al\mbox{.}(2014)]%
        {gonzalez:OSDI:2014:GraphX}
\bibfield{author}{\bibinfo{person}{Joseph~E. Gonzalez},
  \bibinfo{person}{Reynold~S. Xin}, \bibinfo{person}{Ankur Dave},
  \bibinfo{person}{Daniel Crankshaw}, \bibinfo{person}{Michael~J. Franklin},
  {and} \bibinfo{person}{Ion Stoica}.} \bibinfo{year}{2014}\natexlab{}.
\newblock \showarticletitle{GraphX: Graph Processing in a Distributed Dataflow
  Framework}. In \bibinfo{booktitle}{\emph{Symposium on Operating Systems
  Design and Implementation}} \emph{(\bibinfo{series}{OSDI '14})}.
  \bibinfo{publisher}{{USENIX} Assoc.}, \bibinfo{pages}{599--613}.
\newblock


\bibitem[Guerriero et~al\mbox{.}(2021)]%
        {guerriero:TOSEM:2021:StreamGen}
\bibfield{author}{\bibinfo{person}{Michele Guerriero},
  \bibinfo{person}{Damian~Andrew Tamburri}, {and} \bibinfo{person}{Elisabetta
  Di~Nitto}.} \bibinfo{year}{2021}\natexlab{}.
\newblock \showarticletitle{StreamGen: Model-Driven Development of Distributed
  Streaming Applications}.
\newblock \bibinfo{journal}{\emph{ACM Transactions on Software Engineering and
  Methodology}} \bibinfo{volume}{30}, \bibinfo{number}{1}, Article
  \bibinfo{articleno}{1} (\bibinfo{year}{2021}), \bibinfo{numpages}{30}~pages.
\newblock


\bibitem[Hirzel et~al\mbox{.}(2014)]%
        {hirzel:CSur:2014:Catalog}
\bibfield{author}{\bibinfo{person}{Martin Hirzel}, \bibinfo{person}{Robert
  Soul\'{e}}, \bibinfo{person}{Scott Schneider}, \bibinfo{person}{Bu\u{g}ra
  Gedik}, {and} \bibinfo{person}{Robert Grimm}.}
  \bibinfo{year}{2014}\natexlab{}.
\newblock \showarticletitle{A Catalog of Stream Processing Optimizations}.
\newblock \bibinfo{journal}{\emph{ACM Comput. Surv.}} \bibinfo{volume}{46},
  \bibinfo{number}{4}, Article \bibinfo{articleno}{46} (\bibinfo{year}{2014}).
\newblock


\bibitem[Hoozemans et~al\mbox{.}(2021)]%
        {hoozemans:CSM:2021:FPGA}
\bibfield{author}{\bibinfo{person}{Joost Hoozemans}, \bibinfo{person}{Johan
  Peltenburg}, \bibinfo{person}{Fabian Nonnemacher}, \bibinfo{person}{Akos
  Hadnagy}, \bibinfo{person}{Zaid Al-Ars}, {and} \bibinfo{person}{H.~Peter
  Hofstee}.} \bibinfo{year}{2021}\natexlab{}.
\newblock \showarticletitle{FPGA Acceleration for Big Data Analytics:
  Challenges and Opportunities}.
\newblock \bibinfo{journal}{\emph{Circuits and Systems Magazine}}
  \bibinfo{volume}{21}, \bibinfo{number}{2} (\bibinfo{year}{2021}),
  \bibinfo{pages}{30--47}.
\newblock


\bibitem[Huang et~al\mbox{.}(2019)]%
        {huang:ATC:2019:Tangram}
\bibfield{author}{\bibinfo{person}{Yuzhen Huang}, \bibinfo{person}{Xiao Yan},
  \bibinfo{person}{Guanxian Jiang}, \bibinfo{person}{Tatiana Jin},
  \bibinfo{person}{James Cheng}, \bibinfo{person}{An Xu},
  \bibinfo{person}{Zhanhan Liu}, {and} \bibinfo{person}{Shuo Tu}.}
  \bibinfo{year}{2019}\natexlab{}.
\newblock \showarticletitle{Tangram: Bridging Immutable and Mutable
  Abstractions for Distributed Data Analytics}. In
  \bibinfo{booktitle}{\emph{Proc of the USENIX Annual Technical Conf}}
  \emph{(\bibinfo{series}{ATC '19})}. \bibinfo{publisher}{USENIX},
  \bibinfo{pages}{191–205}.
\newblock


\bibitem[Isard et~al\mbox{.}(2007)]%
        {isard:EuroSys:2007:Dryad}
\bibfield{author}{\bibinfo{person}{Michael Isard}, \bibinfo{person}{Mihai
  Budiu}, \bibinfo{person}{Yuan Yu}, \bibinfo{person}{Andrew Birrell}, {and}
  \bibinfo{person}{Dennis Fetterly}.} \bibinfo{year}{2007}\natexlab{}.
\newblock \showarticletitle{Dryad: Distributed Data-Parallel Programs from
  Sequential Building Blocks}. In \bibinfo{booktitle}{\emph{Proc of the
  European Conf on Computer Systems}} \emph{(\bibinfo{series}{EuroSys '07})}.
  \bibinfo{publisher}{ACM}, \bibinfo{pages}{59–72}.
\newblock


\bibitem[Jensen et~al\mbox{.}(2017)]%
        {jensen:TKDE:2017:time_series}
\bibfield{author}{\bibinfo{person}{S\o{}ren~Kejser Jensen},
  \bibinfo{person}{Torben~Bach Pedersen}, {and} \bibinfo{person}{Christian
  Thomsen}.} \bibinfo{year}{2017}\natexlab{}.
\newblock \showarticletitle{Time Series Management Systems: A Survey}.
\newblock \bibinfo{journal}{\emph{IEEE Transactions on Knowledge and Data
  Engineering}} \bibinfo{volume}{29}, \bibinfo{number}{11}
  (\bibinfo{year}{2017}), \bibinfo{pages}{2581--2600}.
\newblock


\bibitem[Jiang et~al\mbox{.}(2020)]%
        {jiang:VLDB:2020:Hologres}
\bibfield{author}{\bibinfo{person}{Xiaowei Jiang}, \bibinfo{person}{Yuejun Hu},
  \bibinfo{person}{Yu Xiang}, \bibinfo{person}{Guangran Jiang},
  \bibinfo{person}{Xiaojun Jin}, \bibinfo{person}{Chen Xia},
  \bibinfo{person}{Weihua Jiang}, \bibinfo{person}{Jun Yu},
  \bibinfo{person}{Haitao Wang}, \bibinfo{person}{Yuan Jiang},
  \bibinfo{person}{Jihong Ma}, \bibinfo{person}{Li Su}, {and}
  \bibinfo{person}{Kai Zeng}.} \bibinfo{year}{2020}\natexlab{}.
\newblock \showarticletitle{Alibaba Hologres: A Cloud-Native Service for Hybrid
  Serving/Analytical Processing}.
\newblock \bibinfo{journal}{\emph{Proc of VLDB}} \bibinfo{volume}{13},
  \bibinfo{number}{12} (\bibinfo{year}{2020}), \bibinfo{pages}{3272–3284}.
\newblock


\bibitem[Kleppmann(2016)]%
        {kleppmann:2016:data-intensive}
\bibfield{author}{\bibinfo{person}{Martin Kleppmann}.}
  \bibinfo{year}{2016}\natexlab{}.
\newblock \bibinfo{booktitle}{\emph{Designing Data-Intensive Applications: The
  Big Ideas Behind Reliable, Scalable, and Maintainable Systems}}.
\newblock \bibinfo{publisher}{O'Reilly}.
\newblock


\bibitem[Kreps et~al\mbox{.}(2011)]%
        {kreps:NetDB:2011:kafka}
\bibfield{author}{\bibinfo{person}{Jay Kreps}, \bibinfo{person}{Neha Narkhede},
  \bibinfo{person}{Jun Rao}, {et~al\mbox{.}}} \bibinfo{year}{2011}\natexlab{}.
\newblock \showarticletitle{Kafka: A distributed messaging system for log
  processing}. In \bibinfo{booktitle}{\emph{Proc of the Intl Workshop on
  Networking meets Databases}} \emph{(\bibinfo{series}{NetDB})}.
  \bibinfo{publisher}{USENIX}, \bibinfo{pages}{1--7}.
\newblock


\bibitem[Kulkarni et~al\mbox{.}(2015)]%
        {kulkarni:SIGMOD:2015:Twitter_Heron}
\bibfield{author}{\bibinfo{person}{Sanjeev Kulkarni}, \bibinfo{person}{Nikunj
  Bhagat}, \bibinfo{person}{Maosong Fu}, \bibinfo{person}{Vikas Kedigehalli},
  \bibinfo{person}{Christopher Kellogg}, \bibinfo{person}{Sailesh Mittal},
  \bibinfo{person}{Jignesh~M. Patel}, \bibinfo{person}{Karthik Ramasamy}, {and}
  \bibinfo{person}{Siddarth Taneja}.} \bibinfo{year}{2015}\natexlab{}.
\newblock \showarticletitle{Twitter Heron: Stream Processing at Scale}. In
  \bibinfo{booktitle}{\emph{Proc of the Intl Conf on Management of Data}}
  \emph{(\bibinfo{series}{SIGMOD '15})}. \bibinfo{publisher}{ACM},
  \bibinfo{pages}{239--250}.
\newblock


\bibitem[Lakshman and Malik(2010)]%
        {lakshman:SIGOPS:2010:cassandra}
\bibfield{author}{\bibinfo{person}{Avinash Lakshman} {and}
  \bibinfo{person}{Prashant Malik}.} \bibinfo{year}{2010}\natexlab{}.
\newblock \showarticletitle{Cassandra: A Decentralized Structured Storage
  System}.
\newblock \bibinfo{journal}{\emph{SIGOPS Operating Systems Review}}
  \bibinfo{volume}{44}, \bibinfo{number}{2} (\bibinfo{year}{2010}),
  \bibinfo{pages}{35--40}.
\newblock


\bibitem[Lee et~al\mbox{.}(2021)]%
        {lee:VLDB:2021:RateupDB}
\bibfield{author}{\bibinfo{person}{Rubao Lee}, \bibinfo{person}{Minghong Zhou},
  \bibinfo{person}{Chi Li}, \bibinfo{person}{Shenggang Hu},
  \bibinfo{person}{Jianping Teng}, \bibinfo{person}{Dongyang Li}, {and}
  \bibinfo{person}{Xiaodong Zhang}.} \bibinfo{year}{2021}\natexlab{}.
\newblock \showarticletitle{The Art of Balance: A RateupDB Experience of
  Building a CPU/GPU Hybrid Database Product}.
\newblock \bibinfo{journal}{\emph{Proc of VLDB}} \bibinfo{volume}{14},
  \bibinfo{number}{12} (\bibinfo{year}{2021}), \bibinfo{pages}{2999–3013}.
\newblock
\showISSN{2150-8097}


\bibitem[Levandoski et~al\mbox{.}(2011)]%
        {levandoski:CIDR:2011:Deuteronomy}
\bibfield{author}{\bibinfo{person}{Justin Levandoski}, \bibinfo{person}{David
  Lomet}, \bibinfo{person}{}, {and} \bibinfo{person}{Kevin~Keliang Zhao}.}
  \bibinfo{year}{2011}\natexlab{}.
\newblock \showarticletitle{Deuteronomy: Transaction Support for Cloud Data}.
  In \bibinfo{booktitle}{\emph{Proc of the Conf on Innovative Data Systems
  Research}} \emph{(\bibinfo{series}{CIDR '11})}.
  \bibinfo{publisher}{www.crdrdb.org}.
\newblock


\bibitem[Lin(2017)]%
        {lin:InternetComputing:2017:LambdaKappa}
\bibfield{author}{\bibinfo{person}{Jimmy Lin}.}
  \bibinfo{year}{2017}\natexlab{}.
\newblock \showarticletitle{The Lambda and the Kappa}.
\newblock \bibinfo{journal}{\emph{IEEE Internet Computing}}
  \bibinfo{volume}{21}, \bibinfo{number}{5} (\bibinfo{year}{2017}),
  \bibinfo{pages}{60--66}.
\newblock


\bibitem[Liu et~al\mbox{.}(2021)]%
        {liu:VLDB:2021:Zen}
\bibfield{author}{\bibinfo{person}{Gang Liu}, \bibinfo{person}{Leying Chen},
  {and} \bibinfo{person}{Shimin Chen}.} \bibinfo{year}{2021}\natexlab{}.
\newblock \showarticletitle{Zen: A High-Throughput Log-Free OLTP Engine for
  Non-Volatile Main Memory}.
\newblock \bibinfo{journal}{\emph{Proc. VLDB}} \bibinfo{volume}{14},
  \bibinfo{number}{5} (\bibinfo{year}{2021}), \bibinfo{pages}{835–848}.
\newblock


\bibitem[Low et~al\mbox{.}(2012)]%
        {low:VLDB:2012:GraphLab}
\bibfield{author}{\bibinfo{person}{Yucheng Low}, \bibinfo{person}{Danny
  Bickson}, \bibinfo{person}{Joseph Gonzalez}, \bibinfo{person}{Carlos
  Guestrin}, \bibinfo{person}{Aapo Kyrola}, {and} \bibinfo{person}{Joseph~M.
  Hellerstein}.} \bibinfo{year}{2012}\natexlab{}.
\newblock \showarticletitle{Distributed GraphLab: A Framework for Machine
  Learning and Data Mining in the Cloud}.
\newblock \bibinfo{journal}{\emph{Proc of VLDB}} \bibinfo{volume}{5},
  \bibinfo{number}{8} (\bibinfo{year}{2012}), \bibinfo{pages}{716–727}.
\newblock


\bibitem[Lu and Holubov\'{a}(2019)]%
        {lu:CSur:2019:multi-model}
\bibfield{author}{\bibinfo{person}{Jiaheng Lu} {and} \bibinfo{person}{Irena
  Holubov\'{a}}.} \bibinfo{year}{2019}\natexlab{}.
\newblock \showarticletitle{Multi-Model Databases: A New Journey to Handle the
  Variety of Data}.
\newblock \bibinfo{journal}{\emph{ACM Comput. Surv.}} \bibinfo{volume}{52},
  \bibinfo{number}{3}, Article \bibinfo{articleno}{55} (\bibinfo{year}{2019}).
\newblock


\bibitem[Macedo and Oliveira(2011)]%
        {macedo:2011:Redis}
\bibfield{author}{\bibinfo{person}{Tiago Macedo} {and} \bibinfo{person}{Fred
  Oliveira}.} \bibinfo{year}{2011}\natexlab{}.
\newblock \bibinfo{booktitle}{\emph{Redis cookbook: Practical techniques for
  fast data manipulation}}.
\newblock \bibinfo{publisher}{O'Reilly}.
\newblock


\bibitem[Maiyya et~al\mbox{.}(2019)]%
        {maiyya:VLDB:19:unifying}
\bibfield{author}{\bibinfo{person}{Sujaya Maiyya}, \bibinfo{person}{Faisal
  Nawab}, \bibinfo{person}{Divy Agrawal}, {and} \bibinfo{person}{Amr~El
  Abbadi}.} \bibinfo{year}{2019}\natexlab{}.
\newblock \showarticletitle{Unifying Consensus and Atomic Commitment for
  Effective Cloud Data Management}.
\newblock \bibinfo{journal}{\emph{Proc of VLDB}} \bibinfo{volume}{12},
  \bibinfo{number}{5} (\bibinfo{year}{2019}), \bibinfo{pages}{611--623}.
\newblock


\bibitem[Malewicz et~al\mbox{.}(2010)]%
        {malewicz:SIGMOD:2010:Pregel}
\bibfield{author}{\bibinfo{person}{Grzegorz Malewicz},
  \bibinfo{person}{Matthew~H. Austern}, \bibinfo{person}{Aart~J.C Bik},
  \bibinfo{person}{James~C. Dehnert}, \bibinfo{person}{Ilan Horn},
  \bibinfo{person}{Naty Leiser}, {and} \bibinfo{person}{Grzegorz Czajkowski}.}
  \bibinfo{year}{2010}\natexlab{}.
\newblock \showarticletitle{Pregel: A System for Large-Scale Graph Processing}.
  In \bibinfo{booktitle}{\emph{Proc of the Intl Conf on Management of Data}}
  \emph{(\bibinfo{series}{SIGMOD '10})}. \bibinfo{publisher}{ACM},
  \bibinfo{pages}{135–146}.
\newblock


\bibitem[Malviya et~al\mbox{.}(2014)]%
        {malviya:ICDE:14:rethinking}
\bibfield{author}{\bibinfo{person}{Nirmesh Malviya}, \bibinfo{person}{Ariel
  Weisberg}, \bibinfo{person}{Samuel Madden}, {and} \bibinfo{person}{Michael
  Stonebraker}.} \bibinfo{year}{2014}\natexlab{}.
\newblock \showarticletitle{Rethinking main memory {OLTP} recovery}. In
  \bibinfo{booktitle}{\emph{Proceeding of the Intl Conf on Data Engineering}}
  \emph{(\bibinfo{series}{ICDE '14})}. \bibinfo{publisher}{IEEE},
  \bibinfo{pages}{604--615}.
\newblock


\bibitem[McCune et~al\mbox{.}(2015)]%
        {mcCune:CSur:2015:TLaV}
\bibfield{author}{\bibinfo{person}{Robert~Ryan McCune}, \bibinfo{person}{Tim
  Weninger}, {and} \bibinfo{person}{Greg Madey}.}
  \bibinfo{year}{2015}\natexlab{}.
\newblock \showarticletitle{Thinking Like a Vertex: A Survey of Vertex-Centric
  Frameworks for Large-Scale Distributed Graph Processing}.
\newblock \bibinfo{journal}{\emph{ACM Comput. Surv.}} \bibinfo{volume}{48},
  \bibinfo{number}{2}, Article \bibinfo{articleno}{25} (\bibinfo{year}{2015}).
\newblock


\bibitem[Meng et~al\mbox{.}(2016)]%
        {meng:JMLR:2016:MLlib}
\bibfield{author}{\bibinfo{person}{Xiangrui Meng}, \bibinfo{person}{Joseph
  Bradley}, \bibinfo{person}{Burak Yavuz}, \bibinfo{person}{Evan Sparks},
  \bibinfo{person}{Shivaram Venkataraman}, \bibinfo{person}{Davies Liu},
  \bibinfo{person}{Jeremy Freeman}, \bibinfo{person}{DB Tsai},
  \bibinfo{person}{Manish Amde}, \bibinfo{person}{Sean Owen},
  \bibinfo{person}{Doris Xin}, \bibinfo{person}{Reynold Xin},
  \bibinfo{person}{Michael~J. Franklin}, \bibinfo{person}{Reza Zadeh},
  \bibinfo{person}{Matei Zaharia}, {and} \bibinfo{person}{Ameet Talwalkar}.}
  \bibinfo{year}{2016}\natexlab{}.
\newblock \showarticletitle{MLlib: Machine Learning in Apache Spark}.
\newblock \bibinfo{journal}{\emph{Journal of Machine Learning Research}}
  \bibinfo{volume}{17}, \bibinfo{number}{1} (\bibinfo{year}{2016}),
  \bibinfo{pages}{1235–1241}.
\newblock


\bibitem[Mohan et~al\mbox{.}(1992)]%
        {mohan:TODS:92:aries}
\bibfield{author}{\bibinfo{person}{C. Mohan}, \bibinfo{person}{Don Haderle},
  \bibinfo{person}{Bruce~G. Lindsay}, \bibinfo{person}{Hamid Pirahesh}, {and}
  \bibinfo{person}{Peter~M. Schwarz}.} \bibinfo{year}{1992}\natexlab{}.
\newblock \showarticletitle{{ARIES:} {A} Transaction Recovery Method Supporting
  Fine-Granularity Locking and Partial Rollbacks Using Write-Ahead Logging}.
\newblock \bibinfo{journal}{\emph{Trans on Database Systems}}
  \bibinfo{volume}{17}, \bibinfo{number}{1} (\bibinfo{year}{1992}),
  \bibinfo{pages}{94--162}.
\newblock


\bibitem[Mozafari et~al\mbox{.}(2017)]%
        {barzan:CIDR:17:SnappyData}
\bibfield{author}{\bibinfo{person}{Barzan Mozafari}, \bibinfo{person}{Jags
  Ramnarayan}, \bibinfo{person}{Sudhir Menon}, \bibinfo{person}{Yogesh
  Mahajan}, \bibinfo{person}{Soubhik Chakraborty}, \bibinfo{person}{Hemant
  Bhanawat}, {and} \bibinfo{person}{Kishor Bachhav}.}
  \bibinfo{year}{2017}\natexlab{}.
\newblock \showarticletitle{SnappyData: {A} Unified Cluster for Streaming,
  Transactions and Interactice Analytics}. In \bibinfo{booktitle}{\emph{Proc of
  the Conf on Innovative Data Systems Research}} \emph{(\bibinfo{series}{CIDR
  '17})}. \bibinfo{publisher}{www.cidrdb.org}.
\newblock


\bibitem[Murray et~al\mbox{.}(2013)]%
        {murray:SOSP:2013:Naiad}
\bibfield{author}{\bibinfo{person}{Derek~G. Murray}, \bibinfo{person}{Frank
  McSherry}, \bibinfo{person}{Rebecca Isaacs}, \bibinfo{person}{Michael Isard},
  \bibinfo{person}{Paul Barham}, {and} \bibinfo{person}{Mart\'{\i}n Abadi}.}
  \bibinfo{year}{2013}\natexlab{}.
\newblock \showarticletitle{Naiad: A Timely Dataflow System}. In
  \bibinfo{booktitle}{\emph{Proc of the Symposium on Operating Systems
  Principles}} \emph{(\bibinfo{series}{SOSP '13})}. \bibinfo{publisher}{ACM},
  \bibinfo{pages}{439–455}.
\newblock


\bibitem[Murray et~al\mbox{.}(2011)]%
        {murray:NSDI:2011:CIEL}
\bibfield{author}{\bibinfo{person}{Derek~G. Murray}, \bibinfo{person}{Malte
  Schwarzkopf}, \bibinfo{person}{Christopher Smowton}, \bibinfo{person}{Steven
  Smith}, \bibinfo{person}{Anil Madhavapeddy}, {and} \bibinfo{person}{Steven
  Hand}.} \bibinfo{year}{2011}\natexlab{}.
\newblock \showarticletitle{CIEL: A Universal Execution Engine for Distributed
  Data-Flow Computing}. In \bibinfo{booktitle}{\emph{Proc of the Conf on
  Networked Systems Design and Implementation}}
  \emph{(\bibinfo{series}{NSDI'11})}. \bibinfo{publisher}{USENIX Assoc.},
  \bibinfo{pages}{113–126}.
\newblock


\bibitem[Nishtala et~al\mbox{.}(2013)]%
        {nishtala:NSDI:2013:Memcached}
\bibfield{author}{\bibinfo{person}{Rajesh Nishtala}, \bibinfo{person}{Hans
  Fugal}, \bibinfo{person}{Steven Grimm}, \bibinfo{person}{Marc Kwiatkowski},
  \bibinfo{person}{Herman Lee}, \bibinfo{person}{Harry~C. Li},
  \bibinfo{person}{Ryan McElroy}, \bibinfo{person}{Mike Paleczny},
  \bibinfo{person}{Daniel Peek}, \bibinfo{person}{Paul Saab},
  \bibinfo{person}{David Stafford}, \bibinfo{person}{Tony Tung}, {and}
  \bibinfo{person}{Venkateshwaran Venkataramani}.}
  \bibinfo{year}{2013}\natexlab{}.
\newblock \showarticletitle{Scaling Memcache at Facebook}. In
  \bibinfo{booktitle}{\emph{Proc of the Conf on Networked Systems Design and
  Implementation}} \emph{(\bibinfo{series}{NSDI '13})}.
  \bibinfo{publisher}{USENIX Assoc.}, \bibinfo{pages}{385–398}.
\newblock


\bibitem[Noghabi et~al\mbox{.}(2017)]%
        {noghabi:VLDB:2017:Samza}
\bibfield{author}{\bibinfo{person}{Shadi~A. Noghabi}, \bibinfo{person}{Kartik
  Paramasivam}, \bibinfo{person}{Yi Pan}, \bibinfo{person}{Navina Ramesh},
  \bibinfo{person}{Jon Bringhurst}, \bibinfo{person}{Indranil Gupta}, {and}
  \bibinfo{person}{Roy~H. Campbell}.} \bibinfo{year}{2017}\natexlab{}.
\newblock \showarticletitle{Samza: Stateful Scalable Stream Processing at
  LinkedIn}.
\newblock \bibinfo{journal}{\emph{Proc of VLDB}} \bibinfo{volume}{10},
  \bibinfo{number}{12} (\bibinfo{year}{2017}), \bibinfo{pages}{1634–1645}.
\newblock


\bibitem[Pelkonen et~al\mbox{.}(2015)]%
        {pelkonen:VLDB:2015:Gorilla}
\bibfield{author}{\bibinfo{person}{Tuomas Pelkonen}, \bibinfo{person}{Scott
  Franklin}, \bibinfo{person}{Justin Teller}, \bibinfo{person}{Paul Cavallaro},
  \bibinfo{person}{Qi Huang}, \bibinfo{person}{Justin Meza}, {and}
  \bibinfo{person}{Kaushik Veeraraghavan}.} \bibinfo{year}{2015}\natexlab{}.
\newblock \showarticletitle{Gorilla: A Fast, Scalable, in-Memory Time Series
  Database}.
\newblock \bibinfo{journal}{\emph{Proc of VLDB}} \bibinfo{volume}{8},
  \bibinfo{number}{12} (\bibinfo{year}{2015}), \bibinfo{pages}{1816–1827}.
\newblock


\bibitem[Peng and Dabek(2010)]%
        {peng:OSDI:2010:LargeScale}
\bibfield{author}{\bibinfo{person}{Daniel Peng} {and} \bibinfo{person}{Frank
  Dabek}.} \bibinfo{year}{2010}\natexlab{}.
\newblock \showarticletitle{Large-Scale Incremental Processing Using
  Distributed Transactions and Notifications}. In
  \bibinfo{booktitle}{\emph{Proc of the Conf on Operating Systems Design and
  Implementation}} \emph{(\bibinfo{series}{OSDI'10})}.
  \bibinfo{publisher}{USENIX Assoc.}, \bibinfo{pages}{251–264}.
\newblock


\bibitem[R\"{o}ger and Mayer(2019)]%
        {roger:CSur:2019:ParallelizationElasticity}
\bibfield{author}{\bibinfo{person}{Henriette R\"{o}ger} {and}
  \bibinfo{person}{Ruben Mayer}.} \bibinfo{year}{2019}\natexlab{}.
\newblock \showarticletitle{A Comprehensive Survey on Parallelization and
  Elasticity in Stream Processing}.
\newblock \bibinfo{journal}{\emph{ACM Comput. Surv.}} \bibinfo{volume}{52},
  \bibinfo{number}{2} (\bibinfo{year}{2019}).
\newblock


\bibitem[Sakr et~al\mbox{.}(2021)]%
        {sakr:CACM:2021:Big_Graphs}
\bibfield{author}{\bibinfo{person}{Sherif Sakr}, \bibinfo{person}{Angela
  Bonifati}, \bibinfo{person}{Hannes Voigt}, \bibinfo{person}{Alexandru Iosup},
  \bibinfo{person}{Khaled Ammar}, \bibinfo{person}{Renzo Angles},
  \bibinfo{person}{Walid Aref}, \bibinfo{person}{Marcelo Arenas},
  \bibinfo{person}{Maciej Besta}, \bibinfo{person}{Peter~A. Boncz},
  \bibinfo{person}{Khuzaima Daudjee}, \bibinfo{person}{Emanuele~Della Valle},
  \bibinfo{person}{Stefania Dumbrava}, \bibinfo{person}{Olaf Hartig},
  \bibinfo{person}{Bernhard Haslhofer}, \bibinfo{person}{Tim Hegeman},
  \bibinfo{person}{Jan Hidders}, \bibinfo{person}{Katja Hose},
  \bibinfo{person}{Adriana Iamnitchi}, \bibinfo{person}{Vasiliki Kalavri},
  \bibinfo{person}{Hugo Kapp}, \bibinfo{person}{Wim Martens},
  \bibinfo{person}{M.~Tamer \"{O}zsu}, \bibinfo{person}{Eric Peukert},
  \bibinfo{person}{Stefan Plantikow}, \bibinfo{person}{Mohamed Ragab},
  \bibinfo{person}{Matei~R. Ripeanu}, \bibinfo{person}{Semih Salihoglu},
  \bibinfo{person}{Christian Schulz}, \bibinfo{person}{Petra Selmer},
  \bibinfo{person}{Juan~F. Sequeda}, \bibinfo{person}{Joshua Shinavier},
  \bibinfo{person}{G\'{a}bor Sz\'{a}rnyas}, \bibinfo{person}{Riccardo
  Tommasini}, \bibinfo{person}{Antonino Tumeo}, \bibinfo{person}{Alexandru
  Uta}, \bibinfo{person}{Ana~Lucia Varbanescu}, \bibinfo{person}{Hsiang-Yun
  Wu}, \bibinfo{person}{Nikolay Yakovets}, \bibinfo{person}{Da Yan}, {and}
  \bibinfo{person}{Eiko Yoneki}.} \bibinfo{year}{2021}\natexlab{}.
\newblock \showarticletitle{The Future is Big Graphs: A Community View on Graph
  Processing Systems}.
\newblock \bibinfo{journal}{\emph{Commun. ACM}} \bibinfo{volume}{64},
  \bibinfo{number}{9} (\bibinfo{year}{2021}), \bibinfo{pages}{62–71}.
\newblock


\bibitem[Sax et~al\mbox{.}(2018)]%
        {sax:BIRTE:2018:StreamsTables}
\bibfield{author}{\bibinfo{person}{Matthias~J. Sax}, \bibinfo{person}{Guozhang
  Wang}, \bibinfo{person}{Matthias Weidlich}, {and}
  \bibinfo{person}{Johann-Christoph Freytag}.} \bibinfo{year}{2018}\natexlab{}.
\newblock \showarticletitle{Streams and Tables: Two Sides of the Same Coin}. In
  \bibinfo{booktitle}{\emph{Proc of the Intl Workshop on Real-Time Business
  Intelligence and Analytics}} \emph{(\bibinfo{series}{BIRTE '18})}.
  \bibinfo{publisher}{ACM}, Article \bibinfo{articleno}{1}.
\newblock


\bibitem[Shah and Salles(2018)]%
        {shah:SIGMOD:2018:reactors}
\bibfield{author}{\bibinfo{person}{Vivek Shah} {and} \bibinfo{person}{Marcos
  Antonio~Vaz Salles}.} \bibinfo{year}{2018}\natexlab{}.
\newblock \showarticletitle{Reactors: {A} Case for Predictable, Virtualized
  Actor Database Systems}. In \bibinfo{booktitle}{\emph{Proc of the Intl Conf
  on Management of Data}} \emph{(\bibinfo{series}{SIGMOD '18})}.
  \bibinfo{publisher}{{ACM}}, \bibinfo{pages}{259--274}.
\newblock


\bibitem[Shao et~al\mbox{.}(2013)]%
        {shao:SIGMOD:2013:Trinity}
\bibfield{author}{\bibinfo{person}{Bin Shao}, \bibinfo{person}{Haixun Wang},
  {and} \bibinfo{person}{Yatao Li}.} \bibinfo{year}{2013}\natexlab{}.
\newblock \showarticletitle{Trinity: A Distributed Graph Engine on a Memory
  Cloud}. In \bibinfo{booktitle}{\emph{Proc of the Intl Conf on Management of
  Data}} \emph{(\bibinfo{series}{SIGMOD '13})}. \bibinfo{publisher}{ACM},
  \bibinfo{pages}{505–516}.
\newblock


\bibitem[Shapiro et~al\mbox{.}(2011)]%
        {shapiro:SSS:2011:CRDT}
\bibfield{author}{\bibinfo{person}{Marc Shapiro}, \bibinfo{person}{Nuno
  Pregui\c{c}a}, \bibinfo{person}{Carlos Baquero}, {and} \bibinfo{person}{Marek
  Zawirski}.} \bibinfo{year}{2011}\natexlab{}.
\newblock \showarticletitle{Conflict-Free Replicated Data Types}. In
  \bibinfo{booktitle}{\emph{Proc of the Intl Conf on Stabilization, Safety, and
  Security of Distributed Systems}} \emph{(\bibinfo{series}{SSS'11})}.
  \bibinfo{publisher}{Springer-Verlag}, \bibinfo{pages}{386–400}.
\newblock


\bibitem[Shi et~al\mbox{.}(2016)]%
        {shi:IoT:2016:Edge}
\bibfield{author}{\bibinfo{person}{Weisong Shi}, \bibinfo{person}{Jie Cao},
  \bibinfo{person}{Quan Zhang}, \bibinfo{person}{Youhuizi Li}, {and}
  \bibinfo{person}{Lanyu Xu}.} \bibinfo{year}{2016}\natexlab{}.
\newblock \showarticletitle{Edge Computing: Vision and Challenges}.
\newblock \bibinfo{journal}{\emph{Internet of Things Journal}}
  \bibinfo{volume}{3}, \bibinfo{number}{5} (\bibinfo{year}{2016}),
  \bibinfo{pages}{637--646}.
\newblock


\bibitem[Shute et~al\mbox{.}(2013)]%
        {shute:VLDB:2013:F1}
\bibfield{author}{\bibinfo{person}{Jeff Shute}, \bibinfo{person}{Radek
  Vingralek}, \bibinfo{person}{Bart Samwel}, \bibinfo{person}{Ben Handy},
  \bibinfo{person}{Chad Whipkey}, \bibinfo{person}{Eric Rollins},
  \bibinfo{person}{Mircea Oancea}, \bibinfo{person}{Kyle Littlefield},
  \bibinfo{person}{David Menestrina}, \bibinfo{person}{Stephan Ellner},
  \bibinfo{person}{John Cieslewicz}, \bibinfo{person}{Ian Rae},
  \bibinfo{person}{Traian Stancescu}, {and} \bibinfo{person}{Himani Apte}.}
  \bibinfo{year}{2013}\natexlab{}.
\newblock \showarticletitle{F1: A Distributed SQL Database That Scales}.
\newblock \bibinfo{journal}{\emph{Proc of VLDB}} \bibinfo{volume}{6},
  \bibinfo{number}{11} (\bibinfo{year}{2013}), \bibinfo{pages}{1068–1079}.
\newblock
\showISSN{2150-8097}


\bibitem[Srinivasan et~al\mbox{.}(2016)]%
        {srinivasan:VLDB:2016:Aerospike}
\bibfield{author}{\bibinfo{person}{V. Srinivasan}, \bibinfo{person}{Brian
  Bulkowski}, \bibinfo{person}{Wei-Ling Chu}, \bibinfo{person}{Sunil
  Sayyaparaju}, \bibinfo{person}{Andrew Gooding}, \bibinfo{person}{Rajkumar
  Iyer}, \bibinfo{person}{Ashish Shinde}, {and} \bibinfo{person}{Thomas
  Lopatic}.} \bibinfo{year}{2016}\natexlab{}.
\newblock \showarticletitle{Aerospike: Architecture of a Real-Time Operational
  DBMS}.
\newblock \bibinfo{journal}{\emph{Proc of VLDB}} \bibinfo{volume}{9},
  \bibinfo{number}{13} (\bibinfo{year}{2016}), \bibinfo{pages}{1389–1400}.
\newblock


\bibitem[Stonebraker(2010)]%
        {stonebraker:CACM:2010:SQLvsNoSQL}
\bibfield{author}{\bibinfo{person}{Michael Stonebraker}.}
  \bibinfo{year}{2010}\natexlab{}.
\newblock \showarticletitle{SQL Databases V. NoSQL Databases}.
\newblock \bibinfo{journal}{\emph{Commun. ACM}} \bibinfo{volume}{53},
  \bibinfo{number}{4} (\bibinfo{year}{2010}), \bibinfo{pages}{10--11}.
\newblock


\bibitem[Stonebraker(2012)]%
        {stonebraker:CACM:2012:NewSQL}
\bibfield{author}{\bibinfo{person}{Michael Stonebraker}.}
  \bibinfo{year}{2012}\natexlab{}.
\newblock \showarticletitle{{New Opportunities for New SQL}}.
\newblock \bibinfo{journal}{\emph{Commun. ACM}} \bibinfo{volume}{55},
  \bibinfo{number}{11} (\bibinfo{year}{2012}), \bibinfo{pages}{10--11}.
\newblock


\bibitem[Stonebraker and Cetintemel(2005)]%
        {stonebraker:ICDE:2005:one_size_fits_all}
\bibfield{author}{\bibinfo{person}{Michael Stonebraker} {and}
  \bibinfo{person}{Ugur Cetintemel}.} \bibinfo{year}{2005}\natexlab{}.
\newblock \showarticletitle{"One Size Fits All": An Idea Whose Time Has Come
  and Gone}. In \bibinfo{booktitle}{\emph{Proc of the Intl Conf on Data
  Engineering}} \emph{(\bibinfo{series}{ICDE '05})}. \bibinfo{publisher}{IEEE},
  \bibinfo{pages}{2--11}.
\newblock


\bibitem[Stonebraker et~al\mbox{.}(2007)]%
        {stonebraker:VLDB:2007:hstore}
\bibfield{author}{\bibinfo{person}{Michael Stonebraker},
  \bibinfo{person}{Samuel Madden}, \bibinfo{person}{Daniel~J Abadi},
  \bibinfo{person}{Stavros Harizopoulos}, \bibinfo{person}{Nabil Hachem}, {and}
  \bibinfo{person}{Pat Helland}.} \bibinfo{year}{2007}\natexlab{}.
\newblock \showarticletitle{The end of an architectural era (It's time for a
  complete rewrite)}. In \bibinfo{booktitle}{\emph{Proc of VLDB}}
  \emph{(\bibinfo{series}{VLDB '07})}. \bibinfo{publisher}{VLDB Endow.},
  \bibinfo{pages}{1150--1160}.
\newblock


\bibitem[Stonebraker and Weisberg(2013)]%
        {stonebraker:IEEEB:2013:voltdb}
\bibfield{author}{\bibinfo{person}{Michael Stonebraker} {and}
  \bibinfo{person}{Ariel Weisberg}.} \bibinfo{year}{2013}\natexlab{}.
\newblock \showarticletitle{The VoltDB Main Memory {DBMS}}.
\newblock \bibinfo{journal}{\emph{{IEEE} Data Engineering Bulletin}}
  \bibinfo{volume}{36}, \bibinfo{number}{2} (\bibinfo{year}{2013}),
  \bibinfo{pages}{21--27}.
\newblock


\bibitem[Taft et~al\mbox{.}(2020)]%
        {taft:SIGMOD:2020:CockroachDB}
\bibfield{author}{\bibinfo{person}{Rebecca Taft}, \bibinfo{person}{Irfan
  Sharif}, \bibinfo{person}{Andrei Matei}, \bibinfo{person}{Nathan
  VanBenschoten}, \bibinfo{person}{Jordan Lewis}, \bibinfo{person}{Tobias
  Grieger}, \bibinfo{person}{Kai Niemi}, \bibinfo{person}{Andy Woods},
  \bibinfo{person}{Anne Birzin}, \bibinfo{person}{Raphael Poss},
  \bibinfo{person}{Paul Bardea}, \bibinfo{person}{Amruta Ranade},
  \bibinfo{person}{Ben Darnell}, \bibinfo{person}{Bram Gruneir},
  \bibinfo{person}{Justin Jaffray}, \bibinfo{person}{Lucy Zhang}, {and}
  \bibinfo{person}{Peter Mattis}.} \bibinfo{year}{2020}\natexlab{}.
\newblock \showarticletitle{CockroachDB: The Resilient Geo-Distributed SQL
  Database}. In \bibinfo{booktitle}{\emph{Proc of the Intl Conf on Management
  of Data}} \emph{(\bibinfo{series}{SIGMOD '20})}. \bibinfo{publisher}{ACM},
  \bibinfo{pages}{1493–1509}.
\newblock


\bibitem[Teixeira et~al\mbox{.}(2015)]%
        {teixeira:SOSP:2015:Arabesque}
\bibfield{author}{\bibinfo{person}{Carlos H.~C. Teixeira},
  \bibinfo{person}{Alexandre~J. Fonseca}, \bibinfo{person}{Marco Serafini},
  \bibinfo{person}{Georgos Siganos}, \bibinfo{person}{Mohammed~J. Zaki}, {and}
  \bibinfo{person}{Ashraf Aboulnaga}.} \bibinfo{year}{2015}\natexlab{}.
\newblock \showarticletitle{Arabesque: A System for Distributed Graph Mining}.
  In \bibinfo{booktitle}{\emph{Proc of the Symposium on Operating Systems
  Principles}} \emph{(\bibinfo{series}{SOSP '15})}. \bibinfo{publisher}{ACM},
  \bibinfo{pages}{425–440}.
\newblock


\bibitem[Thomson et~al\mbox{.}(2012)]%
        {thomson:SIGMOD:2012:Calvin}
\bibfield{author}{\bibinfo{person}{Alexander Thomson},
  \bibinfo{person}{Thaddeus Diamond}, \bibinfo{person}{Shu-Chun Weng},
  \bibinfo{person}{Kun Ren}, \bibinfo{person}{Philip Shao}, {and}
  \bibinfo{person}{Daniel~J. Abadi}.} \bibinfo{year}{2012}\natexlab{}.
\newblock \showarticletitle{Calvin: Fast Distributed Transactions for
  Partitioned Database Systems}. In \bibinfo{booktitle}{\emph{Proc of the Intl
  Conf on Management of Data}} \emph{(\bibinfo{series}{SIGMOD '12})}.
  \bibinfo{publisher}{ACM}, \bibinfo{pages}{1--12}.
\newblock


\bibitem[To et~al\mbox{.}(2018)]%
        {to:VLDBJ:2018:survey_state}
\bibfield{author}{\bibinfo{person}{Quoc-Cuong To}, \bibinfo{person}{Juan Soto},
  {and} \bibinfo{person}{Volker Markl}.} \bibinfo{year}{2018}\natexlab{}.
\newblock \showarticletitle{A Survey of State Management in Big Data Processing
  Systems}.
\newblock \bibinfo{journal}{\emph{The VLDB Journal}} \bibinfo{volume}{27},
  \bibinfo{number}{6} (\bibinfo{year}{2018}), \bibinfo{pages}{847–872}.
\newblock


\bibitem[Toshniwal et~al\mbox{.}(2014)]%
        {toshniwal:SIGMOD:2014:Storm}
\bibfield{author}{\bibinfo{person}{Ankit Toshniwal}, \bibinfo{person}{Siddarth
  Taneja}, \bibinfo{person}{Amit Shukla}, \bibinfo{person}{Karthik Ramasamy},
  \bibinfo{person}{Jignesh~M. Patel}, \bibinfo{person}{Sanjeev Kulkarni},
  \bibinfo{person}{Jason Jackson}, \bibinfo{person}{Krishna Gade},
  \bibinfo{person}{Maosong Fu}, \bibinfo{person}{Jake Donham},
  \bibinfo{person}{Nikunj Bhagat}, \bibinfo{person}{Sailesh Mittal}, {and}
  \bibinfo{person}{Dmitriy Ryaboy}.} \bibinfo{year}{2014}\natexlab{}.
\newblock \showarticletitle{Storm@Twitter}. In \bibinfo{booktitle}{\emph{Proc
  of the Intl Conf on Management of Data}} \emph{(\bibinfo{series}{SIGMOD
  '14})}. \bibinfo{publisher}{ACM}, \bibinfo{pages}{147--156}.
\newblock


\bibitem[Vavilapalli et~al\mbox{.}(2013)]%
        {vavilapalli:SOCC:2013:YARN}
\bibfield{author}{\bibinfo{person}{Vinod~Kumar Vavilapalli},
  \bibinfo{person}{Arun~C. Murthy}, \bibinfo{person}{Chris Douglas},
  \bibinfo{person}{Sharad Agarwal}, \bibinfo{person}{Mahadev Konar},
  \bibinfo{person}{Robert Evans}, \bibinfo{person}{Thomas Graves},
  \bibinfo{person}{Jason Lowe}, \bibinfo{person}{Hitesh Shah},
  \bibinfo{person}{Siddharth Seth}, \bibinfo{person}{Bikas Saha},
  \bibinfo{person}{Carlo Curino}, \bibinfo{person}{Owen O'Malley},
  \bibinfo{person}{Sanjay Radia}, \bibinfo{person}{Benjamin Reed}, {and}
  \bibinfo{person}{Eric Baldeschwieler}.} \bibinfo{year}{2013}\natexlab{}.
\newblock \showarticletitle{Apache Hadoop YARN: Yet Another Resource
  Negotiator}. In \bibinfo{booktitle}{\emph{Proc of the Annual Symposium on
  Cloud Computing}} \emph{(\bibinfo{series}{SOCC '13})}.
  \bibinfo{publisher}{ACM}, Article \bibinfo{articleno}{5}.
\newblock


\bibitem[Verbitski et~al\mbox{.}(2017)]%
        {verbitski:SIGMOD:2017:Aurora}
\bibfield{author}{\bibinfo{person}{Alexandre Verbitski},
  \bibinfo{person}{Anurag Gupta}, \bibinfo{person}{Debanjan Saha},
  \bibinfo{person}{Murali Brahmadesam}, \bibinfo{person}{Kamal Gupta},
  \bibinfo{person}{Raman Mittal}, \bibinfo{person}{Sailesh Krishnamurthy},
  \bibinfo{person}{Sandor Maurice}, \bibinfo{person}{Tengiz Kharatishvili},
  {and} \bibinfo{person}{Xiaofeng Bao}.} \bibinfo{year}{2017}\natexlab{}.
\newblock \showarticletitle{Amazon Aurora: Design Considerations for High
  Throughput Cloud-Native Relational Databases}. In
  \bibinfo{booktitle}{\emph{Proc of the Intl Conf on Management of Data}}
  \emph{(\bibinfo{series}{SIGMOD '17})}. \bibinfo{publisher}{ACM},
  \bibinfo{pages}{1041–1052}.
\newblock


\bibitem[Viotti and Vukoli\'{c}(2016)]%
        {viotti:CSUR:2016:Consistency}
\bibfield{author}{\bibinfo{person}{Paolo Viotti} {and} \bibinfo{person}{Marko
  Vukoli\'{c}}.} \bibinfo{year}{2016}\natexlab{}.
\newblock \showarticletitle{Consistency in Non-Transactional Distributed
  Storage Systems}.
\newblock \bibinfo{journal}{\emph{ACM Comput. Surv.}} \bibinfo{volume}{49},
  \bibinfo{number}{1}, Article \bibinfo{articleno}{19} (\bibinfo{year}{2016}).
\newblock


\bibitem[Visheratin et~al\mbox{.}(2020)]%
        {visheratin:ATC:2020:Peregreen}
\bibfield{author}{\bibinfo{person}{Alexander Visheratin},
  \bibinfo{person}{Alexey Struckov}, \bibinfo{person}{Semen Yufa},
  \bibinfo{person}{Alexey Muratov}, \bibinfo{person}{Denis Nasonov},
  \bibinfo{person}{Nikolay Butakov}, \bibinfo{person}{Yury Kuznetsov}, {and}
  \bibinfo{person}{Michael May}.} \bibinfo{year}{2020}\natexlab{}.
\newblock \showarticletitle{Peregreen {\textendash} modular database for
  efficient storage of historical time series in cloud environments}. In
  \bibinfo{booktitle}{\emph{Proc of the USENIX Annual Technical Conf}}
  \emph{(\bibinfo{series}{ATC '20})}. \bibinfo{publisher}{USENIX},
  \bibinfo{pages}{589--601}.
\newblock


\bibitem[Zaharia et~al\mbox{.}(2010)]%
        {zaharia:EuroSys:2010:Delay}
\bibfield{author}{\bibinfo{person}{Matei Zaharia}, \bibinfo{person}{Dhruba
  Borthakur}, \bibinfo{person}{Joydeep Sen~Sarma}, \bibinfo{person}{Khaled
  Elmeleegy}, \bibinfo{person}{Scott Shenker}, {and} \bibinfo{person}{Ion
  Stoica}.} \bibinfo{year}{2010}\natexlab{}.
\newblock \showarticletitle{Delay Scheduling: A Simple Technique for Achieving
  Locality and Fairness in Cluster Scheduling}. In
  \bibinfo{booktitle}{\emph{Proc of the European Conf on Computer Systems}}
  \emph{(\bibinfo{series}{EuroSys '10})}. \bibinfo{publisher}{ACM},
  \bibinfo{pages}{265–278}.
\newblock


\bibitem[Zaharia et~al\mbox{.}(2013)]%
        {zaharia:SOSP:2013:Discretized_Streams}
\bibfield{author}{\bibinfo{person}{Matei Zaharia}, \bibinfo{person}{Tathagata
  Das}, \bibinfo{person}{Haoyuan Li}, \bibinfo{person}{Timothy Hunter},
  \bibinfo{person}{Scott Shenker}, {and} \bibinfo{person}{Ion Stoica}.}
  \bibinfo{year}{2013}\natexlab{}.
\newblock \showarticletitle{Discretized Streams: Fault-tolerant Streaming
  Computation at Scale}. In \bibinfo{booktitle}{\emph{Proc of the Symposium on
  Operating Systems Principles}} \emph{(\bibinfo{series}{SOSP '13})}.
  \bibinfo{publisher}{ACM}, \bibinfo{pages}{423--438}.
\newblock


\bibitem[Zaharia et~al\mbox{.}(2016)]%
        {zaharia:CACM:2016:spark}
\bibfield{author}{\bibinfo{person}{Matei Zaharia}, \bibinfo{person}{Reynold~S.
  Xin}, \bibinfo{person}{Patrick Wendell}, \bibinfo{person}{Tathagata Das},
  \bibinfo{person}{Michael Armbrust}, \bibinfo{person}{Ankur Dave},
  \bibinfo{person}{Xiangrui Meng}, \bibinfo{person}{Josh Rosen},
  \bibinfo{person}{Shivaram Venkataraman}, \bibinfo{person}{Michael~J.
  Franklin}, \bibinfo{person}{Ali Ghodsi}, \bibinfo{person}{Joseph Gonzalez},
  \bibinfo{person}{Scott Shenker}, {and} \bibinfo{person}{Ion Stoica}.}
  \bibinfo{year}{2016}\natexlab{}.
\newblock \showarticletitle{Apache Spark: A Unified Engine for Big Data
  Processing}.
\newblock \bibinfo{journal}{\emph{Commun. ACM}} \bibinfo{volume}{59},
  \bibinfo{number}{11} (\bibinfo{year}{2016}), \bibinfo{pages}{56--65}.
\newblock


\bibitem[Zhou et~al\mbox{.}(2021)]%
        {zhou:SIGMOD:2021:FoundationDB}
\bibfield{author}{\bibinfo{person}{Jingyu Zhou}, \bibinfo{person}{Meng Xu},
  \bibinfo{person}{Alexander Shraer}, \bibinfo{person}{Bala Namasivayam},
  \bibinfo{person}{Alex Miller}, \bibinfo{person}{Evan Tschannen},
  \bibinfo{person}{Steve Atherton}, \bibinfo{person}{Andrew~J Beamon},
  \bibinfo{person}{Rusty Sears}, \bibinfo{person}{John Leach}, {et~al\mbox{.}}}
  \bibinfo{year}{2021}\natexlab{}.
\newblock \showarticletitle{FoundationDB: A Distributed Unbundled Transactional
  Key Value Store}. In \bibinfo{booktitle}{\emph{Proc of the Intl Conf on
  Management of Data}} \emph{(\bibinfo{series}{SIGMOD '21})}.
  \bibinfo{publisher}{ACM}, \bibinfo{pages}{135–146}.
\newblock


\bibitem[Zhu et~al\mbox{.}(2019)]%
        {zhu:ToS:2019:Solar}
\bibfield{author}{\bibinfo{person}{Tao Zhu}, \bibinfo{person}{Zhuoyue Zhao},
  \bibinfo{person}{Feifei Li}, \bibinfo{person}{Weining Qian},
  \bibinfo{person}{Aoying Zhou}, \bibinfo{person}{Dong Xie},
  \bibinfo{person}{Ryan Stutsman}, \bibinfo{person}{Haining Li}, {and}
  \bibinfo{person}{Huiqi Hu}.} \bibinfo{year}{2019}\natexlab{}.
\newblock \showarticletitle{Solar: Toward a Shared-Everything Database on
  Distributed Log-Structured Storage}.
\newblock \bibinfo{journal}{\emph{Trans on Storage}} \bibinfo{volume}{15},
  \bibinfo{number}{2} (\bibinfo{year}{2019}).
\newblock


\bibitem[Ziegler et~al\mbox{.}(2022)]%
        {ziegler:SIGMOD:2022:ScaleStore}
\bibfield{author}{\bibinfo{person}{Tobias Ziegler}, \bibinfo{person}{Carsten
  Binnig}, {and} \bibinfo{person}{Viktor Leis}.}
  \bibinfo{year}{2022}\natexlab{}.
\newblock \showarticletitle{ScaleStore: A Fast and Cost-Efficient Storage
  Engine Using DRAM, NVMe, and RDMA}. In \bibinfo{booktitle}{\emph{Proc of the
  Intl Conf on Management of Data}} \emph{(\bibinfo{series}{SIGMOD '22})}.
  \bibinfo{publisher}{ACM}, \bibinfo{pages}{685–699}.
\newblock


\end{thebibliography}
\end{document}